\documentclass{aa}  
\usepackage{silence}
\usepackage{graphicx}
\usepackage{txfonts}
\usepackage{booktabs}
\usepackage{lscape}
\usepackage{longtable}
\usepackage[colorlinks=true,     linkcolor=blue, citecolor=blue, filecolor=blue, urlcolor=blue]{hyperref}
\usepackage{xparse}
\usepackage[normalem]{ulem}
\usepackage{multirow}

\usepackage{listings}

\begin{document} 

   \title{The Gas-Phase Mass–Metallicity Relation of Dwarf Galaxies Across Large-Scale Environments Using the CAVITY Parent Sample}
   \titlerunning{The MZR of dwarf galaxies in cosmic web}

   \author{Bahar Bidaran\inst{\ref{ugr1}}
   \and Salvador Duarte Puertas \inst{\ref{ugr1},\ref{ugr2},\ref{laval}}
    \and Isabel Pérez \inst{\ref{ugr1},\ref{ugr2}}
       \and Almudena Zurita\inst{\ref{ugr1},\ref{ugr2}}  
    \and Daniel Espada\inst{\ref{ugr1},\ref{ugr2}}
    \and María Argudo-Fernández\inst{\ref{ugr1},\ref{ugr2}}
    \and Rubén García-Benito \inst{\ref{iaa}}
   \and Laura Sánchez-Menguiano\inst{\ref{ugr1},\ref{ugr2}}
    \and Simon Verley\inst{\ref{ugr1},\ref{ugr2}}
     \and Sebastián F. Sánchez\inst{\ref{iac}}
    \and Jes\'us Falc\'on-Barroso\inst{\ref{iac},\ref{ull}}
    \and Anna Ferr\'e-Mateu\inst{\ref{iac},\ref{ull}}
    \and Pedro Villalba-Gonzalez\inst{\ref{BC}}
    \and Andoni Jiménez\inst{\ref{ugr1}}
    \and Reynier F. Peletier\inst{\ref{kapteyn}}
    \and Tomás Ruiz-Lara\inst{\ref{ugr1},\ref{ugr2}}
}
   \institute{
   Dpto. de F\'{\i}sica Te\'orica y del Cosmos, Facultad de Ciencias (Edificio Mecenas), Universidad de Granada, E-18071, Granada, Spain\label{ugr1}
    \and Instituto Carlos I de F\'\i sica Te\'orica y Computacional, Universidad de Granada, E-18071, Granada, Spain\label{ugr2}
    \and D\'epartement de Physique, de G\'enie Physique et d’Optique, Universit\'e Laval, and Centre de Recherche en Astrophysique du Qu\'ebec (CRAQ), Québec, QC, G1V 0A6, Canada\label{laval}
    \and Instituto de Astrof\'isica de Andaluc\'ia - CSIC, Glorieta de la Astronomía s/n, 18008 Granada, Spain\label{iaa}
        \and Instituto de Astrof\'isica de Canarias, c/V\'ia L\'actea s/n, E-38205, La Laguna, Tenerife, Spain\label{iac}
            \and Departamento de Astrof\'isica, Universidad de La Laguna, E-38206, La Laguna, Tenerife, Spain\label{ull}
            \and Department of Physics and Astronomy, University of British Columbia, Vancouver, BC V6T 1Z1, Canada\label{BC}
                \and Kapteyn Astronomical Institute, University of Groningen, PO Box 800, 9700 AV Groningen, The Netherlands\label{kapteyn}
    }
    
   \date{Received Month Day, Year; accepted Month Day, Year}

  \abstract
  { {The gas-phase mass–metallicity relation (MZR) of galaxies shows a noticeable break in slope and an increased scatter at low stellar masses, suggesting that the physical processes governing chemical enrichment differ between dwarf and high-mass systems. Dwarf galaxies, in particular, are highly susceptible to both internal and environmental mechanisms due to their shallow potential wells. }}
  {{The primary aim of this work is to assess whether a single, universal MZR can describe dwarf galaxies across diverse large-scale environments, or whether systematic environmental variations emerge. To probe these, we examine the MZR and star formation rate (SFR) of dwarf galaxies with stellar masses in the range of 8.9 < log(M$_{\star}$/M$_{\odot}$)\,<\,9.5.  }}
    {Using optical spectra from the Sloan Digital Sky Survey, we measured the fluxes of key emission lines via the pyPipe3D full spectral fitting pipeline. Aperture-corrected fluxes, along with multiple metallicity indicators and calibrations, were used to derive the MZR and the SFR for 353, 311, and 22 dwarf galaxies located in voids, filaments, and clusters, respectively. }
     {We find a systematic variation in the MZR slope, steeper in voids (0.28\,$\pm$\,0.03) and progressively flatter in clusters (0.17\,$\pm$\,0.08), indicating a dependence of the MZR on the large-scale environment in this mass regime. When galaxies are separated by local density, no significant differences are observed between isolated and non-isolated dwarfs in voids. Isolated dwarf galaxies in filaments also exhibit properties similar to those of their counterparts in voids. However, non-isolated filament galaxies exhibit similar MZR slopes comparable to those of cluster dwarfs {and flatter slopes than their counterparts in voids.}}
   {{We report both large- and local-scale environmental dependencies in the gas-phase metallicity and in the slope of the MZR for dwarf galaxies. Consistent with the general consensus on the pre-processing of galaxies in filaments, our results indicate that the influence of the local environment becomes increasingly significant within the filamentary regions of the cosmic web, affecting the chemical enrichment and star formation activity of low-mass systems. These findings further suggest that a portion of the scatter commonly observed in the MZR of dwarf galaxies arises from environmental effects.}
}


   \maketitle

\section{Introduction}\label{Introduction}

The evolutionary history of galaxies is encoded in the chemical enrichment level of their interstellar medium (ISM). Upon their death, stars enrich their surrounding ISM with the metals produced via {nucleosynthesis and released through stellar winds and SN,} which are subsequently incorporated into later generations of stars. In parallel, the inflow of pristine gas from the {galaxy} surroundings, as well as the gas outflow due to SN feedback, stellar winds, and active galactic nuclei (AGNs), contribute to shaping the chemical composition of the ISM \citep[e.g.][]{2025Curti}. This continuing interplay forms the baryonic cycle and is reflected in fundamental scaling relations, such as the mass-metallicity relation (MZR). The MZR is a positive tight \citep[$\sim$0.1 dex scatter; e.g.][]{2025Curti} correlation between stellar mass and metallicity (of stars and gas) of galaxies \citep[e.g.][]{1979Lequeux,2004Tremonti,2013Andrews}. This trend persists even at redshifts ($z$) of 6 to 8, suggesting that the interplay between internal secular processes and gas flows was already influencing galaxy evolution at early epochs \citep[e.g. ][]{2014Zahid, 2019Cresci,  2023Nakajima, 2024Marszewski}. The MZR demonstrates that, to the first order, the metallicity of a galaxy strongly correlates with its stellar mass and reveals a noticeable break in the slope around log(M$_{\star}$/M$_{\odot}$)\,$\sim$\,10, with the relation becoming significantly steeper at lower masses \citep{2014Zahid,2016Guo}. 

Galaxies in the low-mass regime (log(M$_{\star}$/M$_{\odot}$)\,$\leq$\,9.5), commonly referred to as dwarf galaxies, are the most abundant population in the Universe. While the MZR of massive galaxies has been extensively studied, much less is known about its details in dwarf galaxies. For instance,  several studies have mentioned the significant scatter observed in the MZR, both in the gas-phase and stellar components \citep[e.g.][]{2004Tremonti, 2018Lian,2022ApJ...933...44C}, particularly within the stellar mass range typical of dwarf galaxies, and suggested that it may be related to the dependence of the MZR on secondary parameters such as star formation rate \citep[SFR; e.g.][]{2008Ellison, 2010Mannucci, 2019ApJ...878L...6S}, stellar age \citep[e.g.][]{2015Lian}, cold gas content \citep[e.g.][]{2020DeLucia}, or the environment \citep[e.g. ][]{2014Peng}. The scatter {in the low mass range} remains substantial even at $z$ = 2-5 \citep[e.g.][]{2005Savaglio, 2013Zahid,2023Li}. 

The observed change in slope and increased scatter in the MZR, a behaviour also reproduced in cosmological simulations \citep[e.g.][]{2024Marszewski}, provide compelling evidence that the dominant physical processes governing the chemical enrichment of the ISM and, consequently, galaxy evolution, differ between dwarf and massive galaxies. Simulations that include physically motivated supernova (SN) feedback models \citep{1986Dekel} demonstrate that dwarf galaxies, due to their shallow gravitational potentials and low-density ISM \citep{2012Hopkins}, can lose a significant fraction of their enriched gas through internal feedback. This loss is driven by a combination of ultraviolet radiation pressure, stellar winds, photoionisation heating, and SN-driven outflows \citep[e.g.][]{1974Larson, 2022Fraser-McKelvie}. Massive galaxies with deeper potential wells can retain their metal-enriched gas more efficiently \citep{2012Hopkins, 2019Roberts-Borsani}. In parallel, numerical simulations have shown that the efficiency of gas accretion onto galaxies smoothly changes around log(M$_{\star}$/M$_{\odot}$)\,$\sim$\,10.0 \citep{2005MNRAS.363....2K,2006Dekel, 2011MNRAS.414.2458V}. In more massive galaxies, infalling gas is typically shock-heated, leading to longer cooling times and less efficient accretion \citep[e.g.][]{2003Birnboim, 2005MNRAS.363....2K}. In contrast, dwarf galaxies are predominantly fed by cold-mode accretion, which allows for more efficient and continuous gas inflow. Besides, unlike their massive counterparts, the gas flows and evolution of dwarf galaxies can be significantly influenced by their surrounding environment.

\setlength{\tabcolsep}{15.pt}
\begin{table*}
\caption{\label{Samples_tab} Sample breakdown in void, filament, and cluster environments showing the number of galaxies retained after applying different criteria: stellar mass and local environment (Section~\ref{Sample}),  AGN/star-forming, S/N, and main sequence conditions in Section~\ref{analysis2}.}
\centering
\begin{tabular}{llccc}
\toprule
Condition & Subcategory & Void & Filament & Cluster \\
\midrule
Total number of galaxies & -- & 4866 & 15000 & 6189 \\
\hline
Galaxies with $8.9 \leq \log(M_\star/M_\odot) \leq 9.5$ & -- & 1754 & 4473 & 161 \\
\hline
Local environment & Isolated & 487 & 383 & -- \\
                           & Non-isolated & 400 & 450 & 161 \\
\hline

Final sample & Isolated & 252 & 193 & -- \\
                      & Non-isolated & 101 & 118 & 22 \\
\bottomrule
\end{tabular} \\
\vspace{0.2cm} 
\noindent
\end{table*}

The shallow potential wells of dwarf galaxies make them highly sensitive to external mechanisms that can significantly affect their gas content, SFR, and the chemical enrichment \citep{2014Boselli, 2016Williamson}. In galaxy clusters and groups, dwarf galaxies can begin to lose their cold gas content due to hydrodynamical interactions with the hot and dense intracluster medium (ICM), even out to distances of approximately three virial radii \citep{2014Cen}. These processes include thermal evaporation \citep{1977Cowie} and ram pressure stripping \citep[RPS;][]{1972Gunn, 2007Tonnesen}. In addition, \cite{2003Okamoto} and \cite{2005Lanzoni} show that galaxies plunging into the hot intergalactic medium (IGM) can undergo a passive process known as starvation \citep[e.g.][]{1980Larson}. In this scenario, star formation declines not due to gas removal, but rather because the accretion of fresh, pristine gas is significantly suppressed, leading to a more rapid quenching compared to galaxies in lower-density environments \citep[e.g.][]{2002Bekki, 2008MNRAS.387...79V, 2017MNRAS.466.3460V, 2024Baker}.

To date, the impact of the environment on the gas-phase metallicity of dwarf galaxies in clusters has been explored in several studies, yielding conflicting results. For example, investigations of irregular dwarf galaxies (dIrrs) in the Virgo cluster have not revealed any clear environmental trend \citep{2003Lee, 2007Vaduvescu}. More systematic analyses based on larger samples of star-forming dwarf galaxies in nearby clusters and groups suggest that these galaxies tend to be more metal-rich than their counterparts in lower-density environments \citep[e.g.][]{1996Skillman,2009Ellison}, and that, on average, the gas-phase metallicity of dwarfs varies depending on the host halo mass. For instance, in the massive Hercules and Coma clusters, \cite{2012Petropoulou} reported that the gas-phase metallicity of dwarf galaxies increases with decreasing cluster-centric distance. A similar trend has been observed for stellar metallicity in cluster dwarfs \citep[e.g.][]{2009Smith}.

Only recently has the dwarf galaxy population in other components of the cosmic web, such as filaments and voids, begun to receive increasing attention. For instance, \citet{2015Darvish} found that star-forming galaxies residing in filaments tend to be, on average, more metal-rich than their counterparts in the field\footnote{i.e. galaxies not associated with filaments.}. Similarly, \cite{2019Kraljic,2020Pandey, 2025Zarattini} reported, at a fixed stellar mass, a higher fraction of galaxies with redder stellar colors in filaments compared to the field and \cite{2017Kuutma} observed an increased elliptical-to-spiral ratio near filamentary structures.

In voids, the most underdense regions of the cosmic web occupying vast volumes with typical sizes around 35 h$^{-1}$ Mpc \citep[e.g. ][]{2001Peebles}, galaxies experience more sustained accretion of cold, pristine gas, which fuels ongoing star formation and helps regulate their metallicity, typically keeping it lower \citep{2013MNRAS.430.3017B,2020DeLucia}. \cite{2023A&A...680A.111D} showed that, on average, galaxies in voids and filaments exhibit lower stellar metallicities than their counterparts in clusters, and that in voids galaxies are more pristine, having formed their stars more slowly over cosmic time \citep{2023Natur.619..269D}. In the Lynx-Cancer void, \cite{2011Pustilnik} reported systematically lower gas-phase metallicities, with a typical offset of about 0.15 dex in dwarf galaxies. In contrast, \cite{2015Kreckel} found no significant difference in the gas-phase metallicities between dwarf galaxies in voids and isolated dwarfs situated in average-density environments, suggesting that large-scale structure may not have a strong influence on their chemical evolution, albeit based on a small sample of galaxies \citep[see also][]{2017Douglass}. Yet, \cite{2024A&A...691A.341T} demonstrated that the influence of both local- and large-scale environments is more pronounced in low-mass galaxies \citep[see also][]{2024Conrado,2024RMxAA..60..323S, 2025A&A...695A..84P}. Moreover, based on a sample of galaxies with intermediate to high stellar masses, \citet{2025Molina-Calzada} showed that galaxies residing in filaments exhibit higher gas-phase metallicities and lower SFRs compared to their counterparts in voids.

Current findings regarding the influence of large-scale structure on the chemical enrichment of dwarf galaxies remain both limited and, at times, contradictory, preventing a cohesive understanding of the processes at play. In this study, we extend previous efforts by deriving the MZR for star-forming dwarf galaxies residing in different components of the cosmic web. Specifically, we investigate how large-scale structure and local environmental conditions contribute to the observed scatter in the MZR. Furthermore, we examine whether a single, universal MZR can adequately represent dwarf galaxies across distinct large-scale environments, or if these populations exhibit systematic variations in the slope of the relation.

This paper is organised as follows. In Section \ref{Sample}, we outline the criteria used to construct the samples for this study. In Section \ref{Analysis}, we describe the data and methods used for the main analysis. In Section \ref{result}, we present the results for the MZR and SFRs of dwarf galaxies across different large-scale environments, and in Section \ref{discussion}, we discuss them. We summarise the key points of this work in Section \ref{conclusion}. In this paper, we assume a flat $\Lambda$CDM
cosmology with H$_{0}$\,=\,69.6\,[km~s$^{-1}$ Mpc$^{-1}$], $\Omega_{\rm M}$\,=\,0.286, and $\Omega_{\rm \Lambda}$\,=\,0.714. 
\begin{figure*}
\centering
\includegraphics[width=0.80\textwidth]{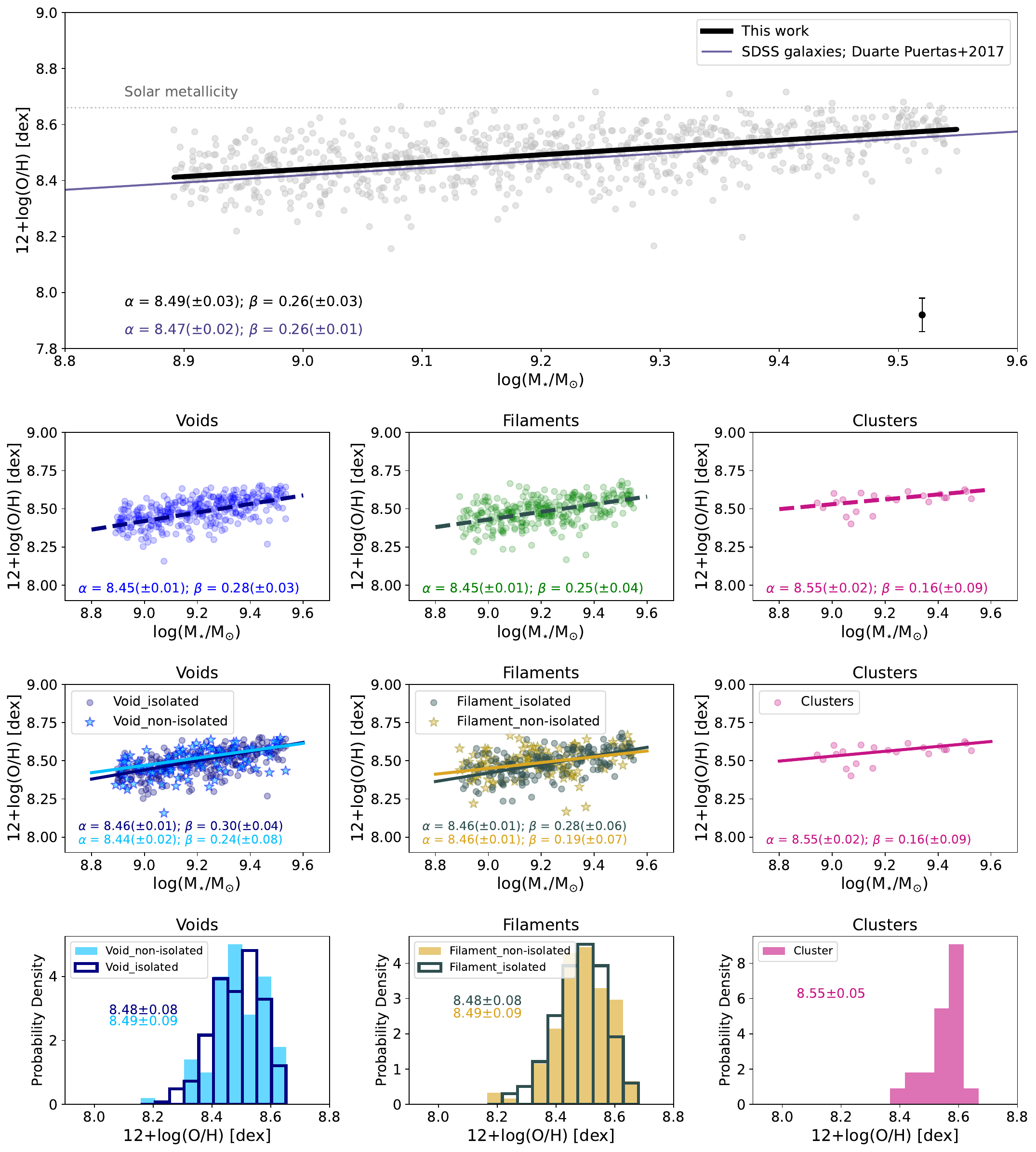}
\caption{MZR for star-forming dwarf galaxies in different parts of the cosmic web. \textit{Top panel:} MZR for dwarf galaxies, based on the combined samples from voids, filaments, and clusters, shown as grey points. The solid black line represents the linear fit to these data, while the solid purple line shows the {N2-based, aperture-corrected MZR derived using the large sample of SDSS galaxies from} \cite{2017A&A...599A..71D}. The intercepts {($\alpha$)}, defined at the mean stellar mass, and slopes {($\beta$)} of both relations are indicated in the lower-left corner of the plot. The uncertainties on individual data points range from 0.01 to 0.06. For reference, the solar gas-phase metallicity, {12+log(O/H)=8.69 dex \citep{2021Asplund}}, is indicated by a horizontal dotted grey line. {In the lower-right corner, the median of the
error bars for the entire sample is shown}. \textit{Second row:} MZR for star-forming dwarf galaxies, separated based on their large-scale environment. From left to right, the MZR of dwarf galaxies in voids, filaments, and clusters are shown in blue, green, and pink data points, respectively. \textit{Third row:} MZR of dwarf galaxies in each sample, separated into isolated and non-isolated systems. In the left-hand panel, non-isolated dwarf galaxies in voids are shown as light-blue data points, while in the middle panel, those in filaments are represented by orange data points. Isolated dwarf galaxies in voids (left-hand panel) and filaments (middle panel) are shown with dark blue and green data points, respectively. In all panels of the second and third rows, the linear fits to the data are shown as dashed lines for the three samples (voids, filaments, and clusters) and as solid lines when separated by local environment. \textit{Fourth row:} Distributions of metallicities for isolated and non-isolated dwarf galaxies. These distributions in voids, filaments, and clusters are shown from left to right, respectively, with colors matching those used in the panels of the third row. } 
\label{Fig2_MZR}
\end{figure*}

\section{Sample}\label{Sample}
The primary aim of this work is to investigate how the evolution of the MZR in star-forming dwarf galaxies is influenced by their location within distinct large-scale structures, namely clusters, filaments, walls, and voids. Given the comparable galaxy number densities observed in cosmic filaments and walls \citep[e.g.][]{2014Cautun}, we refer to these environments collectively as filaments. Therefore, we base the analysis on three primary samples of galaxies in voids, filaments, and clusters.

We constructed the sample of void dwarf galaxies based on the Calar Alto Void Integral-field Treasury surveY (CAVITY\footnote{\url{https://cavity.caha.es/}}) parent sample of 4866 galaxies in 15 nearby voids \citep{2024Perez}. The CAVITY parent sample is a refined and representative version of the \cite{2012Pan} catalogueue of galaxies in 1055 nearby voids. Galaxies in the CAVITY parent sample are entirely confined within the redshift range of 0.005\,$<$\,z\,$<$\,0.050 and the Sloan Digital Sky Survey (SDSS) footprint \citep[for more details see][]{2024Perez}. To cover the other parts of the large-scale structure, we used two catalogueues of \cite{2023Natur.619..269D}, containing 15000 and 6189 galaxies in filaments and clusters, respectively. The cluster sample in \cite{2023Natur.619..269D} is defined based on the \cite{2017Tempel} catalogueue. All three catalogueues are based on the 7th data release of SDSS (SDSS-DR7) and comprise galaxies with similar stellar mass and redshift ranges, equally affected by the SDSS-DR7 magnitude completeness limit at r-Petrosian\,$\le$\,17.77 mag \citep{2002Strauss}.

From these large samples, in the first step, we selected galaxies with 8.9\,$\leq$\,log(M$_{\star}$/M$_{\odot}$)\,$\leq$\,9.5. For this, we used stellar masses provided by MPA-JHU\footnote{available at \url{https://wwwmpa.mpa-garching.mpg.de/SDSS/}} that are estimated based on photometry and are complete in the redshift range of 0.005\,$<z<$\,0.220 \citep{2004Tremonti,2004Brinchmann}. Galaxies with stellar masses below log(M$_{\star}$/M$_{\odot}$)\,$\leq$\,8.9 are excluded from the sample, as the SDSS becomes incomplete at these lower masses in the redshift range investigated here. Furthermore, to exclude misclassified H\,{\scshape ii} regions, we removed targets with major axis ($D_{25}$) $<$ 10 kpc \citep{2025A&A...695A..84P}, following the same approach as in \cite{2024Guo}. These two criteria left us with 1754, 4473, and 161 dwarf galaxies in voids, filaments, and clusters, respectively (see Table~\ref{Samples_tab}). 

To further characterise the local environment of the dwarf galaxies, we subdivided those located in voids and filaments into two categories: isolated and non-isolated. This classification was based on an analysis of the velocity difference-projected distance space around each dwarf galaxy, using data from the NASA-Sloan Atlas (NSA\footnote{\url{https://www.sdss4.org/dr17/manga/manga-target-selection/nsa/}}). Following the criteria established in \cite{2015A&A...578A.110A}, we defined a galaxy as isolated if it had no neighbouring galaxies brighter than M$_{\rm r}$\,$\sim$\,$-$17\,mag within a projected radius of 1.5 Mpc and a line-of-sight velocity difference of $\Delta$V=500 km/s. These are common criteria adopted in different studies \citep[e.g.][]{2012Geha, 2021Dickey}. To avoid uncertainties in the determination of the local environment associated with very nearby galaxies (e.g. caused by large peculiar velocities and unreliable redshift-based distances in the nearby Universe), we excluded systems with redshifts below $z < 0.010$ \citep{2007Verley}. As a result, 487 galaxies in voids and 383 galaxies in filaments satisfied the isolation criteria. By definition, galaxies in clusters are not considered isolated; therefore, no further subdivision is applied to the cluster sample.

The isolated dwarf galaxies in voids and filaments represent a small fraction of the total dwarf galaxy population in the primary samples, within the defined mass range (i.e. 27\% and 8\% in voids and filaments, respectively). This results in large subsamples of non-isolated systems {(i.e. 1267 and 4090 dwarf galaxies in voids and filaments, respectively)}, for which a full analysis would be extensive and complex. {For this reason, we defined a subsample of non-isolated dwarf galaxies for each of the two environments.} We quantified their local surroundings using the local density, defined as the number of neighbouring galaxies within a projected distance of 1.5\,Mpc, based on \cite{2012Geha}, and a relative radial velocity of 500\,km/s, normalised by the corresponding volume \citep[see also][]{2007Verley, 2015A&A...578A.110A}. {For having similar sample sizes to the corresponding isolated subsamples}, a total of 400 and 450 non-isolated dwarf galaxies in voids and filaments were selected, respectively. Details of this sample selection are explained in Appendix~\ref{appendix0}.

All the selected dwarf galaxies, {in both main and comparison samples}, were visually inspected. Interacting dwarfs {(i.e. dwarf galaxies showing clear signs of interaction with other neighbour galaxies or disrupted tidal tails)}, large galaxies with wrong $D_{25}$, and misclassified H\,{\scshape ii} regions were excluded. 

\section{Data and analysis}\label{Analysis}
To measure emission lines, we utilised SDSS-DR7 optical spectra of dwarf galaxies, which were obtained with the 2.5 meter telescope at Apache Point Observatory (APO). Each SDSS spectrum, with a wavelength-dependent spectral resolution and resolving power of R = 1500 at $\lambda$ = 3800 \AA\, captures the integrated light within a 3 arcsecond fiber aperture, corresponding to the central $\sim$ 0.3–1.6 kpc of galaxies in the redshift range 0.01$<$z$<$0.05. We proceeded with their analysis using the steps described below.

\subsection{Fitting emission lines}\label{analysis1}
Each SDSS spectrum was first corrected for Galactic foreground extinction assuming R$_{\rm v}$ = 3.1 and using the \cite{1989Cardelli} Galactic extinction law. We retrieved the $E(B-V)$ values based on the galaxies' {equatorial coordinates} from the NASA/IPAC infrared science archive\footnote{\url{https://irsa.ipac.caltech.edu/frontpage/}}. Since the SDSS spectral resolution varies with wavelength, we convolved each spectrum with a wavelength-dependent kernel to bring the entire spectrum to the lowest resolution in the range. This is done by dividing the spectrum into small wavelength intervals (50 \AA\, windows), calculating the local instrumental dispersion in each window, and convolving each segment with a Gaussian kernel whose width compensates for the difference between the local and worst dispersion. This approach yields a final spectrum with a constant instrumental dispersion, which is essential for consistent analysis.

To measure emission line fluxes from SDSS spectra, we employed the full spectral fitting pipeline pyPipe3D \citep{2016RMxAA..52..171S, 2022Lacerda}, which simultaneously fits the stellar continuum and gas emission lines. The stellar continuum is modelled as a linear combination of single stellar population (SSP) models and subtracted to isolate the emission lines. For this purpose, we used a set of 1272 SSP models from \cite{2010Vazdekis, 2015Vazdekis}, based on the MILES stellar library \citep{2006MNRAS.371..703S,2007Cenarro,2011A&A...532A..95F} with a spectral resolution (full width at half maximum; FWHM) of 2.51 \AA. These models are constructed assuming a bi-modal initial mass function (IMF) with BASTI isochrones \citep{2009Pietrinferni} and a slope of 1.3 \citep{1996Vazdekis}, covering ages of 0.03 to 14 Gyr and metallicities ranging from [M/H]\,=\,–2.27 to +0.40\,dex. We performed the fit over the spectral range 380 to 710 nm, which includes the prominent forbidden and Balmer recombination lines, essential for the present analysis. We compared the measured emission lines with those reported for each galaxy in the MPA-JHU catalogueue and found overall good agreement between the two sets of measurements.

We corrected measured emission lines for the intrinsic reddening, assuming the same extinction law and R$_{\nu}$ value as explained above. Here we considered the Balmer decrement (H$_{\alpha}$/H$_{\beta}$)\,=\,2.86 \citep{1984Osterbrock}. To remove galaxies whose spectra could potentially be dominated by energy sources other than star formation, such as AGNs and shocks, we used the Baldwin, Phillips, Terlevich (BPT) diagnostic diagram \citep{1981Baldwin}. We also used the equivalent width (EW) of the H$\alpha$ line measured by pyPipe3D to construct the WHAN (EW$_{\rm H\alpha}$ vs. [NII]/H$\alpha$) diagram \citep{2011CidFernandes}. Taking advantage of kinematics that the code reports, we also explored the WHaD (EW$_{\rm H\alpha}$ vs. velocity dispersion of H$\alpha$) diagram \citep{2024A&A...682A..71S}. Galaxies falling on the AGN side in these three diagnostic tools were removed. Furthermore, to be assured of the robustness of metallicity and SFR estimates, we removed galaxies that did not have signal-to-noise (S/N)\,$>$\,3 in all the key emission lines, namely H$\beta$, [OIII]$\lambda$5007, H$\alpha$, and [NII]$\lambda$6584.

\setlength{\tabcolsep}{15.pt}
\begin{table*}
\caption{\label{Fits} Best-fit slope ($\beta$) and normalisation ($\alpha$), evaluated at the mean stellar mass $<$$\rm log(M_{\star})$$>$, describing the MZR and the SFR-$M_{\star}$ relations based on aperture-corrected values.}
\centering
\begin{tabular}{c c c c c}
\hline
LSS Sample & $\alpha_{\rm MZR, <log(M_{\star})>}$ &  $\beta_{\rm MZR}$& $\alpha_{\rm SFR-M_{\star}, <log(M_{\star})>}$ &$\beta_{\rm SFR-M_{\star}}$\\
\hline
\hline
\texttt{Voids} & 8.45 $\pm$ 0.01 &  0.28 $\pm$ 0.03& -0.58 $\pm$ 0.01 & 0.97 $\pm$ 0.08\\
\texttt{Filaments} & 8.45 $\pm$ 0.01 &  0.25 $\pm$ 0.04& -0.55 $\pm$ 0.01& 0.86 $\pm$ 0.08\\
\texttt{Clusters} & 8.55 $\pm$ 0.02 &  0.16 $\pm$ 0.09& -0.56 $\pm$ 0.06 &0.85$\pm$ 0.30\\
\hline 
\texttt{Voids\_isolated} & 8.48 $\pm$ 0.01 &  0.30 $\pm$ 0.04& -0.57 $\pm$ 0.02& 0.98 $\pm$0.09\\
\texttt{Voids\_non-isolated} & 8.49 $\pm$ 0.02 &  0.24 $\pm$ 0.08& -0.60 $\pm$ 0.02 & 0.93 $\pm$ 0.14\\
\texttt{Filament\_isolated} & 8.48 $\pm$ 0.01 &  0.28 $\pm$ 0.06& -0.52 $\pm$ 0.02& 0.79 $\pm$ 0.10\\
\texttt{Filament\_non-isolated} & 8.49 $\pm$ 0.01 &  0.19 $\pm$ 0.07& -0.60 $\pm$ 0.02 & 0.94 $\pm$ 0.13\\

\hline
\end{tabular}\\
\vspace{0.2cm} 
\noindent

\end{table*}

\subsection{Aperture-corrected metallicity and star formation rate}\label{analysis2}

The 3 arcsecond SDSS fiber diameter implies that galaxies in the local Universe (z < 0.22) are only partially mapped using this instrument. This limitation strongly affects the derivation of {global} quantities, such as the SFR and stellar mass, which depend on the area of the galaxy analysed. In addition, the radius covered by the SDSS fibre for each galaxy is different. To mitigate this systematic effect, we corrected all emission lines involved in the metallicity and SFR determinations for aperture effects, ensuring consistency in the portion of the galaxy analysed \citep[see][for further details]{2016ApJ...826...71I, 2022A&A...666A.186D}. For completeness, in Appendix \ref{appendix1}, we repeat the analysis using non–aperture-corrected values to demonstrate that this correction does not affect the trends discussed below.

To trace the gas-phase metallicity of dwarf galaxies, we used the oxygen abundance, expressed as 12$+$log(O/H). In the absence of deep spectroscopic data for these dwarf galaxies, which would allow for the detection of [OIII]$\lambda$4363 and metallicity measurements via the direct method, we rely instead on strong line methods based on calibrations of ratios of prominent emission lines. 

Among the choices we had, the N2 ([NII]$\lambda$6584/H$\alpha$) indicator was selected as the primary metallicity tracer in this study \citep{1979Pagel, 1994Calzetti}. The N2 measurements in this work lie well within the valid range of –2.5\,$<$\,N2\,$<$\,–0.3 \citep{2004Pettini}, supporting the robustness of the adopted approach. The N2 diagnostic offers a straightforward method for estimating gas-phase metallicity, particularly in low-resolution spectra, by using emission lines that are both strong and closely spaced in wavelength. This proximity reduces the impact of reddening corrections, enhancing the robustness of the derived metallicities \citep[e.g.][]{2021Zurita}. Moreover, relying on other emission lines, such as [OII]$\lambda$3727, would significantly reduce the sample size because of the high S/N required for accurate measurements.  In addition, the use of N2 ensures the reproducibility of the results. For completeness, other indicators, specifically O3N2 {and R23}, are also used, wherever applicable, to estimate the metallicities of galaxies in these three samples and to cross-check the trends discussed in the subsequent sections. The details of these measurements and the corresponding comparisons are provided in the Appendix \ref{appendix1}.

The aperture-corrected gas-phase metallicities reported in this study for star-forming dwarf galaxies in void, filament, and cluster environments were calculated following the linear calibration proposed by \cite{2004Pettini}:
\begin{equation}
\rm 12+log(O/H) = 8.90 + 0.57 \times N{\rm 2} \quad .
\end{equation}

The concept of the fundamental metallicity relation (FMR) was first introduced by \cite{2010Mannucci, 2010A&A...521L..53L}, who showed that the MZR observed in the local Universe is a consequence of a more general three-parameter relation linking stellar mass, gas-phase metallicity, and SFR. This relation is thought to arise primarily from the balance between the inflow of pristine gas and the outflow of metal-enriched material. Therefore, no comprehensive interpretation of the MZR can be drawn without examining the SFR–M$_{\rm \star}$\, relation. The aperture-corrected SFR has been derived in different studies \citep[e.g.][]{2003ApJ...599..971H,2004Brinchmann,2007ApJS..173..267S,2017A&A...599A..71D}. In this work, we estimated the SFR of each dwarf galaxy, following the aperture-free SFR–M$_\star$ relation proposed by \cite{2009Kennicutt}: 
\begin{equation}\label{SFR_eq}
\rm SFR(M_{\odot} yr^{-1}) = L_{H\alpha}/\eta_{H\alpha}, 
\end{equation}

\setlength{\parindent}{0pt}where $\eta_{H\alpha}$ is the aperture-correction factor applied to the observed H$\alpha$ luminosity, defined as the ratio between the total (aperture-free)  H$\alpha$ luminosity and the luminosity measured within the spectroscopic aperture. The correction is derived based on \cite{2017A&A...599A..71D} refinement of \cite{2003ApJ...599..971H} calibration and the H$\alpha$ luminosity ($L_{\rm H\alpha}$) was estimated based on aperture corrected values.
To ensure that the results are not contaminated by quiescent or starburst galaxies, both of which can exhibit unusually low or high gas-phase metallicities and alter the MZR, we adopted a similar approach to that of \cite{2020Bluck} for removing these outliers. Specifically, we included only galaxies with SFR offsets with respect to the main sequence \citep[i.e.][]{2017A&A...599A..71D} in the range of -0.5$<$$\Delta$\,SFR\,$<$0.5. Galaxies with -0.5$>$$\Delta$SFR typically fall in the green valley or are already quiescent, while those with $\Delta$SFR$>$0.5 are predominantly classified as starbursts.

The resulting sample after removal of low-S/N galaxies, AGN candidates, and quenched or star-bursting ones, comprising both isolated and non-isolated subsamples, consists of 353 dwarf galaxies in voids, 311 in filaments, and 22 in clusters (see Table~\ref{Samples_tab}). {The low number of cluster dwarf galaxies in the final sample is consistent with expectations, given that gas-rich, star-forming dwarfs exhibiting strong emission lines are uncommon in dense environments.}

\section{Results}\label{result}
\subsection{Gas-phase metallicity}\label{result1}

In Fig. \ref{Fig2_MZR} we present the MZR based on metallicities derived as described in Section \ref{analysis2}. In each panel of this figure, we fit a linear model to the data, representing the MZR within the stellar mass and environmental ranges under study (for more details see Appendix~\ref{Appendix11}). These fits are based on the low-mass regime (i.e. log(M$_{\star}$/M$_{\odot}$)$<$10.0), where previous studies have also reported a linear relation \citep[e.g.][]{2024Scholte}. Best-fit normalisation ($\alpha$), defined at the mean stellar mass $<$log$(\rm M_{\star})$$>$, and slope ($\beta$) parameters are reported within each panel as well as in Table \ref{Fits}.

In the first panel of Fig.~\ref{Fig2_MZR}, we show the MZR for dwarf galaxies, combining samples from voids, filaments, and clusters (in grey). The solid black line indicates the best-fit MZR for the full sample. For comparison, we include the MZR {for the large sample ($\sim$194000) of SDSS galaxies from \cite{2017A&A...599A..71D}}, derived using the N2 indicator and with the same calibration as the sample of voids, filaments and clusters. {Most of the measured metallicities} lie below the solar values of {12+log(O/H)\,=\,8.69 \citep{2021Asplund}}.

\begin{figure*}
\centering
\includegraphics[width=0.85\textwidth]{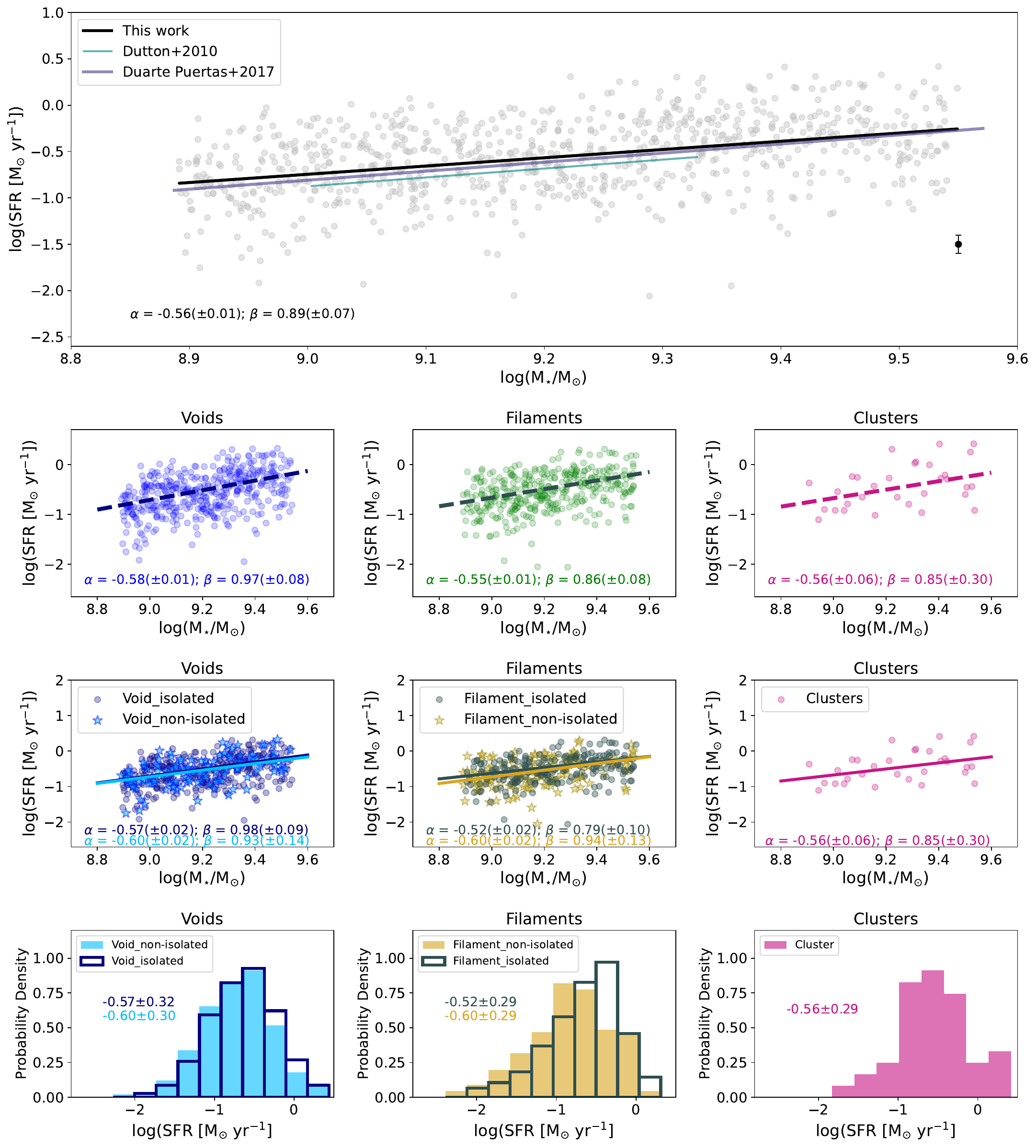}
\caption{SFR-$M_{\star}$ for dwarf galaxies in different parts of the cosmic web. \textit{Top panel:} SFR-$M_{\star}$ for dwarf galaxies based on the combined samples from voids, filaments, and clusters, shown as grey points. The intercepts and slopes are indicated in the lower-left
corner of the plot. The uncertainties in log(SFR) on individual data points on logarithmic scales range {from 0.02 to 0.1}. The solid purple and blue lines show the SFR-$M_{\star}$ relations by \cite{2017A&A...599A..71D} and \cite{2010Dutton}, respectively. \textit{Second to Fourth rows:} Same as in Fig.~\ref{Fig2_MZR} but for SFR values.} 
\label{Fig3_SFR}
\end{figure*}

In the second row, we separate dwarf galaxies given their location within the large-scale structure. From left to right, we show dwarf galaxies in voids (blue), filaments (green), and clusters (pink). Overall, galaxies in voids and filaments exhibit similar trends, while cluster dwarfs display a clear offset towards higher metallicities, indicating that they are generally more metal-rich than their counterparts in less dense environments. We also find a variation in the slope of the MZR, which changes from a steeper relation (0.28$\pm$0.03) in voids to a flatter one (0.16$\pm$0.08) in clusters, suggesting a possible dependence on the large-scale environment. 

In the third row of Fig. \ref{Fig2_MZR}, we further distinguish between isolated and non-isolated systems. Isolated dwarf galaxies in voids and filaments are shown as dark blue and dark green points, respectively, while non-isolated dwarfs in these environments are represented in light blue and orange. The gradual transition from a steep MZR in voids to a flatter relation in the higher-density cluster environments is also evident when considering these subsamples. Within both voids and filaments, non-isolated dwarfs exhibit a flatter MZR compared to their isolated counterparts (see Table~\ref{Fits}). The most notable change in slope occurs within the filament population. While isolated filament dwarfs show a  MZR slope consistent with that of void galaxies, non-isolated filament dwarfs display a slope similar to that found in cluster environments. Even among isolated systems, void dwarfs tend to have a slightly steeper MZR than those in filaments, although the difference is not statistically significant. Despite these variations in slope, the average gas-phase metallicities of isolated and non-isolated dwarfs in both voids and filaments converge around $\sim$8.48 dex, which remains lower than the average value observed in clusters (8.55 dex; the fourth row of Fig.~\ref{Fig2_MZR}).

We repeated this analysis for metallicities based on alternative indicators (i.e. O3N2 {and R23}) and calibrations \citep{2013Marino}. The corresponding slopes are listed in Table \ref{Other calibrators and indicators}. {The N2 values we measured are systematically 0.03 dex higher than those based on the O3N2 and {0.45 dex lower than those based on the R23}.} The trends described above remain consistent when these alternative indicators are used. We also repeated the analysis without imposing the S/N requirement and without using aperture-corrected values (Table~\ref{Aperture_effect_tab}), and obtained similar results in both cases. To the extent permitted by the data and its characteristics, these tests confirm that the observed trends are robust and not influenced by aperture corrections or by the adopted metallicity indicators or calibrations.

We also examined potential correlations between the gas-phase metallicity of dwarf galaxies and their position within the voids. To quantify this, we measured the distance of each galaxy from the centre of the void, approximating the void as a sphere with a volume equivalent to the actual void \citep{2024Perez}. The analysis revealed no significant correlation between the metallicity and the distance to the void centre. {Likewise, we found no meaningful correlation between the gas-phase metallicity of non-isolated void dwarf galaxies and their local volume density. }

\begin{figure*}
\centering
\includegraphics[width=0.8\textwidth]{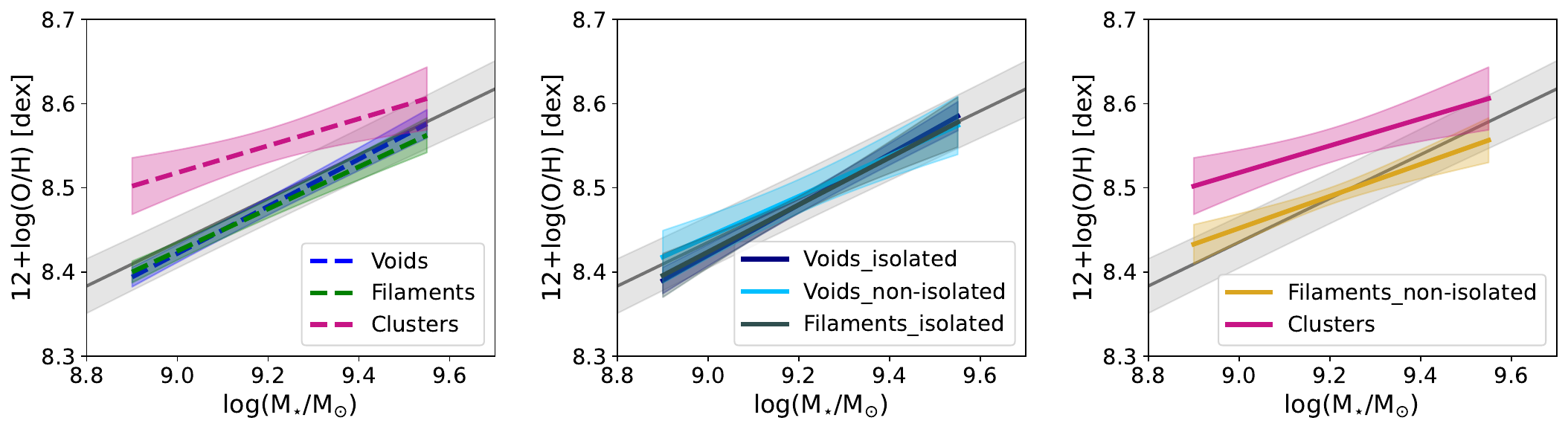}
\includegraphics[width=0.8\textwidth]{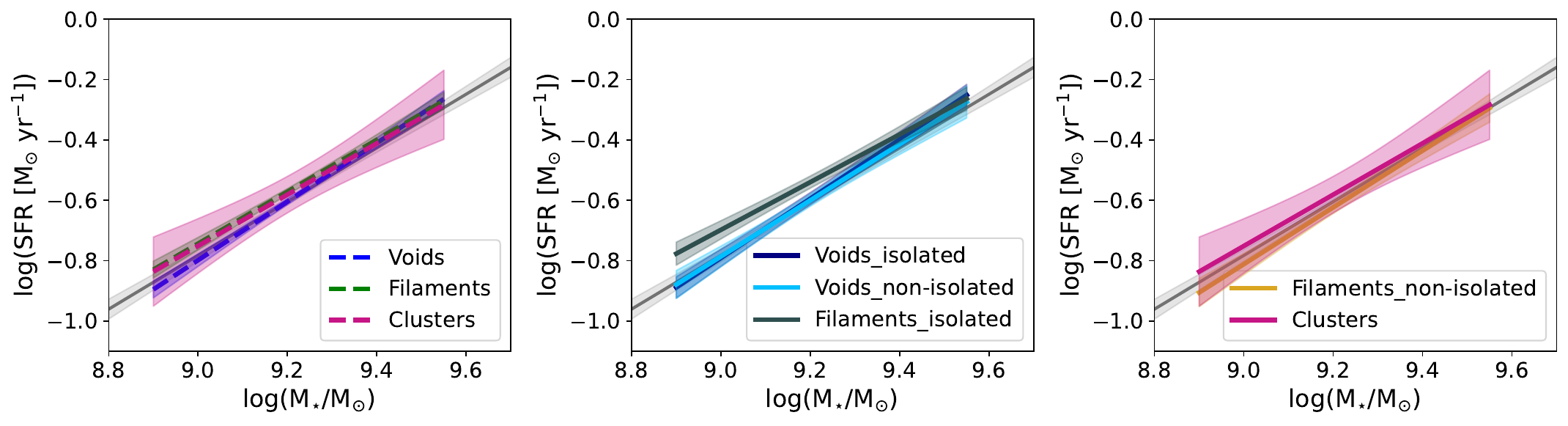}
\caption{\textit{Top row: } MZR for dwarf galaxies in different environments of the cosmic web, grouped according to the MZR slope measured in this study. \textit{Left-hand panel:} MZR for dwarf galaxies in voids, filaments, and clusters. \textit{Middle panel:} MZR for subsamples with similar steep slopes (i.e. $\sim$ 0.27), consisting of isolated dwarf galaxies in voids and filaments, as well as non-isolated dwarf galaxies in voids. \textit{Right-hand panel:} MZR for subsamples with flatter slopes (i.e. $\sim$ 0.17), consisting of non-isolated dwarf galaxies in filaments and cluster galaxies. All relations are color-coded as in Fig.\ref{Fig2_MZR}. The shaded regions around the best-fit lines indicate the 95 percent confidence intervals, derived from the distribution of fitted slopes and intercepts via bootstrap resampling (see Section~\ref{result1}). \textit{Bottom row:} Same as the top row but for the SFR-$M_{\star}$ relation. {In all panels, the solid black line represents the global MZR (top panels) and the SFR-$M_{\star}$ relation (bottom panels) fitted to the full sample of dwarf galaxies. }} 
\label{Fig4_slopes}
\end{figure*}

\subsection{Star formation rate }\label{result2}

In Fig.~\ref{Fig3_SFR}, we present the SFR-$M_{\star}$ relation for dwarf galaxies in voids, filaments, and clusters, following the same layout as Fig.~\ref{Fig2_MZR}. All SFR values are aperture-corrected. Linear fits to the SFR-$M_{\star}$ relations are obtained using the same method described in Appendix~\ref{Appendix11}. Similar to Fig.~\ref{Fig2_MZR}, the intercepts and slopes of the fits are also reported in each corresponding panel as well as in Table~\ref{Fits}.  

The top panel shows the relation for the combined sample of void, filament, and cluster galaxies. The SFR-$M_{\star}$ relation we reported for the sample (solid black line) is in good agreement with the aperture-corrected SFR-$M_{\star}$ relation for SDSS galaxies based on \cite{2003ApJ...599..971H} reported by \cite{2017A&A...599A..71D} (solid purple line) and is systematically above the relation reported by \cite{2010Dutton} (solid blue line); however still within the uncertainty range of both fits. 

In the second row of Fig.~\ref{Fig3_SFR} we present the SFR-$M_{\star}$ relation for dwarf galaxies in voids, filaments, and clusters, separately. Similar to the metallicity, we found no differences between the SFR-$M_{\star}$ relation in voids and filament dwarf galaxies. The relation for cluster dwarf galaxies shows substantial scatter but remains in good agreement with that for void and filament galaxies. Similarly, we compare this relation for isolated and non-isolated dwarf galaxies in the third row of Fig.~\ref{Fig3_SFR}. We found no differences in the SFR-$M_{\star}$ relation for isolated and non-isolated dwarf galaxies in voids and filaments. 

In the bottom row, we compare the overall distributions of galaxies across different local and large-scale environments, with mean values indicated in each panel. No significant difference is observed between isolated and non-isolated dwarf galaxies in voids. However, non-isolated dwarfs in filaments show a shift towards lower SFRs. The systematic offset between isolated and non-isolated dwarf galaxies in filaments raises the concern that the SFR–$M_{\star}$ relations reported above may be influenced by low statistics or different underlying stellar mass distribution. To investigate the latter, we performed a Kolmogorov–Smirnov (KS) test on the stellar mass distributions of the two filament subsamples. The test yields a KS statistic of 0.25 and a p-value of 3.5$\times$10$^{-6}$, indicating a statistically significant difference between the two distributions. Consequently, the observed SFR differences between isolated and non-isolated dwarf galaxies in filaments may, at least in part, be driven by differences in their stellar mass distributions. 

We repeated the main analysis of this work in Appendix~\ref{AppendixAB} using the full comparison sample of 1267 and 4090 star-forming dwarf galaxies in voids and filaments, respectively. Even with the improved statistics, the KS test still indicates a statistically significant difference between the stellar mass distributions of isolated and non-isolated dwarf galaxies in filaments. This trend is largely driven by the stellar mass distribution of isolated filament dwarf galaxies, which, given the isolation criteria applied, constitutes the full isolated sample in the parent catalogue and cannot be improved by increasing statistics. With the larger sample size, we find no significant difference in the SFR–$M_{\star}$ relation between non-isolated dwarf galaxies in voids and filaments. We found that the non-isolated dwarf galaxies in voids show higher SFR than their isolated counterparts; however, the large scatter among the isolated galaxies prevents us from drawing firm conclusions. Furthermore, the expanded statistics do not alter the MZR trends reported in Section~\ref{result1}.

\section{Discussion}\label{discussion}
The MZR is a cumulative relation that reflects how chemical enrichment in galaxies is governed by physical processes that establish a balance between gas flows and star formation. The MZR becomes notably steeper below a turnover mass of log(M$_{\star}$/M$_{\odot}$)$\sim$10, suggesting that the baryonic cycle in low-mass galaxies is governed by different processes than in more massive ones \citep[e.g.][]{2014Zahid, 2016Guo}. In this regime, the shallow gravitational potential well of dwarf galaxies renders them highly sensitive not only to internal feedback but also to environmental influences. Thus, environmental conditions are expected to leave measurable imprints on the metallicity of dwarf galaxies.

In this work, we demonstrate a systematic difference in the gas-phase metallicity of dwarf galaxies across cosmic environments. In the top rows of Fig.~\ref{Fig4_slopes}, we summarise the fits to the MZR and the SFR-$M_{\star}$ relation that we have measured on different samples from Section \ref{result1}. The comparison between three components of the cosmic web in the left-hand panel shows that cluster dwarfs (pink profile) are consistently more metal-rich than their counterparts in voids and filaments (blue and green lines, respectively).
In the middle and right-hand panels of Fig.~\ref{Fig4_slopes}, we separate the samples according to the MZR slopes derived in this study (see also Table~\ref{Fits}). The middle panels show the subsamples (i.e. dwarf galaxies in voids and isolated dwarfs in filaments) with steeper slopes, while the right-hand panels present those with flatter slopes (i.e. non-isolated dwarfs in filaments and those in clusters). By probing the local environment (within projected distances of 1.5 Mpc and $\Delta$V = 500 km/s), we find no significant differences between dwarfs in voids and isolated ones in filaments. Yet, non-isolated dwarfs in filaments already show MZR slopes comparable to those in clusters. 

\begin{figure}
\includegraphics[width=0.9\columnwidth]{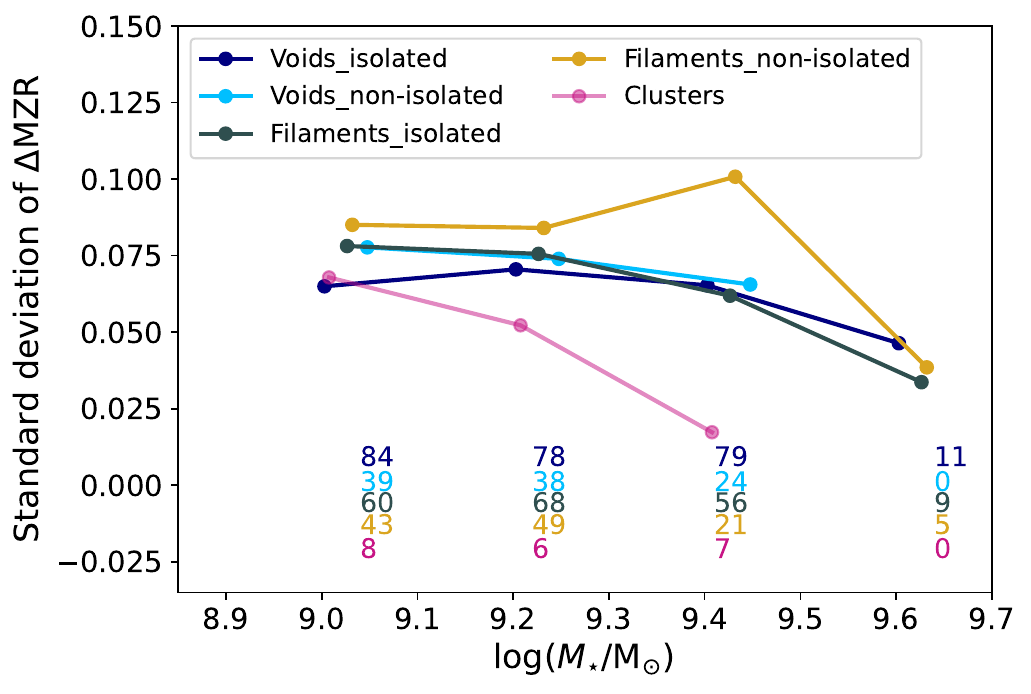}
\centering
\caption{{Standard deviation of the metallicity residuals ($\Delta MZR$) computed in stellar-mass bins for each environment. The galaxy counts per mass bin for each subsample are shown along the bottom of the plot.}} 
\label{standard deviation}
\end{figure}

{In Fig.~\ref{standard deviation}, we show the standard deviation of the metallicity residuals ($\Delta MZR$) computed in stellar-mass bins for each environment. {The reference relation is the MZR fitted to the full sample (i.e. y=8.49+0.26x) in Fig.~\ref{Fig2_MZR}, which we adopt as a common baseline, rather than environment-specific relations, in order to account for the possible impact of both local and large-scale environments in the observed scatter.} We exclude cluster dwarfs from this interpretation, as small-number statistics dominate their apparent behaviour and are therefore unreliable. Within each large-scale structure, dwarf galaxies residing in locally dense environments exhibit larger scatter around the MZR than those in less dense regions, and filament dwarfs show greater scatter than void dwarfs. We note that the standard deviation lies well within the typical systematic uncertainty of $\pm$0.2\,dex in metallicity derived using the N2 index \citep{2004Pettini}. While this limits the statistical certainty, the observed trend suggests that environmental effects contribute, at least in part, to the intrinsic scatter of the MZR at dwarf galaxy masses.} In the following, we interpret these results under the assumption of a constant IMF\footnote{The IMF defines the distribution of stellar masses at birth and influences the total metal yield returned to the ISM via stellar evolution.}.

\subsection{Voids}\label{discussion1}

Star-forming main-sequence dwarf galaxies in voids exhibit low metallicity and a steep MZR slope. These results are consistent with what \cite{2011Pustilnik} reported for a small sample of void dwarf galaxies. The range of absolute values and slopes measured for MZR and SFR-$M_{\star}$ relations for isolated and non-isolated dwarf galaxies in this environment is also very similar, indicating little to no apparent effect from the local environment in these least dense regions of the cosmic web. 

Voids are largely free from the complex environmental mechanisms, such as RPS and starvation. The cosmological hydrodynamical models of \cite{2008Finlator}, which incorporate different stellar feedback models, suggest that the ISM metal enrichment of galaxies in the absence of environmental mechanisms, is mainly governed by an "equilibrium" state between the inflow of metal-poor gas \citep[which dilutes the ISM;][]{2009Tolstoy}, star formation, and the outflow of metal-enriched gas driven by SN \citep[e.g.][]{1974Larson,2004DeLucia}. Cold-gas inflow rate is expected to be higher at $z$=0 for dwarf galaxies in voids than in higher-density regions of the cosmic web \citep[e.g.][]{2011Cen}. On the other hand, in void environments, dwarf galaxies are often unable to retain their outflowing gas, primarily because of their shallow gravitational potential wells. In other words, much of the metal-enriched material produced in these galaxies escapes the system powered by the energy deposited by the stellar winds and SN in the ISM (i.e. energy-driven wind scenario).  In the absence of environmental effects, \cite{2008Finlator} models yield Z $ \propto$ M$_\star^ {0.3}$ for the case of an energy-driven wind scenario \citep[i.e. V$ \propto$v$_{circ}^{-2}$ winds][]{2025Curti}, which is in good agreement with the measurements for isolated dwarf galaxies in voids ($\beta$ = 0.30 $\pm$0.04). The combined effects of higher cold-gas inflow and metal-enriched outflows naturally lead to lower gas-phase metallicities in void dwarfs.

One would expect non-isolated dwarf galaxies to be more easily displaced from equilibrium, for instance, through gas removal by the impact of the local environment. Both observational and theoretical studies suggest that galaxies in high-density environments, or those that are not isolated, tend to exhibit higher levels of metal enrichment \citep[e.g.][]{2012Petropoulou, 2012Pasquali}. However, in this work, we find no significant differences in gas-phase metallicities or SFRs between isolated and non-isolated dwarf galaxies in voids at fixed stellar mass. The MZR slope measured for non-isolated void dwarfs ($\beta$ = 0.24$\pm$0.08) is also in good agreement with the predictions of the "energy-driven wind" scenario from cosmological hydrodynamical simulations \citep{2008Finlator}. This suggests that, despite being non-isolated, void galaxies reside in environments that are not dense enough to trigger environmental mechanisms, such as RPS, strangulation, or tidal interactions, that would disrupt the equilibrium between gas accretion, star formation, and outflows. {Similarly, irrespective of their local environment, these galaxies may experience more sustained access to the gas inflows that fuel star formation, a trend supported by Illustris cosmological simulations, which show that the connectivity of low-mass galaxies to gas inflows depends on large-scale structure, with the lowest connectivity occurring in high-density environments \citep{2021A&A...649A.117G}. }This interpretation is further supported by the analysis of the H\,\textsc{i} content of void dwarf galaxies shown in Appendix~\ref{AppendixB}, which, although based on a smaller number of galaxies in both the isolated and non-isolated samples, reveals no systematic differences between the two.

\subsection{Filaments}\label{discussion2}
Unlike in voids, filament dwarf galaxies exhibit a more pronounced difference in the MZR slope between isolated and non-isolated systems, suggesting a potential transition in the influence or efficiency of local environmental mechanisms in filamentary regions. The MZR slope we measure for isolated filament dwarfs ($\beta$\,=\,0.28$\pm$0.06) closely matches that of isolated void galaxies and aligns well with predictions from models adopting an energy-driven wind scenario \citep{2008Finlator}. One possibility is that some of these systems are originally void galaxies now being gravitationally drawn towards filamentary structures, but have yet to undergo environmental transformation in filaments. Alternatively, the similarity between isolated filament dwarfs and those in voids (isolated and non-isolated) could be due to their location in low-density regions within the filamentary structure (e.g. away from the filamentary spine), where environment-driven mechanisms are not strong enough to significantly affect their properties \citep[e.g.][]{2017Chen,2017Kuutma,2021Winkel}. {The latter is a probable since \cite{2015A&A...578A.110A} have shown that about 60\% of isolated galaxies in filaments are mainly located in the outer part of these structures.}

In contrast, non-isolated dwarf galaxies in filaments exhibit a noticeably flatter MZR slope than isolated filament dwarfs, or void galaxies in any local environment (see Table~\ref{Fits}), resembling the trend observed for dwarf galaxies in clusters. These galaxies also tend to exhibit slightly higher metallicities and lower SFR than their isolated filament counterparts. Based on these trends, two possible scenarios may explain the findings for the MZR of these systems within the filamentary structure:
\begin{itemize}
    \item A significant fraction of non-isolated dwarf galaxies may reside within galaxy groups embedded in filaments. These group environments are known to drive environmental processing through mechanisms such as tidal interactions, mergers, strangulation, and RPS, all of which can influence ISM chemical enrichment \citep[e.g.][]{2019Sarron, 2025Finn}.
    \item Observational studies show that galaxy populations become progressively redder and more dominated by early-type morphologies as they approach the filament spine, where the local galaxy density increases \citep[e.g.][]{2016MNRAS.455..127M, 2017Kuutma, 2019Sarron}. Simulations also predict a rise in the number density of red, early-type galaxies towards filament nodes \citep[e.g.][]{2018Kraljic,2023Hasan,2024Bulichi}. Therefore, it is plausible that the non-isolated filament dwarfs discussed here are located near filament spines or nodes, where interactions with the denser environment have significantly impacted their gas-phase metallicity and SFR.
\end{itemize}

In higher-density environments, several authors \citep[e.g.][]{2011MNRAS.416.1354D} suggest that higher gas-phase metallicity in dwarf galaxies is a consequence of the environmental dependence of wind recycling. In low-density environments, such as voids, metal-enriched material can sufficiently escape the potential well of dwarf galaxies. However, in higher-density environments, stellar winds are slowed down, leading to faster recycling (see Section~\ref{discussion3}). This effect can be approximated by the no-wind case depicted in \cite{2011MNRAS.416.1354D} for which a MZR slope of 0.17 is expected. The predicted MZR slope is very similar to what we have measured for non-isolated dwarf galaxies in filaments ($\beta$ =\,0.19\, $\pm$0.07), and together with lower SFR, implies that in parts of the filaments, the environment is already strong enough to leave imprints on the ISM chemical enrichment of dwarf galaxies.

{The cosmic web is a dynamic structure in which galaxies, driven primarily by gravity, migrate from voids into filaments and eventually into clusters. Therefore, it is highly plausible that a present-day cluster dwarf galaxy retains imprints of environmental conditions it experienced in the past \citep[i.e. pre-processing;][]{2004Fujita}.  The present findings are consistent with and extend the results of  \cite{2015Darvish} and \cite{2021Chung, 2023Chung} regarding the pre-processing of cluster dwarf galaxies within filamentary structures before their infall onto clusters by revealing clear imprints of the local environment in filaments on both the MZR and SFR of dwarf galaxies. This suggests that the "nurture" effect of the large-scale environment on dwarf galaxies begins within the filamentary structures, particularly in locally denser regions such as near spines and nodes, before they enter the cluster region \citep[see also][]{2019AragonCalvo}.}

\subsection{Clusters}\label{discussion3}
The highest metallicities and the flattest MZR slope in the present study are measured for dwarf galaxies residing in clusters. The slope we derive for this population ($\beta$ = 0.16 $\pm$ 0.09), although based on a small number of cluster dwarf galaxies, is in good agreement with what \cite{2012Petropoulou} reported for dwarf galaxies in the central regions of the Coma cluster and theoretical predictions of \cite{2011MNRAS.416.1354D} for no stellar wind scenario, suggesting efficient recycling of enriched gas in these galaxies.

Dwarf galaxies in clusters are subject to a variety of environmental processes that act concurrently but over different timescales. Upon becoming satellites of a massive cluster, the inflow of pristine gas is suppressed, limiting the availability of fresh material for star formation; a process known as starvation or strangulation \citep[e.g.][]{1980Larson, 2002Bekki, 2008MNRAS.387...79V, 2015Peng}. Simultaneously, as these galaxies travel at high velocities through the dense and hot ICM, they undergo hydrodynamical effects such as thermal evaporation \citep{1977Cowie} and RPS \citep{1972Gunn}. Ram pressure, in particular, efficiently removes the cold gas content of a galaxy, directly impacting its fuel for future star formation. In addition to stripping, ram pressure can also compress the remaining gas, triggering or enhancing star formation by up to a factor of 2, particularly in the inner disk where the gravitational potential is deeper \citep[][]{1998Fujita,1999Fujita, 2001Vollmer}. This phenomenon has been observed among cluster dwarf galaxies. For example, \citet{2022Bidaran, 2023Bidaran} found a correlation between the recent peak of star formation and the infall time for a sample of nine recently accreted dwarf galaxies in Virgo. This enhanced star formation accelerates the chemical enrichment of the ISM and exhausts the remaining gas. At the same time, the hot, dense halo of the cluster inhibits outflows, which would otherwise escape the shallow potential wells of dwarf galaxies \citep[][]{2005Schindler,2006Kapferer,2009Kapferer}. Consequently, cluster dwarfs experience a brief but intense period of metal retention and rapid enrichment, followed by quenching.

These findings are consistent with those of \citet{2012Petropoulou}, who report an MZR slope of $\sim$\,0.19$\pm$\,0.03 for dwarf galaxies in the central regions of the Coma cluster. They also identify a systematic steepening of the MZR from the cluster core to its outskirts, reflecting changes in the environmental density \citep[see also][]{2016Williamson}. Although we do not have information on the spatial position or infall time of the cluster dwarf galaxies studied in this work, on a larger scale, we have observed a similar trend that, in dwarf galaxies metallicity increases and MZR slope becomes flatter as we move from under dense regions such as voids towards higher-density regions of the cosmic web (e.g. parts of filaments and clusters).

As noted in Section~\ref{analysis2}, from an initial sample of 161 cluster dwarf galaxies, only 22 satisfied the selection criteria (i.e. S/N\,$>$\,3 in all relevant emission lines and ongoing star formation as the dominant ionisation mechanism). This is expected, since a significant fraction of cluster dwarfs are quiescent and lack detectable emission lines. Theoretical models predict that cluster dwarfs lose their gas within one pericentre passage and are fully quenched within $\sim$5 Gyr after infall into the cluster \citep[e.g.][]{2020Rhee}. Hence, the star-forming dwarfs analysed here are special cases, in the sense that they are likely recent arrivals undergoing their final episode of star formation before transitioning to quiescence. This partly explains the scatter of SFR values, which is possibly enhanced due to ram pressure, compared to their counterparts in filaments and voids.

\section{Conclusion}\label{conclusion}
In this work, we investigated how the location within the cosmic web influences the evolution of the MZR in star-forming dwarf galaxies across three distinct environments: clusters, filaments, and voids. We examined the MZR both as a function of large-scale environment and local galaxy density. The results indicate that dwarf galaxies in denser environments, such as clusters and specific regions of filaments, tend to exhibit higher gas-phase metallicities than those in the sparest regions. Consequently, the MZR is flatter for cluster galaxies and non-isolated dwarfs in filaments, whereas star-forming dwarfs in voids and isolated ones in filaments follow a steeper relation. These findings suggest that part of the scatter commonly reported in the MZR of local dwarf galaxies could be due to environmental effects. The main results of this study can be summarised as follows:
\begin{itemize}
    \item Dwarf galaxies in clusters are systematically more metal-rich than their counterparts in filaments and voids. This result is consistent with \citet{2023A&A...680A.111D}, who reported a similar trend in stellar metallicities for galaxies of comparable stellar masses. These findings suggest that cluster dwarf galaxies are more chemically evolved than those in the least dense regions of the cosmic web, likely due to combinations of environmental effects that perturb the equilibrium governing gas accretion, star formation, and metal-enriched outflows in dwarf galaxies.
    \item Isolated dwarf galaxies in voids and filaments and non-isolated dwarf galaxies in voids exhibit a similar MZR slope ($\beta \sim 0.28$), consistent with the predictions of hydrodynamical simulations by \citet{2011MNRAS.416.1354D} for the scenario of energy-driven winds in galaxies that are in equilibrium between gas inflow, star formation, and outflow. This suggests that, although non-isolated dwarf galaxies in voids may experience some level of environmental influence, the local densities they inhabit are not sufficiently high to leave detectable imprints on their gas-phase metallicities. The results on the SFRs and H\,\textsc{i} content of void dwarf galaxies support this picture. {Our findings are in line with \cite{2025A&A...695A.256A}, who reported that galaxies with 9.0\,$<$log(M$_{\star}$/M$_{\odot}$)\,$<$\,10.0 exhibit a steeper MZR slope when isolated compared to those in pairs or triplets.}
    \item Non-isolated dwarf galaxies in filaments and those in clusters exhibit a similar MZR slope ($\beta \sim 0.17$), consistent with the no-wind case described by \citet{2011MNRAS.416.1354D}. The flatter MZR and systematically higher metallicities observed for dwarfs in high-density environments may result from the truncation of pristine gas inflows, which would otherwise dilute the ISM, as well as from the suppression of outflowing winds, both of which are expected to be more effective in dense regions \citep[e.g.][]{2005Schindler}. In addition,  environmental mechanisms such as RPS cannot only remove cold gas from satellites but also induce star formation episodes, thereby accelerating chemical enrichment through faster recycling.
    \item {The trends we observe in the MZR are consistent with those found in the SFR–$M_{\star}$ relation. In particular, in voids, the SFRs of isolated and non-isolated dwarf galaxies are similar within the uncertainties, mirroring the behaviour seen in the MZR. We cannot draw firm conclusions regarding the apparent difference in SFRs between isolated and non-isolated dwarf galaxies in filaments, where the latter appear lower, as this trend likely reflects differences in the underlying stellar mass distributions. A similar analysis applied to a larger sample of dwarf galaxies would be valuable to clarify these effects.}
 
\end{itemize}

{The results of this work are broadly consistent with the perspective offered by \cite{2025Molina-Calzada}, who show that the filament environment systematically influences the gas-phase metallicity of massive galaxies. This analysis relies on nitrogen-sensitive indicators (N2), which can be easily affected by ionisation sources other than star formation \citep[e.g.][]{2005MNRAS.361.1063P}. Although careful sample selection minimised AGN contamination, further exploration using additional metallicity indicators would strengthen these results. Further progress will benefit from deeper surveys with improved sensitivity to low-mass galaxies and enhanced reconstruction of three-dimensional structures, such as cosmic web filaments. Likewise, forthcoming results from current IFU surveys, including CAVITY, which will provide resolved metallicity gradients, will be crucial for testing these trends and developing a more complete view of the baryonic cycle in galaxies within the context of large-scale structure.}

\begin{acknowledgements}
{We thank the referee for the relevant suggestions that have certainly improved the final version. }We thank Fabio Bresolin for helpful discussions and suggestions. BB, SDP, IP, AZ, LSM, and SV acknowledge financial support from the Grant AST22-4.4, funded by Consejería de Universidad, Investigación e Innovación (Junta de Andalucía) and Gobierno de España and Unión Europea – NextGenerationEU, and by the research projects PID2020-113689GB-I00, PID2020-114414GB-I00, and PID2023-149578NB-I00 funded by the Spanish Ministry of Science and Innovation (MCIN/AEI/10.13039/501100011033) and by FEDER/UE. BB and LSM acknowledge support by the Munich Institute for Astro-, Particle and BioPhysics (MIAPbP), which is funded by the Deutsche Forschungsgemeinschaft (DFG, German Research Foundation) under Germany´s Excellence Strategy – EXC-2094 – 390783311. DE acknowledges support from a Beatriz Galindo senior fellowship (BG20/00224) from the Spanish Ministry of Science and Innovation. MAF acknowledges support from the Emergia program (EMERGIA$20_38888$) from Consejería de Universidad, Investigación e Innovación de la Junta de Andalucía. RGB acknowledges financial support from the Severo Ochoa grant CEX2021-001131-S funded by MCIN AEI / 10.13039/501100011033 and PID2022-141755NB-I00. JFB acknowledges support from the PID2022-140869NBI00 grant from the Spanish Ministry of Science and Innovation. AFM has received support from RYC2021-031099-I and PID2024-162088NB-I00 from the MICIN/AEI/10.13039/501100011033/ and UE/NextGenerationEU/PRTR. PVG acknowledges that the project that gave rise to these results received the support of a fellowship from “la Caixa” Foundation (ID 100010434), with the fellowship code B005800. TRL acknowledges support from Juan de la Cierva fellowship (IJC2020-043742-I) and Ramón y Cajal fellowship (RYC2023-043063-I, financed by MCIU/AEI/10.13039/501100011033 and by the FSE+). 
This research has made use of the NASA/IPAC Extragalactic Database, operated by the Jet Propulsion Laboratory
of the California Institute of Technology, under contract with the National Aeronautics and Space Administration. Funding for SDSS-III has been provided by the Alfred P. Sloan Foundation, the Participating Institutions, the National Science Foundation, and the U.S. Department of Energy Office of Science. The SDSS-III web site is \href{http://www.sdss3.org/}{http://www.sdss3.org/}. The SDSS-IV site is \href{http://www.sdss.org}{http://www.sdss.org}.

\end{acknowledgements}

\bibliographystyle{aa}
\bibliography{aa58889-26}

@ARTICLE{2023Natur.619..269D,
       author = {{Dom{\'\i}nguez-G{\'o}mez}, Jes{\'u}s and {P{\'e}rez}, Isabel and {Ruiz-Lara}, Tom{\'a}s and {Peletier}, Reynier F. and {S{\'a}nchez-Bl{\'a}zquez}, Patricia and {Lisenfeld}, Ute and {Falc{\'o}n-Barroso}, Jes{\'u}s and {Alc{\'a}zar-Laynez}, Manuel and {Argudo-Fern{\'a}ndez}, Mar{\'\i}a and {Bl{\'a}zquez-Calero}, Guillermo and {Courtois}, H{\'e}l{\`e}ne and {Duarte Puertas}, Salvador and {Espada}, Daniel and {Florido}, Estrella and {Garc{\'\i}a-Benito}, Rub{\'e}n and {Jim{\'e}nez}, Andoni and {Kreckel}, Kathryn and {Rela{\~n}o}, M{\'o}nica and {S{\'a}nchez-Menguiano}, Laura and {van der Hulst}, Thijs and {van de Weygaert}, Rien and {Verley}, Simon and {Zurita}, Almudena},
        title = "{Galaxies in voids assemble their stars slowly}",
      journal = {\nat},
         year = 2023,
        month = jul,
       volume = {619},
       number = {7969},
        pages = {269-271},
          doi = {10.1038/s41586-023-06109-1},
archivePrefix = {arXiv},
       eprint = {2306.16818},
 primaryClass = {astro-ph.GA},
       adsurl = {https://ui.adsabs.harvard.edu/abs/2023Natur.619..269D}
}

@ARTICLE{2014Cautun,
       author = {{Cautun}, Marius and {van de Weygaert}, Rien and {Jones}, Bernard J.~T. and {Frenk}, Carlos S.},
        title = "{Evolution of the cosmic web}",
      journal = {\mnras},
         year = 2014,
        month = jul,
       volume = {441},
       number = {4},
        pages = {2923-2973},
          doi = {10.1093/mnras/stu768},
archivePrefix = {arXiv},
       eprint = {1401.7866},
 primaryClass = {astro-ph.CO},
       adsurl = {https://ui.adsabs.harvard.edu/abs/2014MNRAS.441.2923C}
}

@ARTICLE{2009Smith,
       author = {{Smith}, Russell J. and {Lucey}, John R. and {Hudson}, Michael J. and {Allanson}, Steven P. and {Bridges}, Terry J. and {Hornschemeier}, Ann E. and {Marzke}, Ronald O. and {Miller}, Neal A.},
        title = "{A spectroscopic survey of dwarf galaxies in the Coma cluster: stellar populations, environment and downsizing}",
      journal = {\mnras},
         year = 2009,
        month = feb,
       volume = {392},
       number = {4},
        pages = {1265-1294},
          doi = {10.1111/j.1365-2966.2008.14180.x},
archivePrefix = {arXiv},
       eprint = {0810.5558},
 primaryClass = {astro-ph},
       adsurl = {https://ui.adsabs.harvard.edu/abs/2009MNRAS.392.1265S}
}

@ARTICLE{2011Pustilnik,
       author = {{Pustilnik}, S.~A. and {Tepliakova}, A.~L. and {Kniazev}, A. Yu.},
        title = "{Study of galaxies in the Lynx-Cancer void. II. Element abundances}",
      journal = {Astrophysical Bulletin},
         year = 2011,
        month = jul,
       volume = {66},
       number = {3},
        pages = {255-292},
          doi = {10.1134/S1990341311030011},
archivePrefix = {arXiv},
       eprint = {1108.4850},
 primaryClass = {astro-ph.CO},
       adsurl = {https://ui.adsabs.harvard.edu/abs/2011AstBu..66..255P}
}

@ARTICLE{2015Darvish,
       author = {{Darvish}, Behnam and {Mobasher}, Bahram and {Sobral}, David and {Hemmati}, Shoubaneh and {Nayyeri}, Hooshang and {Shivaei}, Irene},
        title = "{Spectroscopic Study of Star-forming Galaxies in Filaments and the Field at z \raisebox{-0.5ex}\textasciitilde 0.5: Evidence for Environmental Dependence of Electron Density}",
      journal = {\apj},
         year = 2015,
        month = dec,
       volume = {814},
       number = {2},
          eid = {84},
        pages = {84},
          doi = {10.1088/0004-637X/814/2/84},
archivePrefix = {arXiv},
       eprint = {1510.05009},
 primaryClass = {astro-ph.GA},
       adsurl = {https://ui.adsabs.harvard.edu/abs/2015ApJ...814...84D}
}

@ARTICLE{2023Nakajima,
       author = {{Nakajima}, Kimihiko and {Ouchi}, Masami and {Isobe}, Yuki and {Harikane}, Yuichi and {Zhang}, Yechi and {Ono}, Yoshiaki and {Umeda}, Hiroya and {Oguri}, Masamune},
        title = "{JWST Census for the Mass-Metallicity Star Formation Relations at z = 4-10 with Self-consistent Flux Calibration and Proper Metallicity Calibrators}",
      journal = {\apjs},
         year = 2023,
        month = dec,
       volume = {269},
       number = {2},
          eid = {33},
        pages = {33},
          doi = {10.3847/1538-4365/acd556},
archivePrefix = {arXiv},
       eprint = {2301.12825},
 primaryClass = {astro-ph.GA},
       adsurl = {https://ui.adsabs.harvard.edu/abs/2023ApJS..269...33N}
}

@ARTICLE{2025Curti,
       author = {{Curti}, Mirko},
        title = "{The Chemical Evolution of Galaxies}",
      journal = {arXiv e-prints},
         year = 2025,
        month = apr,
          eid = {arXiv:2504.08933},
        pages = {arXiv:2504.08933},
          doi = {10.48550/arXiv.2504.08933},
archivePrefix = {arXiv},
       eprint = {2504.08933},
 primaryClass = {astro-ph.GA},
       adsurl = {https://ui.adsabs.harvard.edu/abs/2025arXiv250408933C}
}

@ARTICLE{2023Bidaran,
       author = {{Bidaran}, Bahar and {La Barbera}, Francesco and {Pasquali}, Anna and {van de Ven}, Glenn and {Peletier}, Reynier and {Falc{\'o}n-Barroso}, Jesus and {Gadotti}, Dimitri A. and {Sybilska}, Agnieszka and {Grebel}, Eva K.},
        title = "{On the accretion of a new group of galaxies onto Virgo - III. The stellar population radial gradients of dEs}",
      journal = {\mnras},
         year = 2023,
        month = nov,
       volume = {525},
       number = {3},
        pages = {4329-4346},
          doi = {10.1093/mnras/stad2546},
archivePrefix = {arXiv},
       eprint = {2308.16768},
 primaryClass = {astro-ph.GA},
       adsurl = {https://ui.adsabs.harvard.edu/abs/2023MNRAS.525.4329B}
}

@ARTICLE{2022Bidaran,
       author = {{Bidaran}, Bahar and {La Barbera}, Francesco and {Pasquali}, Anna and {Peletier}, Reynier and {van de Ven}, Glenn and {Grebel}, Eva K. and {Falc{\'o}n-Barroso}, Jesus and {Sybilska}, Agnieszka and {Gadotti}, Dimitri A. and {Coccato}, Lodovico},
        title = "{On the accretion of a new group of galaxies on to Virgo - II. The effect of pre-processing on the stellar population content of dEs}",
      journal = {\mnras},
         year = 2022,
        month = sep,
       volume = {515},
       number = {3},
        pages = {4622-4638},
          doi = {10.1093/mnras/stac2005},
archivePrefix = {arXiv},
       eprint = {2207.06977},
 primaryClass = {astro-ph.GA},
       adsurl = {https://ui.adsabs.harvard.edu/abs/2022MNRAS.515.4622B}
}

@ARTICLE{2014Peng,
       author = {{Peng}, Ying-jie and {Maiolino}, Roberto},
        title = "{The dependence of the galaxy mass-metallicity relation on environment and the implied metallicity of the IGM}",
      journal = {\mnras},
         year = 2014,
        month = feb,
       volume = {438},
       number = {1},
        pages = {262-270},
          doi = {10.1093/mnras/stt2175},
archivePrefix = {arXiv},
       eprint = {1311.1816},
 primaryClass = {astro-ph.CO},
       adsurl = {https://ui.adsabs.harvard.edu/abs/2014MNRAS.438..262P}
}

@ARTICLE{2013Zahid,
       author = {{Zahid}, H. Jabran and {Geller}, Margaret J. and {Kewley}, Lisa J. and {Hwang}, Ho Seong and {Fabricant}, Daniel G. and {Kurtz}, Michael J.},
        title = "{The Chemical Evolution of Star-forming Galaxies over the Last 11 Billion Years}",
      journal = {\apjl},
         year = 2013,
        month = jul,
       volume = {771},
       number = {2},
          eid = {L19},
        pages = {L19},
          doi = {10.1088/2041-8205/771/2/L19},
archivePrefix = {arXiv},
       eprint = {1303.5987},
 primaryClass = {astro-ph.CO},
       adsurl = {https://ui.adsabs.harvard.edu/abs/2013ApJ...771L..19Z}
}

@ARTICLE{2005Savaglio,
       author = {{Savaglio}, S. and {Glazebrook}, K. and {Le Borgne}, D. and {Juneau}, S. and {Abraham}, R.~G. and {Chen}, H. -W. and {Crampton}, D. and {McCarthy}, P.~J. and {Carlberg}, R.~G. and {Marzke}, R.~O. and {Roth}, K. and {J{\o}rgensen}, I. and {Murowinski}, R.},
        title = "{The Gemini Deep Deep Survey. VII. The Redshift Evolution of the Mass-Metallicity Relation}",
      journal = {\apj},
         year = 2005,
        month = dec,
       volume = {635},
       number = {1},
        pages = {260-279},
          doi = {10.1086/497331},
archivePrefix = {arXiv},
       eprint = {astro-ph/0508407},
 primaryClass = {astro-ph},
       adsurl = {https://ui.adsabs.harvard.edu/abs/2005ApJ...635..260S}
}

@ARTICLE{2019Cresci,
       author = {{Cresci}, G. and {Mannucci}, F. and {Curti}, M.},
        title = "{Fundamental metallicity relation in CALIFA, SDSS-IV MaNGA, and high-z galaxies}",
      journal = {\aap},
         year = 2019,
        month = jul,
       volume = {627},
          eid = {A42},
        pages = {A42},
          doi = {10.1051/0004-6361/201834637},
archivePrefix = {arXiv},
       eprint = {1811.06015},
 primaryClass = {astro-ph.GA},
       adsurl = {https://ui.adsabs.harvard.edu/abs/2019A&A...627A..42C}
}

@ARTICLE{2022A&A...666A.186D,
       author = {{Duarte Puertas}, S. and {Vilchez}, J.~M. and {Iglesias-P{\'a}ramo}, J. and {Moll{\'a}}, M. and {P{\'e}rez-Montero}, E. and {Kehrig}, C. and {Pilyugin}, L.~S. and {Zinchenko}, I.~A.},
        title = "{Mass-metallicity and star formation rate in galaxies: A complex relation tuned to stellar age}",
      journal = {\aap},
         year = 2022,
        month = oct,
       volume = {666},
          eid = {A186},
        pages = {A186},
          doi = {10.1051/0004-6361/202141571},
archivePrefix = {arXiv},
       eprint = {2205.01203},
 primaryClass = {astro-ph.GA},
       adsurl = {https://ui.adsabs.harvard.edu/abs/2022A&A...666A.186D}
}

@ARTICLE{2017A&A...599A..71D,
       author = {{Duarte Puertas}, S. and {Vilchez}, J.~M. and {Iglesias-P{\'a}ramo}, J. and {Kehrig}, C. and {P{\'e}rez-Montero}, E. and {Rosales-Ortega}, F.~F.},
        title = "{Aperture-free star formation rate of SDSS star-forming galaxies}",
      journal = {\aap},
         year = 2017,
        month = mar,
       volume = {599},
          eid = {A71},
        pages = {A71},
          doi = {10.1051/0004-6361/201629044},
archivePrefix = {arXiv},
       eprint = {1611.07935},
 primaryClass = {astro-ph.GA},
       adsurl = {https://ui.adsabs.harvard.edu/abs/2017A&A...599A..71D}
}

@ARTICLE{2003ApJ...599..971H,
       author = {{Hopkins}, A.~M. and {Miller}, C.~J. and {Nichol}, R.~C. and {Connolly}, A.~J. and {Bernardi}, M. and {G{\'o}mez}, P.~L. and {Goto}, T. and {Tremonti}, C.~A. and {Brinkmann}, J. and {Ivezi{\'c}}, {\v{Z}}. and {Lamb}, D.~Q.},
        title = "{Star Formation Rate Indicators in the Sloan Digital Sky Survey}",
      journal = {\apj},
         year = 2003,
        month = dec,
       volume = {599},
       number = {2},
        pages = {971-991},
          doi = {10.1086/379608},
archivePrefix = {arXiv},
       eprint = {astro-ph/0306621},
 primaryClass = {astro-ph},
       adsurl = {https://ui.adsabs.harvard.edu/abs/2003ApJ...599..971H}
}

@ARTICLE{2007ApJS..173..267S,
       author = {{Salim}, Samir and {Rich}, R. Michael and {Charlot}, St{\'e}phane and {Brinchmann}, Jarle and {Johnson}, Benjamin D. and {Schiminovich}, David and {Seibert}, Mark and {Mallery}, Ryan and {Heckman}, Timothy M. and {Forster}, Karl and {Friedman}, Peter G. and {Martin}, D. Christopher and {Morrissey}, Patrick and {Neff}, Susan G. and {Small}, Todd and {Wyder}, Ted K. and {Bianchi}, Luciana and {Donas}, Jos{\'e} and {Lee}, Young-Wook and {Madore}, Barry F. and {Milliard}, Bruno and {Szalay}, Alex S. and {Welsh}, Barry Y. and {Yi}, Sukyoung K.},
        title = "{UV Star Formation Rates in the Local Universe}",
      journal = {\apjs},
         year = 2007,
        month = dec,
       volume = {173},
       number = {2},
        pages = {267-292},
          doi = {10.1086/519218},
archivePrefix = {arXiv},
       eprint = {0704.3611},
 primaryClass = {astro-ph},
       adsurl = {https://ui.adsabs.harvard.edu/abs/2007ApJS..173..267S}
}

@ARTICLE{2020Bluck,
       author = {{Bluck}, Asa F.~L. and {Maiolino}, Roberto and {S{\'a}nchez}, Sebastian F. and {Ellison}, Sara L. and {Thorp}, Mallory D. and {Piotrowska}, Joanna M. and {Teimoorinia}, Hossen and {Bundy}, Kevin A.},
        title = "{Are galactic star formation and quenching governed by local, global, or environmental phenomena?}",
      journal = {\mnras},
         year = 2020,
        month = feb,
       volume = {492},
       number = {1},
        pages = {96-139},
          doi = {10.1093/mnras/stz3264},
archivePrefix = {arXiv},
       eprint = {1911.08857},
 primaryClass = {astro-ph.GA},
       adsurl = {https://ui.adsabs.harvard.edu/abs/2020MNRAS.492...96B}
}

@ARTICLE{2009Kennicutt,
       author = {{Kennicutt}, Jr., Robert C. and {Hao}, Cai-Na and {Calzetti}, Daniela and {Moustakas}, John and {Dale}, Daniel A. and {Bendo}, George and {Engelbracht}, Charles W. and {Johnson}, Benjamin D. and {Lee}, Janice C.},
        title = "{Dust-corrected Star Formation Rates of Galaxies. I. Combinations of H{\ensuremath{\alpha}} and Infrared Tracers}",
      journal = {\apj},
         year = 2009,
        month = oct,
       volume = {703},
       number = {2},
        pages = {1672-1695},
          doi = {10.1088/0004-637X/703/2/1672},
archivePrefix = {arXiv},
       eprint = {0908.0203},
 primaryClass = {astro-ph.CO},
       adsurl = {https://ui.adsabs.harvard.edu/abs/2009ApJ...703.1672K}
}

@ARTICLE{1994Calzetti,
       author = {{Calzetti}, Daniela and {Kinney}, Anne L. and {Storchi-Bergmann}, Thaisa},
        title = "{Dust Extinction of the Stellar Continua in Starburst Galaxies: The Ultraviolet and Optical Extinction Law}",
      journal = {\apj},
         year = 1994,
        month = jul,
       volume = {429},
        pages = {582},
          doi = {10.1086/174346},
       adsurl = {https://ui.adsabs.harvard.edu/abs/1994ApJ...429..582C}
}

@ARTICLE{1979Pagel,
       author = {{Pagel}, B.~E.~J. and {Edmunds}, M.~G. and {Blackwell}, D.~E. and {Chun}, M.~S. and {Smith}, G.},
        title = "{On the composition of H II regions in southern galaxies - I. NGC 300 and 1365.}",
      journal = {\mnras},
         year = 1979,
        month = oct,
       volume = {189},
        pages = {95-113},
          doi = {10.1093/mnras/189.1.95},
       adsurl = {https://ui.adsabs.harvard.edu/abs/1979MNRAS.189...95P}
}

@ARTICLE{2004Pettini,
       author = {{Pettini}, Max and {Pagel}, Bernard E.~J.},
        title = "{[OIII]/[NII] as an abundance indicator at high redshift}",
      journal = {\mnras},
         year = 2004,
        month = mar,
       volume = {348},
       number = {3},
        pages = {L59-L63},
          doi = {10.1111/j.1365-2966.2004.07591.x},
archivePrefix = {arXiv},
       eprint = {astro-ph/0401128},
 primaryClass = {astro-ph},
       adsurl = {https://ui.adsabs.harvard.edu/abs/2004MNRAS.348L..59P}
}

@ARTICLE{2021Zurita,
       author = {{Zurita}, A. and {Florido}, E. and {Bresolin}, F. and {P{\'e}rez-Montero}, E. and {P{\'e}rez}, I.},
        title = "{Bar effect on gas-phase abundance gradients. I. Data sample and chemical abundances}",
      journal = {\mnras},
         year = 2021,
        month = jan,
       volume = {500},
       number = {2},
        pages = {2359-2379},
          doi = {10.1093/mnras/staa2246},
archivePrefix = {arXiv},
       eprint = {2007.12289},
 primaryClass = {astro-ph.GA},
       adsurl = {https://ui.adsabs.harvard.edu/abs/2021MNRAS.500.2359Z}
}

@ARTICLE{2015Kreckel,
       author = {{Kreckel}, K. and {Croxall}, K. and {Groves}, B. and {van de Weygaert}, R. and {Pogge}, R.~W.},
        title = "{The Metallicity of Void Dwarf Galaxies}",
      journal = {\apjl},
         year = 2015,
        month = jan,
       volume = {798},
       number = {1},
          eid = {L15},
        pages = {L15},
          doi = {10.1088/2041-8205/798/1/L15},
archivePrefix = {arXiv},
       eprint = {1410.5821},
 primaryClass = {astro-ph.GA},
       adsurl = {https://ui.adsabs.harvard.edu/abs/2015ApJ...798L..15K}
}

@ARTICLE{2012Petropoulou,
       author = {{Petropoulou}, V. and {V{\'\i}lchez}, J. and {Iglesias-P{\'a}ramo}, J.},
        title = "{Environmental Effects on the Metal Enrichment of Low-mass Galaxies in Nearby Clusters}",
      journal = {\apj},
         year = 2012,
        month = apr,
       volume = {749},
       number = {2},
          eid = {133},
        pages = {133},
          doi = {10.1088/0004-637X/749/2/133},
archivePrefix = {arXiv},
       eprint = {1202.4164},
 primaryClass = {astro-ph.CO},
       adsurl = {https://ui.adsabs.harvard.edu/abs/2012ApJ...749..133P}
}

@ARTICLE{2009Ellison,
       author = {{Ellison}, Sara L. and {Simard}, Luc and {Cowan}, Nicolas B. and {Baldry}, Ivan K. and {Patton}, David R. and {McConnachie}, Alan W.},
        title = "{The mass-metallicity relation in galaxy clusters: the relative importance of cluster membership versus local environment}",
      journal = {\mnras},
         year = 2009,
        month = jul,
       volume = {396},
       number = {3},
        pages = {1257-1272},
          doi = {10.1111/j.1365-2966.2009.14817.x},
archivePrefix = {arXiv},
       eprint = {0903.4684},
 primaryClass = {astro-ph.CO},
       adsurl = {https://ui.adsabs.harvard.edu/abs/2009MNRAS.396.1257E}
}

@ARTICLE{2007Vaduvescu,
       author = {{Vaduvescu}, Ovidiu and {McCall}, Marshall L. and {Richer}, Michael G.},
        title = "{Chemical Properties of Star-Forming Dwarf Galaxies}",
      journal = {\aj},
         year = 2007,
        month = aug,
       volume = {134},
       number = {2},
        pages = {604-616},
          doi = {10.1086/518865},
archivePrefix = {arXiv},
       eprint = {0704.2705},
 primaryClass = {astro-ph},
       adsurl = {https://ui.adsabs.harvard.edu/abs/2007AJ....134..604V}
}

@ARTICLE{2003Lee,
       author = {{Lee}, Henry and {Grebel}, E.~K. and {Hodge}, P.~W.},
        title = "{Nebular abundances of nearby southern dwarf galaxies}",
      journal = {\aap},
         year = 2003,
        month = apr,
       volume = {401},
        pages = {141-159},
          doi = {10.1051/0004-6361:20030101},
archivePrefix = {arXiv},
       eprint = {astro-ph/0301492},
 primaryClass = {astro-ph},
       adsurl = {https://ui.adsabs.harvard.edu/abs/2003A&A...401..141L}
}

@ARTICLE{2024Conrado,
       author = {{Conrado}, Ana M. and {Gonz{\'a}lez Delgado}, Rosa M. and {Garc{\'\i}a-Benito}, Rub{\'e}n and {P{\'e}rez}, Isabel and {Verley}, Simon and {Ruiz-Lara}, Tom{\'a}s and {S{\'a}nchez-Menguiano}, Laura and {Duarte Puertas}, Salvador and {Jim{\'e}nez}, Andoni and {Dom{\'\i}nguez-G{\'o}mez}, Jes{\'u}s and {Espada}, Daniel and {Argudo-Fern{\'a}ndez}, Mar{\'\i}a and {Alc{\'a}zar-Laynez}, Manuel and {Bl{\'a}zquez-Calero}, Guillermo and {Bidaran}, Bahar and {Zurita}, Almudena and {Peletier}, Reynier and {Torres-R{\'\i}os}, Gloria and {Florido}, Estrella and {Rodr{\'\i}guez Mart{\'\i}nez}, M{\'o}nica and {del Moral-Castro}, Ignacio and {van de Weygaert}, Rien and {Falc{\'o}n-Barroso}, Jes{\'u}s and {Lugo-Aranda}, Alejandra Z. and {S{\'a}nchez}, Sebasti{\'a}n F. and {van der Hulst}, Thijs and {Courtois}, H{\'e}l{\`e}ne M. and {Ferr{\'e}-Mateu}, Anna and {S{\'a}nchez-Bl{\'a}zquez}, Patricia and {Rom{\'a}n}, Javier and {Aceituno}, Jes{\'u}s},
        title = "{The CAVITY project: The spatially resolved stellar population properties of galaxies in voids}",
      journal = {\aap},
         year = 2024,
        month = jul,
       volume = {687},
          eid = {A98},
        pages = {A98},
          doi = {10.1051/0004-6361/202449414},
archivePrefix = {arXiv},
       eprint = {2404.10823},
 primaryClass = {astro-ph.GA},
       adsurl = {https://ui.adsabs.harvard.edu/abs/2024A&A...687A..98C}
}

@ARTICLE{2012Pan,
       author = {{Pan}, Danny C. and {Vogeley}, Michael S. and {Hoyle}, Fiona and {Choi}, Yun-Young and {Park}, Changbom},
        title = "{Cosmic voids in Sloan Digital Sky Survey Data Release 7}",
      journal = {\mnras},
         year = 2012,
        month = apr,
       volume = {421},
       number = {2},
        pages = {926-934},
          doi = {10.1111/j.1365-2966.2011.20197.x},
archivePrefix = {arXiv},
       eprint = {1103.4156},
 primaryClass = {astro-ph.CO},
       adsurl = {https://ui.adsabs.harvard.edu/abs/2012MNRAS.421..926P}
}

@ARTICLE{2024Perez,
       author = {{P{\'e}rez}, I. and {Verley}, S. and {S{\'a}nchez-Menguiano}, L. and {Ruiz-Lara}, T. and {Garc{\'\i}a-Benito}, R. and {Duarte Puertas}, S. and {Jim{\'e}nez}, A. and {Dom{\'\i}nguez-G{\'o}mez}, J. and {Espada}, D. and {Peletier}, R.~F. and {Rom{\'a}n}, J. and {Rodr{\'\i}guez}, M.~I. and {Argudo-Fern{\'a}ndez}, M. and {Torres-R{\'\i}os}, G. and {Bidaran}, B. and {Alc{\'a}zar-Laynez}, M. and {van de Weygaert}, R. and {S{\'a}nchez}, S.~F. and {Lisenfeld}, U. and {Zurita}, A. and {Florido}, E. and {van der Hulst}, J.~M. and {Bl{\'a}zquez-Calero}, G. and {Villalba-Gonz{\'a}lez}, P. and {del Moral-Castro}, I. and {S{\'a}nchez Alarc{\'o}n}, P. and {Lugo-Aranda}, A. and {Walo-Mart{\'\i}n}, D. and {Conrado}, A. and {Gonz{\'a}lez Delgado}, R. and {Falc{\'o}n-Barroso}, J. and {Ferr{\'e}-Mateu}, A. and {Hern{\'a}ndez-S{\'a}nchez}, M. and {Awad}, P. and {Kreckel}, K. and {Courtois}, H. and {Espada-Miura}, R. and {Rela{\~n}o}, M. and {Galbany}, L. and {S{\'a}nchez-Bl{\'a}zquez}, P. and {P{\'e}rez-Montero}, E. and {S{\'a}nchez-Portal}, M. and {Bongiovanni}, A. and {Planelles}, S. and {Quilis}, V. and {Weijmans}, A. and {Raj}, M.~A. and {Arag{\'o}n-Calvo}, M.~A. and {Azzaro}, M. and {Bergond}, G. and {Blazek}, M. and {Cikota}, S. and {Fern{\'a}ndez-Mart{\'\i}n}, A. and {Gardini}, A. and {Guijarro}, A. and {Hermelo}, I. and {Mart{\'\i}n}, P. and {Vico Linares}, J.~I.},
        title = "{CAVITY, Calar Alto Void Integral-field Treasury surveY and project extension}",
      journal = {\aap},
         year = 2024,
        month = sep,
       volume = {689},
          eid = {A213},
        pages = {A213},
          doi = {10.1051/0004-6361/202449749},
archivePrefix = {arXiv},
       eprint = {2405.04217},
 primaryClass = {astro-ph.GA},
       adsurl = {https://ui.adsabs.harvard.edu/abs/2024A&A...689A.213P}
}

@ARTICLE{2002Strauss,
       author = {{Strauss}, Michael A. and {Weinberg}, David H. and {Lupton}, Robert H. and {Narayanan}, Vijay K. and {Annis}, James and {Bernardi}, Mariangela and {Blanton}, Michael and {Burles}, Scott and {Connolly}, A.~J. and {Dalcanton}, Julianne and {Doi}, Mamoru and {Eisenstein}, Daniel and {Frieman}, Joshua A. and {Fukugita}, Masataka and {Gunn}, James E. and {Ivezi{\'c}}, {\v{Z}}eljko and {Kent}, Stephen and {Kim}, Rita S.~J. and {Knapp}, G.~R. and {Kron}, Richard G. and {Munn}, Jeffrey A. and {Newberg}, Heidi Jo and {Nichol}, R.~C. and {Okamura}, Sadanori and {Quinn}, Thomas R. and {Richmond}, Michael W. and {Schlegel}, David J. and {Shimasaku}, Kazuhiro and {SubbaRao}, Mark and {Szalay}, Alexander S. and {Vanden Berk}, Dan and {Vogeley}, Michael S. and {Yanny}, Brian and {Yasuda}, Naoki and {York}, Donald G. and {Zehavi}, Idit},
        title = "{Spectroscopic Target Selection in the Sloan Digital Sky Survey: The Main Galaxy Sample}",
      journal = {\aj},
         year = 2002,
        month = sep,
       volume = {124},
       number = {3},
        pages = {1810-1824},
          doi = {10.1086/342343},
archivePrefix = {arXiv},
       eprint = {astro-ph/0206225},
 primaryClass = {astro-ph},
       adsurl = {https://ui.adsabs.harvard.edu/abs/2002AJ....124.1810S}
}

@ARTICLE{2015A&A...578A.110A,
       author = {{Argudo-Fern{\'a}ndez}, M. and {Verley}, S. and {Bergond}, G. and {Duarte Puertas}, S. and {Ramos Carmona}, E. and {Sabater}, J. and {Fern{\'a}ndez Lorenzo}, M. and {Espada}, D. and {Sulentic}, J. and {Ruiz}, J.~E. and {Leon}, S.},
        title = "{Catalogues of isolated galaxies, isolated pairs, and isolated triplets in the local Universe}",
      journal = {\aap},
         year = 2015,
        month = jun,
       volume = {578},
          eid = {A110},
        pages = {A110},
          doi = {10.1051/0004-6361/201526016},
archivePrefix = {arXiv},
       eprint = {1504.00117},
 primaryClass = {astro-ph.GA},
       adsurl = {https://ui.adsabs.harvard.edu/abs/2015A&A...578A.110A}
}

@ARTICLE{2004Brinchmann,
       author = {{Brinchmann}, J. and {Charlot}, S. and {White}, S.~D.~M. and {Tremonti}, C. and {Kauffmann}, G. and {Heckman}, T. and {Brinkmann}, J.},
        title = "{The physical properties of star-forming galaxies in the low-redshift Universe}",
      journal = {\mnras},
         year = 2004,
        month = jul,
       volume = {351},
       number = {4},
        pages = {1151-1179},
          doi = {10.1111/j.1365-2966.2004.07881.x},
archivePrefix = {arXiv},
       eprint = {astro-ph/0311060},
 primaryClass = {astro-ph},
       adsurl = {https://ui.adsabs.harvard.edu/abs/2004MNRAS.351.1151B}
}

@ARTICLE{2004Tremonti,
       author = {{Tremonti}, Christy A. and {Heckman}, Timothy M. and {Kauffmann}, Guinevere and {Brinchmann}, Jarle and {Charlot}, St{\'e}phane and {White}, Simon D.~M. and {Seibert}, Mark and {Peng}, Eric W. and {Schlegel}, David J. and {Uomoto}, Alan and {Fukugita}, Masataka and {Brinkmann}, Jon},
        title = "{The Origin of the Mass-Metallicity Relation: Insights from 53,000 Star-forming Galaxies in the Sloan Digital Sky Survey}",
      journal = {\apj},
         year = 2004,
        month = oct,
       volume = {613},
       number = {2},
        pages = {898-913},
          doi = {10.1086/423264},
archivePrefix = {arXiv},
       eprint = {astro-ph/0405537},
 primaryClass = {astro-ph},
       adsurl = {https://ui.adsabs.harvard.edu/abs/2004ApJ...613..898T}
}

@ARTICLE{2024Guo,
       author = {{Guo}, Yan and {Sengupta}, Chandreyee and {Scott}, Tom C. and {Lagos}, Patricio and {Luo}, Yu},
        title = "{Catalogue of nearby blue and near-solar gas metallicity SDSS dwarf galaxies}",
      journal = {\mnras},
         year = 2024,
        month = mar,
       volume = {528},
       number = {4},
        pages = {6593-6607},
          doi = {10.1093/mnras/stae390},
archivePrefix = {arXiv},
       eprint = {2402.13612},
 primaryClass = {astro-ph.GA},
       adsurl = {https://ui.adsabs.harvard.edu/abs/2024MNRAS.528.6593G}
}

@ARTICLE{2017Tempel,
       author = {{Tempel}, E. and {Tuvikene}, T. and {Kipper}, R. and {Libeskind}, N.~I.},
        title = "{Merging groups and clusters of galaxies from the SDSS data. The catalogue of groups and potentially merging systems}",
      journal = {\aap},
         year = 2017,
        month = jun,
       volume = {602},
          eid = {A100},
        pages = {A100},
          doi = {10.1051/0004-6361/201730499},
archivePrefix = {arXiv},
       eprint = {1704.04477},
 primaryClass = {astro-ph.CO},
       adsurl = {https://ui.adsabs.harvard.edu/abs/2017A&A...602A.100T}
}

@ARTICLE{1989Cardelli,
       author = {{Cardelli}, Jason A. and {Clayton}, Geoffrey C. and {Mathis}, John S.},
        title = "{The Relationship between Infrared, Optical, and Ultraviolet Extinction}",
      journal = {\apj},
         year = 1989,
        month = oct,
       volume = {345},
        pages = {245},
          doi = {10.1086/167900},
       adsurl = {https://ui.adsabs.harvard.edu/abs/1989ApJ...345..245C}
}

@ARTICLE{2007Verley,
       author = {{Verley}, S. and {Odewahn}, S.~C. and {Verdes-Montenegro}, L. and {Leon}, S. and {Combes}, F. and {Sulentic}, J. and {Bergond}, G. and {Espada}, D. and {Garc{\'\i}a}, E. and {Lisenfeld}, U. and {Sabater}, J.},
        title = "{The AMIGA sample of isolated galaxies. IV. A catalogue of neighbours around isolated galaxies}",
      journal = {\aap},
         year = 2007,
        month = aug,
       volume = {470},
       number = {2},
        pages = {505-513},
          doi = {10.1051/0004-6361:20077307},
archivePrefix = {arXiv},
       eprint = {0705.0479},
 primaryClass = {astro-ph},
       adsurl = {https://ui.adsabs.harvard.edu/abs/2007A&A...470..505V}
}

@ARTICLE{2022Lacerda,
       author = {{Lacerda}, Eduardo A.~D. and {S{\'a}nchez}, S.~F. and {Mej{\'\i}a-Narv{\'a}ez}, A. and {Camps-Fari{\~n}a}, A. and {Espinosa-Ponce}, C. and {Barrera-Ballesteros}, J.~K. and {Ibarra-Medel}, H. and {Lugo-Aranda}, A.~Z.},
        title = "{pyFIT3D and pyPipe3D - The new version of the integral field spectroscopy data analysis pipeline}",
      journal = {\na},
         year = 2022,
        month = nov,
       volume = {97},
          eid = {101895},
        pages = {101895},
          doi = {10.1016/j.newast.2022.101895},
archivePrefix = {arXiv},
       eprint = {2202.08027},
 primaryClass = {astro-ph.GA},
       adsurl = {https://ui.adsabs.harvard.edu/abs/2022NewA...9701895L}
}

@ARTICLE{2010Vazdekis,
       author = {{Vazdekis}, A. and {S{\'a}nchez-Bl{\'a}zquez}, P. and {Falc{\'o}n-Barroso}, J. and {Cenarro}, A.~J. and {Beasley}, M.~A. and {Cardiel}, N. and {Gorgas}, J. and {Peletier}, R.~F.},
        title = "{Evolutionary stellar population synthesis with MILES - I. The base models and a new line index system}",
      journal = {\mnras},
         year = 2010,
        month = jun,
       volume = {404},
       number = {4},
        pages = {1639-1671},
          doi = {10.1111/j.1365-2966.2010.16407.x},
archivePrefix = {arXiv},
       eprint = {1004.4439},
 primaryClass = {astro-ph.CO},
       adsurl = {https://ui.adsabs.harvard.edu/abs/2010MNRAS.404.1639V}
}

@ARTICLE{2015Vazdekis,
       author = {{Vazdekis}, A. and {Coelho}, P. and {Cassisi}, S. and {Ricciardelli}, E. and {Falc{\'o}n-Barroso}, J. and {S{\'a}nchez-Bl{\'a}zquez}, P. and {La Barbera}, F. and {Beasley}, M.~A. and {Pietrinferni}, A.},
        title = "{Evolutionary stellar population synthesis with MILES - II. Scaled-solar and {\ensuremath{\alpha}}-enhanced models}",
      journal = {\mnras},
         year = 2015,
        month = may,
       volume = {449},
       number = {2},
        pages = {1177-1214},
          doi = {10.1093/mnras/stv151},
archivePrefix = {arXiv},
       eprint = {1504.08032},
 primaryClass = {astro-ph.GA},
       adsurl = {https://ui.adsabs.harvard.edu/abs/2015MNRAS.449.1177V}
}

@ARTICLE{2006MNRAS.371..703S,
       author = {{S{\'a}nchez-Bl{\'a}zquez}, P. and {Peletier}, R.~F. and {Jim{\'e}nez-Vicente}, J. and {Cardiel}, N. and {Cenarro}, A.~J. and {Falc{\'o}n-Barroso}, J. and {Gorgas}, J. and {Selam}, S. and {Vazdekis}, A.},
        title = "{Medium-resolution Isaac Newton Telescope library of empirical spectra}",
      journal = {\mnras},
         year = 2006,
        month = sep,
       volume = {371},
       number = {2},
        pages = {703-718},
          doi = {10.1111/j.1365-2966.2006.10699.x},
archivePrefix = {arXiv},
       eprint = {astro-ph/0607009},
 primaryClass = {astro-ph},
       adsurl = {https://ui.adsabs.harvard.edu/abs/2006MNRAS.371..703S}
}

@ARTICLE{2007Cenarro,
       author = {{Cenarro}, A.~J. and {Peletier}, R.~F. and {S{\'a}nchez-Bl{\'a}zquez}, P. and {Selam}, S.~O. and {Toloba}, E. and {Cardiel}, N. and {Falc{\'o}n-Barroso}, J. and {Gorgas}, J. and {Jim{\'e}nez-Vicente}, J. and {Vazdekis}, A.},
        title = "{Medium-resolution Isaac Newton Telescope library of empirical spectra - II. The stellar atmospheric parameters}",
      journal = {\mnras},
         year = 2007,
        month = jan,
       volume = {374},
       number = {2},
        pages = {664-690},
          doi = {10.1111/j.1365-2966.2006.11196.x},
archivePrefix = {arXiv},
       eprint = {astro-ph/0611618},
 primaryClass = {astro-ph},
       adsurl = {https://ui.adsabs.harvard.edu/abs/2007MNRAS.374..664C}
}

@ARTICLE{2011CidFernandes,
       author = {{Cid Fernandes}, R. and {Stasi{\'n}ska}, G. and {Mateus}, A. and {Vale Asari}, N.},
        title = "{A comprehensive classification of galaxies in the Sloan Digital Sky Survey: how to tell true from fake AGN?}",
      journal = {\mnras},
         year = 2011,
        month = may,
       volume = {413},
       number = {3},
        pages = {1687-1699},
          doi = {10.1111/j.1365-2966.2011.18244.x},
archivePrefix = {arXiv},
       eprint = {1012.4426},
 primaryClass = {astro-ph.CO},
       adsurl = {https://ui.adsabs.harvard.edu/abs/2011MNRAS.413.1687C}
}

@ARTICLE{1981Baldwin,
       author = {{Baldwin}, J.~A. and {Phillips}, M.~M. and {Terlevich}, R.},
        title = "{Classification parameters for the emission-line spectra of extragalactic objects.}",
      journal = {\pasp},
         year = 1981,
        month = feb,
       volume = {93},
        pages = {5-19},
          doi = {10.1086/130766},
       adsurl = {https://ui.adsabs.harvard.edu/abs/1981PASP...93....5B}
}

@ARTICLE{1984Osterbrock,
       author = {{Osterbrock}, D.~E.},
        title = "{Active galactic nuclei.}",
      journal = {\qjras},
         year = 1984,
        month = mar,
       volume = {25},
        pages = {1-18},
       adsurl = {https://ui.adsabs.harvard.edu/abs/1984QJRAS..25....1O}
}

@ARTICLE{2016RMxAA..52..171S,
       author = {{S{\'a}nchez}, S.~F. and {P{\'e}rez}, E. and {S{\'a}nchez-Bl{\'a}zquez}, P. and {Garc{\'\i}a-Benito}, R. and {Ibarra-Mede}, H.~J. and {Gonz{\'a}lez}, J.~J. and {Rosales-Ortega}, F.~F. and {S{\'a}nchez-Menguiano}, L. and {Ascasibar}, Y. and {Bitsakis}, T. and {Law}, D. and {Cano-D{\'\i}az}, M. and {L{\'o}pez-Cob{\'a}}, C. and {Marino}, R.~A. and {Gil de Paz}, A. and {L{\'o}pez-S{\'a}nchez}, A.~R. and {Barrera-Ballesteros}, J. and {Galbany}, L. and {Mast}, D. and {Abril-Melgarejo}, V. and {Roman-Lopes}, A.},
        title = "{Pipe3D, a pipeline to analyze Integral Field Spectroscopy Data: II. Analysis sequence and CALIFA dataproducts}",
      journal = {\rmxaa},
         year = 2016,
        month = apr,
       volume = {52},
        pages = {171-220},
          doi = {10.48550/arXiv.1602.01830},
archivePrefix = {arXiv},
       eprint = {1602.01830},
 primaryClass = {astro-ph.IM},
       adsurl = {https://ui.adsabs.harvard.edu/abs/2016RMxAA..52..171S}
}

@ARTICLE{2009Pietrinferni,
       author = {{Pietrinferni}, Adriano and {Cassisi}, Santi and {Salaris}, Maurizio and {Percival}, Susan and {Ferguson}, Jason W.},
        title = "{A Large Stellar Evolution Database for Population Synthesis Studies. V. Stellar Models and Isochrones with CNONa Abundance Anticorrelations}",
      journal = {\apj},
         year = 2009,
        month = may,
       volume = {697},
       number = {1},
        pages = {275-282},
          doi = {10.1088/0004-637X/697/1/275},
archivePrefix = {arXiv},
       eprint = {0903.0825},
 primaryClass = {astro-ph.SR},
       adsurl = {https://ui.adsabs.harvard.edu/abs/2009ApJ...697..275P}
}

@ARTICLE{1996Vazdekis,
       author = {{Vazdekis}, A. and {Casuso}, E. and {Peletier}, R.~F. and {Beckman}, J.~E.},
        title = "{A New Chemo-evolutionary Population Synthesis Model for Early-Type Galaxies. I. Theoretical Basis}",
      journal = {\apjs},
         year = 1996,
        month = oct,
       volume = {106},
        pages = {307},
          doi = {10.1086/192340},
archivePrefix = {arXiv},
       eprint = {astro-ph/9605112},
 primaryClass = {astro-ph},
       adsurl = {https://ui.adsabs.harvard.edu/abs/1996ApJS..106..307V}
}

@ARTICLE{2011A&A...532A..95F,
       author = {{Falc{\'o}n-Barroso}, J. and {S{\'a}nchez-Bl{\'a}zquez}, P. and {Vazdekis}, A. and {Ricciardelli}, E. and {Cardiel}, N. and {Cenarro}, A.~J. and {Gorgas}, J. and {Peletier}, R.~F.},
        title = "{An updated MILES stellar library and stellar population models}",
      journal = {\aap},
         year = 2011,
        month = aug,
       volume = {532},
          eid = {A95},
        pages = {A95},
          doi = {10.1051/0004-6361/201116842},
archivePrefix = {arXiv},
       eprint = {1107.2303},
 primaryClass = {astro-ph.CO},
       adsurl = {https://ui.adsabs.harvard.edu/abs/2011A&A...532A..95F}
}

@ARTICLE{2016Williamson,
       author = {{Williamson}, David and {Martel}, Hugo and {Romeo}, Alessandro B.},
        title = "{Chemodynamic Evolution of Dwarf Galaxies in Tidal Fields}",
      journal = {\apj},
         year = 2016,
        month = nov,
       volume = {831},
       number = {1},
          eid = {1},
        pages = {1},
          doi = {10.3847/0004-637X/831/1/1},
archivePrefix = {arXiv},
       eprint = {1608.06849},
 primaryClass = {astro-ph.GA},
       adsurl = {https://ui.adsabs.harvard.edu/abs/2016ApJ...831....1W}
}

@ARTICLE{2013Andrews,
       author = {{Andrews}, Brett H. and {Martini}, Paul},
        title = "{The Mass-Metallicity Relation with the Direct Method on Stacked Spectra of SDSS Galaxies}",
      journal = {\apj},
         year = 2013,
        month = mar,
       volume = {765},
       number = {2},
          eid = {140},
        pages = {140},
          doi = {10.1088/0004-637X/765/2/140},
archivePrefix = {arXiv},
       eprint = {1211.3418},
 primaryClass = {astro-ph.CO},
       adsurl = {https://ui.adsabs.harvard.edu/abs/2013ApJ...765..140A}
}

@ARTICLE{2015Lian,
       author = {{Lian}, J.~H. and {Li}, J.~R. and {Yan}, W. and {Kong}, X.},
        title = "{The mass-metallicity relation of Lyman-break analogues and its dependence on galaxy properties}",
      journal = {\mnras},
         year = 2015,
        month = jan,
       volume = {446},
       number = {2},
        pages = {1449-1457},
          doi = {10.1093/mnras/stu2184},
archivePrefix = {arXiv},
       eprint = {1411.6331},
 primaryClass = {astro-ph.GA},
       adsurl = {https://ui.adsabs.harvard.edu/abs/2015MNRAS.446.1449L}
}

@ARTICLE{2005MNRAS.363....2K,
       author = {{Kere{\v{s}}}, Du{\v{s}}an and {Katz}, Neal and {Weinberg}, David H. and {Dav{\'e}}, Romeel},
        title = "{How do galaxies get their gas?}",
      journal = {\mnras},
         year = 2005,
        month = oct,
       volume = {363},
       number = {1},
        pages = {2-28},
          doi = {10.1111/j.1365-2966.2005.09451.x},
archivePrefix = {arXiv},
       eprint = {astro-ph/0407095},
 primaryClass = {astro-ph},
       adsurl = {https://ui.adsabs.harvard.edu/abs/2005MNRAS.363....2K}
}

@ARTICLE{1974Larson,
       author = {{Larson}, Richard B.},
        title = "{Effects of supernovae on the early evolution of galaxies}",
      journal = {\mnras},
         year = 1974,
        month = nov,
       volume = {169},
        pages = {229-246},
          doi = {10.1093/mnras/169.2.229},
       adsurl = {https://ui.adsabs.harvard.edu/abs/1974MNRAS.169..229L}
}

@ARTICLE{2011MNRAS.414.2458V,
       author = {{van de Voort}, Freeke and {Schaye}, Joop and {Booth}, C.~M. and {Haas}, Marcel R. and {Dalla Vecchia}, Claudio},
        title = "{The rates and modes of gas accretion on to galaxies and their gaseous haloes}",
      journal = {\mnras},
         year = 2011,
        month = jul,
       volume = {414},
       number = {3},
        pages = {2458-2478},
          doi = {10.1111/j.1365-2966.2011.18565.x},
archivePrefix = {arXiv},
       eprint = {1011.2491},
 primaryClass = {astro-ph.CO},
       adsurl = {https://ui.adsabs.harvard.edu/abs/2011MNRAS.414.2458V}
}

@ARTICLE{2006Dekel,
       author = {{Dekel}, Avishai and {Birnboim}, Yuval},
        title = "{Galaxy bimodality due to cold flows and shock heating}",
      journal = {\mnras},
         year = 2006,
        month = may,
       volume = {368},
       number = {1},
        pages = {2-20},
          doi = {10.1111/j.1365-2966.2006.10145.x},
archivePrefix = {arXiv},
       eprint = {astro-ph/0412300},
 primaryClass = {astro-ph},
       adsurl = {https://ui.adsabs.harvard.edu/abs/2006MNRAS.368....2D}
}

@ARTICLE{2014Boselli,
       author = {{Boselli}, Alessandro and {Gavazzi}, Giuseppe},
        title = "{On the origin of the faint-end of the red sequence in high-density environments}",
      journal = {\aapr},
         year = 2014,
        month = nov,
       volume = {22},
          eid = {74},
        pages = {74},
          doi = {10.1007/s00159-014-0074-y},
archivePrefix = {arXiv},
       eprint = {1411.5513},
 primaryClass = {astro-ph.GA},
       adsurl = {https://ui.adsabs.harvard.edu/abs/2014A&ARv..22...74B}
}

@ARTICLE{2025A&A...695A..84P,
       author = {{P{\'e}rez}, I. and {Gil}, L. and {Ferr{\'e}-Mateu}, A. and {Torres-R{\'\i}os}, G. and {Zurita}, A. and {Argudo-Fern{\'a}ndez}, M. and {Bidaran}, B. and {S{\'a}nchez-Menguiano}, L. and {Ruiz-Lara}, T. and {Dom{\'\i}nguez-G{\'o}mez}, J. and {Duarte Puertas}, S. and {Espada}, D. and {Falc{\'o}n-Barroso}, J. and {Florido}, E. and {Garc{\'\i}a-Benito}, R. and {Jim{\'e}nez}, A. and {Peletier}, R.~F. and {Rom{\'a}n}, J. and {S{\'a}nchez Alarc{\'o}n}, P. and {S{\'a}nchez-Bl{\'a}zquez}, P. and {V{\'a}squez-Bustos}, P.},
        title = "{Galaxy mass-size segregation in the cosmic web from the CAVITY parent sample}",
      journal = {\aap},
         year = 2025,
        month = mar,
       volume = {695},
          eid = {A84},
        pages = {A84},
          doi = {10.1051/0004-6361/202452514},
archivePrefix = {arXiv},
       eprint = {2501.07345},
 primaryClass = {astro-ph.GA},
       adsurl = {https://ui.adsabs.harvard.edu/abs/2025A&A...695A..84P}
}

@ARTICLE{2023A&A...680A.111D,
       author = {{Dom{\'\i}nguez-G{\'o}mez}, Jes{\'u}s and {P{\'e}rez}, Isabel and {Ruiz-Lara}, Tom{\'a}s and {Peletier}, Reynier F. and {S{\'a}nchez-Bl{\'a}zquez}, Patricia and {Lisenfeld}, Ute and {Bidaran}, Bahar and {Falc{\'o}n-Barroso}, Jes{\'u}s and {Alc{\'a}zar-Laynez}, Manuel and {Argudo-Fern{\'a}ndez}, Mar{\'\i}a and {Bl{\'a}zquez-Calero}, Guillermo and {Courtois}, H{\'e}l{\`e}ne and {Duarte Puertas}, Salvador and {Espada}, Daniel and {Florido}, Estrella and {Garc{\'\i}a-Benito}, Rub{\'e}n and {Jim{\'e}nez}, Andoni and {Kreckel}, Kathryn and {Rela{\~n}o}, M{\'o}nica and {S{\'a}nchez-Menguiano}, Laura and {van der Hulst}, Thijs and {van de Weygaert}, Rien and {Verley}, Simon and {Zurita}, Almudena},
        title = "{Stellar mass-metallicity relation throughout the large-scale structure of the Universe: CAVITY mother sample}",
      journal = {\aap},
         year = 2023,
        month = dec,
       volume = {680},
          eid = {A111},
        pages = {A111},
          doi = {10.1051/0004-6361/202346884},
archivePrefix = {arXiv},
       eprint = {2310.11412},
 primaryClass = {astro-ph.GA},
       adsurl = {https://ui.adsabs.harvard.edu/abs/2023A&A...680A.111D}
}

@ARTICLE{2024A&A...691A.341T,
       author = {{Torres-R{\'\i}os}, G. and {P{\'e}rez}, I. and {Verley}, S. and {Dom{\'\i}nguez-G{\'o}mez}, J. and {Argudo-Fern{\'a}ndez}, M. and {Duarte Puertas}, S. and {Jim{\'e}nez}, A. and {Ruiz-Lara}, T. and {Zurita}, A. and {Bidaran}, B. and {Conrado}, A. and {Espada}, D. and {Garc{\'\i}a-Benito}, R. and {Gonz{\'a}lez Delgado}, R.~M. and {Falc{\'o}n-Barroso}, J. and {Florido}, E. and {S{\'a}nchez-Bl{\'a}zquez}, P. and {S{\'a}nchez-Menguiano}, L.},
        title = "{Effect of the local and large-scale environment on the star formation histories of galaxies}",
      journal = {\aap},
         year = 2024,
        month = nov,
       volume = {691},
          eid = {A341},
        pages = {A341},
          doi = {10.1051/0004-6361/202450675},
archivePrefix = {arXiv},
       eprint = {2410.00959},
 primaryClass = {astro-ph.GA},
       adsurl = {https://ui.adsabs.harvard.edu/abs/2024A&A...691A.341T}
}

@ARTICLE{2017Douglass,
       author = {{Douglass}, Kelly A. and {Vogeley}, Michael S.},
        title = "{Determining the Large-scale Environmental Dependence of Gas-phase Metallicity in Dwarf Galaxies}",
      journal = {\apj},
         year = 2017,
        month = jan,
       volume = {834},
       number = {2},
          eid = {186},
        pages = {186},
          doi = {10.3847/1538-4357/834/2/186},
archivePrefix = {arXiv},
       eprint = {1604.08599},
 primaryClass = {astro-ph.GA},
       adsurl = {https://ui.adsabs.harvard.edu/abs/2017ApJ...834..186D}
}

@ARTICLE{2008MNRAS.387...79V,
       author = {{van den Bosch}, Frank C. and {Aquino}, Daniel and {Yang}, Xiaohu and {Mo}, H.~J. and {Pasquali}, Anna and {McIntosh}, Daniel H. and {Weinmann}, Simone M. and {Kang}, Xi},
        title = "{The importance of satellite quenching for the build-up of the red sequence of present-day galaxies}",
      journal = {\mnras},
         year = 2008,
        month = jun,
       volume = {387},
       number = {1},
        pages = {79-91},
          doi = {10.1111/j.1365-2966.2008.13230.x},
archivePrefix = {arXiv},
       eprint = {0710.3164},
 primaryClass = {astro-ph},
       adsurl = {https://ui.adsabs.harvard.edu/abs/2008MNRAS.387...79V}
}

@ARTICLE{2005Lanzoni,
       author = {{Lanzoni}, B. and {Guiderdoni}, B. and {Mamon}, G.~A. and {Devriendt}, J. and {Hatton}, S.},
        title = "{GALICS- VI. Modelling hierarchical galaxy formation in clusters}",
      journal = {\mnras},
         year = 2005,
        month = aug,
       volume = {361},
       number = {2},
        pages = {369-384},
          doi = {10.1111/j.1365-2966.2005.09252.x},
archivePrefix = {arXiv},
       eprint = {astro-ph/0502490},
 primaryClass = {astro-ph},
       adsurl = {https://ui.adsabs.harvard.edu/abs/2005MNRAS.361..369L}
}

@ARTICLE{2003Okamoto,
       author = {{Okamoto}, Takashi and {Nagashima}, Masahiro},
        title = "{Environmental Effects on Evolution of Cluster Galaxies in a {\ensuremath{\Lambda}}-dominated Cold Dark Matter Universe}",
      journal = {\apj},
         year = 2003,
        month = apr,
       volume = {587},
       number = {2},
        pages = {500-513},
          doi = {10.1086/368251},
archivePrefix = {arXiv},
       eprint = {astro-ph/0108434},
 primaryClass = {astro-ph},
       adsurl = {https://ui.adsabs.harvard.edu/abs/2003ApJ...587..500O}
}

@ARTICLE{2007Tonnesen,
       author = {{Tonnesen}, Stephanie and {Bryan}, Greg L. and {van Gorkom}, J.~H.},
        title = "{Environmentally Driven Evolution of Simulated Cluster Galaxies}",
      journal = {\apj},
         year = 2007,
        month = dec,
       volume = {671},
       number = {2},
        pages = {1434-1445},
          doi = {10.1086/523034},
archivePrefix = {arXiv},
       eprint = {0709.1720},
 primaryClass = {astro-ph},
       adsurl = {https://ui.adsabs.harvard.edu/abs/2007ApJ...671.1434T}
}

@ARTICLE{2014Cen,
       author = {{Cen}, Renyue and {Pop}, Ana Roxana and {Bahcall}, Neta A.},
        title = "{Gas loss in simulated galaxies as they fall into clusters}",
      journal = {Proceedings of the National Academy of Science},
         year = 2014,
        month = jun,
       volume = {111},
       number = {22},
        pages = {7914-7919},
          doi = {10.1073/pnas.1407300111},
archivePrefix = {arXiv},
       eprint = {1405.0537},
 primaryClass = {astro-ph.GA},
       adsurl = {https://ui.adsabs.harvard.edu/abs/2014PNAS..111.7914C}
}

@ARTICLE{2019Roberts-Borsani,
       author = {{Roberts-Borsani}, G.~W. and {Saintonge}, A.},
        title = "{The prevalence and properties of cold gas inflows and outflows around galaxies in the local Universe}",
      journal = {\mnras},
         year = 2019,
        month = jan,
       volume = {482},
       number = {3},
        pages = {4111-4145},
          doi = {10.1093/mnras/sty2824},
archivePrefix = {arXiv},
       eprint = {1807.07575},
 primaryClass = {astro-ph.GA},
       adsurl = {https://ui.adsabs.harvard.edu/abs/2019MNRAS.482.4111R}
}

@ARTICLE{2012Hopkins,
       author = {{Hopkins}, Philip F. and {Quataert}, Eliot and {Murray}, Norman},
        title = "{Stellar feedback in galaxies and the origin of galaxy-scale winds}",
      journal = {\mnras},
         year = 2012,
        month = apr,
       volume = {421},
       number = {4},
        pages = {3522-3537},
          doi = {10.1111/j.1365-2966.2012.20593.x},
archivePrefix = {arXiv},
       eprint = {1110.4638},
 primaryClass = {astro-ph.CO},
       adsurl = {https://ui.adsabs.harvard.edu/abs/2012MNRAS.421.3522H}
}

@ARTICLE{1986Dekel,
       author = {{Dekel}, A. and {Silk}, J.},
        title = "{The Origin of Dwarf Galaxies, Cold Dark Matter, and Biased Galaxy Formation}",
      journal = {\apj},
         year = 1986,
        month = apr,
       volume = {303},
        pages = {39},
          doi = {10.1086/164050},
       adsurl = {https://ui.adsabs.harvard.edu/abs/1986ApJ...303...39D}
}

@ARTICLE{2022Fraser-McKelvie,
       author = {{Fraser-McKelvie}, A. and {Cortese}, L.},
        title = "{Beyond Galaxy Bimodality: The Complex Interplay between Kinematic Morphology and Star Formation in the Local Universe}",
      journal = {\apj},
         year = 2022,
        month = oct,
       volume = {937},
       number = {2},
          eid = {117},
        pages = {117},
          doi = {10.3847/1538-4357/ac874d},
archivePrefix = {arXiv},
       eprint = {2208.01936},
 primaryClass = {astro-ph.GA},
       adsurl = {https://ui.adsabs.harvard.edu/abs/2022ApJ...937..117F}
}

@ARTICLE{2018Lian,
       author = {{Lian}, Jianhui and {Thomas}, Daniel and {Maraston}, Claudia and {Goddard}, Daniel and {Comparat}, Johan and {Gonzalez-Perez}, Violeta and {Ventura}, Paolo},
        title = "{The mass-metallicity relations for gas and stars in star-forming galaxies: strong outflow versus variable IMF}",
      journal = {\mnras},
         year = 2018,
        month = feb,
       volume = {474},
       number = {1},
        pages = {1143-1164},
          doi = {10.1093/mnras/stx2829},
archivePrefix = {arXiv},
       eprint = {1710.11135},
 primaryClass = {astro-ph.GA},
       adsurl = {https://ui.adsabs.harvard.edu/abs/2018MNRAS.474.1143L}
}

@ARTICLE{2023Li,
       author = {{Li}, Mingyu and {Cai}, Zheng and {Bian}, Fuyan and {Lin}, Xiaojing and {Li}, Zihao and {Wu}, Yunjing and {Sun}, Fengwu and {Zhang}, Shiwu and {Golden-Marx}, Emmet and {Sun}, Zechang and {Zou}, Siwei and {Fan}, Xiaohui and {Egami}, Eiichi and {Charlot}, Stephane and {Bruzual}, Gustavo and {Chevallard}, Jacopo},
        title = "{The Mass-Metallicity Relation of Dwarf Galaxies at Cosmic Noon from JWST Observations}",
      journal = {\apjl},
         year = 2023,
        month = sep,
       volume = {955},
       number = {1},
          eid = {L18},
        pages = {L18},
          doi = {10.3847/2041-8213/acf470},
archivePrefix = {arXiv},
       eprint = {2211.01382},
 primaryClass = {astro-ph.GA},
       adsurl = {https://ui.adsabs.harvard.edu/abs/2023ApJ...955L..18L}
}

@ARTICLE{2016Guo,
       author = {{Guo}, Yicheng and {Koo}, David C. and {Lu}, Yu and {Forbes}, John C. and {Rafelski}, Marc and {Trump}, Jonathan R. and {Amor{\'\i}n}, Ricardo and {Barro}, Guillermo and {Dav{\'e}}, Romeel and {Faber}, S.~M. and {Hathi}, Nimish P. and {Yesuf}, Hassen and {Cooper}, Michael C. and {Dekel}, Avishai and {Guhathakurta}, Puragra and {Kirby}, Evan N. and {Koekemoer}, Anton M. and {P{\'e}rez-Gonz{\'a}lez}, Pablo G. and {Lin}, Lihwai and {Newman}, Jeffery A. and {Primack}, Joel R. and {Rosario}, David J. and {Willmer}, Christopher N.~A. and {Yan}, Renbin},
        title = "{Stellar Mass-Gas-phase Metallicity Relation at 0.5 {\ensuremath{\leq}} z {\ensuremath{\leq}} 0.7: A Power Law with Increasing Scatter toward the Low-mass Regime}",
      journal = {\apj},
         year = 2016,
        month = may,
       volume = {822},
       number = {2},
          eid = {103},
        pages = {103},
          doi = {10.3847/0004-637X/822/2/103},
archivePrefix = {arXiv},
       eprint = {1603.04863},
 primaryClass = {astro-ph.GA},
       adsurl = {https://ui.adsabs.harvard.edu/abs/2016ApJ...822..103G}
}

@ARTICLE{1979Lequeux,
       author = {{Lequeux}, J. and {Peimbert}, M. and {Rayo}, J.~F. and {Serrano}, A. and {Torres-Peimbert}, S.},
        title = "{Chemical Composition and Evolution of Irregular and Blue Compact Galaxies}",
      journal = {\aap},
         year = 1979,
        month = dec,
       volume = {80},
        pages = {155},
       adsurl = {https://ui.adsabs.harvard.edu/abs/1979A&A....80..155L}
}

@ARTICLE{2014Zahid,
       author = {{Zahid}, H. Jabran and {Dima}, Gabriel I. and {Kudritzki}, Rolf-Peter and {Kewley}, Lisa J. and {Geller}, Margaret J. and {Hwang}, Ho Seong and {Silverman}, John D. and {Kashino}, Daichi},
        title = "{The Universal Relation of Galactic Chemical Evolution: The Origin of the Mass-Metallicity Relation}",
      journal = {\apj},
         year = 2014,
        month = aug,
       volume = {791},
       number = {2},
          eid = {130},
        pages = {130},
          doi = {10.1088/0004-637X/791/2/130},
archivePrefix = {arXiv},
       eprint = {1404.7526},
 primaryClass = {astro-ph.GA},
       adsurl = {https://ui.adsabs.harvard.edu/abs/2014ApJ...791..130Z}
}

@ARTICLE{1996Skillman,
       author = {{Skillman}, Evan D. and {Kennicutt}, Jr., Robert C. and {Shields}, Gregory A. and {Zaritsky}, Dennis},
        title = "{Chemical Abundances in Virgo Spiral Galaxies. II. Effects of Cluster Environment}",
      journal = {\apj},
         year = 1996,
        month = may,
       volume = {462},
        pages = {147},
          doi = {10.1086/177138},
archivePrefix = {arXiv},
       eprint = {astro-ph/9511019},
 primaryClass = {astro-ph},
       adsurl = {https://ui.adsabs.harvard.edu/abs/1996ApJ...462..147S}
}

@ARTICLE{2020Pandey,
       author = {{Pandey}, Biswajit and {Sarkar}, Suman},
        title = "{Exploring galaxy colour in different environments of the cosmic web with SDSS}",
      journal = {\mnras},
         year = 2020,
        month = nov,
       volume = {498},
       number = {4},
        pages = {6069-6082},
          doi = {10.1093/mnras/staa2772},
archivePrefix = {arXiv},
       eprint = {2002.08400},
 primaryClass = {astro-ph.GA},
       adsurl = {https://ui.adsabs.harvard.edu/abs/2020MNRAS.498.6069P}
}

@ARTICLE{1984Haynes,
       author = {{Haynes}, Martha P. and {Giovanelli}, Ricardo and {Chincarini}, Guido L.},
        title = "{The Influence of Envirionment on the H I Content of Galaxies}",
      journal = {\araa},
         year = 1984,
        month = jan,
       volume = {22},
        pages = {445-470},
          doi = {10.1146/annurev.aa.22.090184.002305},
       adsurl = {https://ui.adsabs.harvard.edu/abs/1984ARA&A..22..445H}
}

@ARTICLE{2009Chung,
       author = {{Chung}, Aeree and {van Gorkom}, J.~H. and {Kenney}, Jeffrey D.~P. and {Crowl}, Hugh and {Vollmer}, Bernd},
        title = "{VLA Imaging of Virgo Spirals in Atomic Gas (VIVA). I. The Atlas and the H I Properties}",
      journal = {\aj},
         year = 2009,
        month = dec,
       volume = {138},
       number = {6},
        pages = {1741-1816},
          doi = {10.1088/0004-6256/138/6/1741},
       adsurl = {https://ui.adsabs.harvard.edu/abs/2009AJ....138.1741C}
}

@ARTICLE{1997Huchtmeier,
       author = {{Huchtmeier}, W.~K. and {Hopp}, U. and {Kuhn}, B.},
        title = "{HI observations of dwarf galaxies in voids.}",
      journal = {\aap},
         year = 1997,
        month = mar,
       volume = {319},
        pages = {67-73},
       adsurl = {https://ui.adsabs.harvard.edu/abs/1997A&A...319...67H}
}

@ARTICLE{2023Chung,
       author = {{Chung}, Jiwon and {Lee}, Joon Hyeop and {Jeong}, Hyunjin and {Kim}, Suk},
        title = "{Witnessing a Transformation to Blue-cored Dwarf Early-type Galaxies in Filaments and the Cluster Outskirts: Gas-phase Abundances and Internal Kinematics Perspectives}",
      journal = {\apj},
         year = 2023,
        month = jun,
       volume = {949},
       number = {2},
          eid = {80},
        pages = {80},
          doi = {10.3847/1538-4357/accae1},
archivePrefix = {arXiv},
       eprint = {2304.02803},
 primaryClass = {astro-ph.GA},
       adsurl = {https://ui.adsabs.harvard.edu/abs/2023ApJ...949...80C}
}

@ARTICLE{2016ApJ...826...71I,
       author = {{Iglesias-P{\'a}ramo}, J. and {V{\'\i}lchez}, J.~M. and {Rosales-Ortega}, F.~F. and {S{\'a}nchez}, S.~F. and {Duarte Puertas}, S. and {Petropoulou}, V. and {Gil de Paz}, A. and {Galbany}, L. and {Moll{\'a}}, M. and {Catal{\'a}n-Torrecilla}, C. and {Castillo Morales}, A. and {Mast}, D. and {Husemann}, B. and {Garc{\'\i}a-Benito}, R. and {Mendoza}, M.~A. and {Kehrig}, C. and {P{\'e}rez-Montero}, E. and {Papaderos}, P. and {Gomes}, J.~M. and {Walcher}, C.~J. and {Gonz{\'a}lez Delgado}, R.~M. and {Marino}, R.~A. and {L{\'o}pez-S{\'a}nchez}, {\'A}. R. and {Ziegler}, B. and {Flores}, H. and {Alves}, J.},
        title = "{Aperture Effects on the Oxygen Abundance Determinations from CALIFA Data}",
      journal = {\apj},
         year = 2016,
        month = jul,
       volume = {826},
       number = {1},
          eid = {71},
        pages = {71},
          doi = {10.3847/0004-637X/826/1/71},
archivePrefix = {arXiv},
       eprint = {1605.03490},
 primaryClass = {astro-ph.GA},
       adsurl = {https://ui.adsabs.harvard.edu/abs/2016ApJ...826...71I}
}

@ARTICLE{2004Fujita,
       author = {{Fujita}, Yutaka},
        title = "{Pre-Processing of Galaxies before Entering a Cluster}",
      journal = {\pasj},
         year = 2004,
        month = feb,
       volume = {56},
        pages = {29-43},
          doi = {10.1093/pasj/56.1.29},
archivePrefix = {arXiv},
       eprint = {astro-ph/0311193},
 primaryClass = {astro-ph},
       adsurl = {https://ui.adsabs.harvard.edu/abs/2004PASJ...56...29F}
}

@ARTICLE{2020DeLucia,
       author = {{De Lucia}, Gabriella and {Xie}, Lizhi and {Fontanot}, Fabio and {Hirschmann}, Michaela},
        title = "{Gas accretion regulates the scatter of the mass-metallicity relation}",
      journal = {\mnras},
         year = 2020,
        month = nov,
       volume = {498},
       number = {3},
        pages = {3215-3227},
          doi = {10.1093/mnras/staa2556},
archivePrefix = {arXiv},
       eprint = {2008.09127},
 primaryClass = {astro-ph.GA},
       adsurl = {https://ui.adsabs.harvard.edu/abs/2020MNRAS.498.3215D}
}

@ARTICLE{2025A&A...695A.256A,
       author = {{Argudo-Fern{\'a}ndez}, M. and {Duarte Puertas}, S. and {Verley}, S.},
        title = "{Fundamental relation in isolated galaxies, pairs, and triplets in the local Universe}",
      journal = {\aap},
         year = 2025,
        month = mar,
       volume = {695},
          eid = {A256},
        pages = {A256},
          doi = {10.1051/0004-6361/202348724},
archivePrefix = {arXiv},
       eprint = {2502.15638},
 primaryClass = {astro-ph.GA},
       adsurl = {https://ui.adsabs.harvard.edu/abs/2025A&A...695A.256A}
}

@ARTICLE{2022ApJ...933...44C,
       author = {{Camps-Fari{\~n}a}, Artemi and {S{\'a}nchez}, Sebasti{\'a}n F. and {Mej{\'\i}a-Narv{\'a}ez}, Alfredo and {Lacerda}, Eduardo and {Carigi}, Leticia and {Bruzual}, Gustavo and {Alvarez-Hurtado}, Paola and {Drory}, Niv and {Lane}, Richard R. and {Boardman}, Nicholas Fraser and {Blanc}, Guillermo A.},
        title = "{Chemical Evolution History of MaNGA Galaxies}",
      journal = {\apj},
         year = 2022,
        month = jul,
       volume = {933},
       number = {1},
          eid = {44},
        pages = {44},
          doi = {10.3847/1538-4357/ac6cea},
archivePrefix = {arXiv},
       eprint = {2203.01159},
 primaryClass = {astro-ph.GA},
       adsurl = {https://ui.adsabs.harvard.edu/abs/2022ApJ...933...44C}
}

@ARTICLE{2024A&A...682A..71S,
       author = {{S{\'a}nchez}, S.~F. and {Lugo-Aranda}, A.~Z. and {S{\'a}nchez Almeida}, J. and {Barrera-Ballesteros}, J.~K. and {Gonzalez-Mart{\'\i}n}, O. and {Salim}, S. and {Agostino}, C.~J.},
        title = "{WHaD diagram: Classifying the ionizing source with one single emission line}",
      journal = {\aap},
         year = 2024,
        month = feb,
       volume = {682},
          eid = {A71},
        pages = {A71},
          doi = {10.1051/0004-6361/202347711},
archivePrefix = {arXiv},
       eprint = {2311.10573},
 primaryClass = {astro-ph.GA},
       adsurl = {https://ui.adsabs.harvard.edu/abs/2024A&A...682A..71S}
}

@ARTICLE{2005MNRAS.361.1063P,
       author = {{P{\'e}rez-Montero}, Enrique and {D{\'\i}az}, Angeles I.},
        title = "{A comparative analysis of empirical calibrators for nebular metallicity}",
      journal = {\mnras},
         year = 2005,
        month = aug,
       volume = {361},
       number = {3},
        pages = {1063-1076},
          doi = {10.1111/j.1365-2966.2005.09263.x},
archivePrefix = {arXiv},
       eprint = {astro-ph/0506344},
 primaryClass = {astro-ph},
       adsurl = {https://ui.adsabs.harvard.edu/abs/2005MNRAS.361.1063P}
}

@ARTICLE{2024RMxAA..60..323S,
       author = {{S{\'a}nchez}, S.~F. and {Garc{\'\i}a-Benito}, R. and {Gonz{\'a}lez Delgado}, R. and {Conrado}, A. and {Perez}, I. and {Lugo-Aranda}, A.~Z. and {S{\'a}nchez-Menguiano}, L. and {Ruiz-Lara}, T. and {Jim{\'e}nez}, A. and {Duarte Puertas}, S. and {Dom{\'\i}nguez-G{\'o}mez}, J. and {Torres-R{\'\i}os}, G. and {Argudo-Fern{\'a}ndez}, M. and {Bl{\'a}zquez-Calero}, G. and {Alc{\'a}zar-Laynez}, M. and {Verley}, S. and {Espada}, D. and {Lisenfeld}, U. and {Zurita}, A. and {Florido}, E. and {Bidaran}, B. and {Villalba-Gonz{\'a}lez}, P. and {Ferr{\'e}-Mateu}, A. and {S{\'a}nchez Alarc{\'o}n}, P.~M. and {Rom{\'a}n}, J. and {del Moral-Castro}, I. and {Ag{\"u}i}, F.},
        title = "{The CAVITY Project: Spatially-Resolved and Characteristic Properties of Galaxies Derived Using pyPipe3D}",
      journal = {\rmxaa},
         year = 2024,
        month = oct,
       volume = {60},
        pages = {323-341},
          doi = {10.22201/ia.01851101p.2024.60.02.10},
       adsurl = {https://ui.adsabs.harvard.edu/abs/2024RMxAA..60..323S}
}

@ARTICLE{2003Birnboim,
       author = {{Birnboim}, Yuval and {Dekel}, Avishai},
        title = "{Virial shocks in galactic haloes?}",
      journal = {\mnras},
         year = 2003,
        month = oct,
       volume = {345},
       number = {1},
        pages = {349-364},
          doi = {10.1046/j.1365-8711.2003.06955.x},
archivePrefix = {arXiv},
       eprint = {astro-ph/0302161},
 primaryClass = {astro-ph},
       adsurl = {https://ui.adsabs.harvard.edu/abs/2003MNRAS.345..349B}
}

@ARTICLE{2021A&A...649A.117G,
       author = {{Gal{\'a}rraga-Espinosa}, Daniela and {Aghanim}, Nabila and {Langer}, Mathieu and {Tanimura}, Hideki},
        title = "{Properties of gas phases around cosmic filaments at z = 0 in the IllustrisTNG simulation}",
      journal = {\aap},
         year = 2021,
        month = may,
       volume = {649},
          eid = {A117},
        pages = {A117},
          doi = {10.1051/0004-6361/202039781},
archivePrefix = {arXiv},
       eprint = {2010.15139},
 primaryClass = {astro-ph.CO},
       adsurl = {https://ui.adsabs.harvard.edu/abs/2021A&A...649A.117G}
}

@ARTICLE{2024Bulichi,
       author = {{Bulichi}, Teodora-Elena and {Dav{\'e}}, Romeel and {Kraljic}, Katarina},
        title = "{How galaxy properties vary with filament proximity in the SIMBA simulations}",
      journal = {\mnras},
         year = 2024,
        month = apr,
       volume = {529},
       number = {3},
        pages = {2595-2610},
          doi = {10.1093/mnras/stae667},
archivePrefix = {arXiv},
       eprint = {2309.03282},
 primaryClass = {astro-ph.GA},
       adsurl = {https://ui.adsabs.harvard.edu/abs/2024MNRAS.529.2595B}
}

@ARTICLE{2019AragonCalvo,
       author = {{Aragon Calvo}, Miguel A. and {Neyrinck}, Mark C. and {Silk}, Joseph},
        title = "{Galaxy Quenching from Cosmic Web Detachment}",
      journal = {The Open Journal of Astrophysics},
         year = 2019,
        month = jul,
       volume = {2},
       number = {1},
          eid = {7},
        pages = {7},
          doi = {10.21105/astro.1697.07881},
archivePrefix = {arXiv},
       eprint = {1607.07881},
 primaryClass = {astro-ph.GA},
       adsurl = {https://ui.adsabs.harvard.edu/abs/2019OJAp....2E...7A}
}

@ARTICLE{2010A&A...521L..53L,
       author = {{Lara-L{\'o}pez}, M.~A. and {Cepa}, J. and {Bongiovanni}, A. and {P{\'e}rez Garc{\'\i}a}, A.~M. and {Ederoclite}, A. and {Casta{\~n}eda}, H. and {Fern{\'a}ndez Lorenzo}, M. and {Povi{\'c}}, M. and {S{\'a}nchez-Portal}, M.},
        title = "{A fundamental plane for field star-forming galaxies}",
      journal = {\aap},
         year = 2010,
        month = oct,
       volume = {521},
          eid = {L53},
        pages = {L53},
          doi = {10.1051/0004-6361/201014803},
archivePrefix = {arXiv},
       eprint = {1005.0509},
 primaryClass = {astro-ph.CO},
       adsurl = {https://ui.adsabs.harvard.edu/abs/2010A&A...521L..53L}
}

@ARTICLE{2019ApJ...878L...6S,
       author = {{S{\'a}nchez Almeida}, J. and {S{\'a}nchez-Menguiano}, L.},
        title = "{The Fundamental Metallicity Relation Emerges from the Local Anti-correlation between Star Formation Rate and Gas-phase Metallicity that Exists in Disk Galaxies}",
      journal = {\apjl},
         year = 2019,
        month = jun,
       volume = {878},
       number = {1},
          eid = {L6},
        pages = {L6},
          doi = {10.3847/2041-8213/ab218d},
archivePrefix = {arXiv},
       eprint = {1905.05826},
 primaryClass = {astro-ph.GA},
       adsurl = {https://ui.adsabs.harvard.edu/abs/2019ApJ...878L...6S}
}

@ARTICLE{2021Chung,
       author = {{Chung}, Jiwon and {Kim}, Suk and {Rey}, Soo-Chang and {Lee}, Youngdae},
        title = "{Star-forming Dwarf Galaxies in Filamentary Structures around the Virgo Cluster: Probing Chemical Pre-processing in Filament Environments}",
      journal = {\apj},
         year = 2021,
        month = dec,
       volume = {923},
       number = {2},
          eid = {235},
        pages = {235},
          doi = {10.3847/1538-4357/ac3002},
archivePrefix = {arXiv},
       eprint = {2110.07836},
 primaryClass = {astro-ph.GA},
       adsurl = {https://ui.adsabs.harvard.edu/abs/2021ApJ...923..235C}
}

@ARTICLE{2010Dutton,
       author = {{Dutton}, Aaron A. and {van den Bosch}, Frank C. and {Dekel}, Avishai},
        title = "{On the origin of the galaxy star-formation-rate sequence: evolution and scatter}",
      journal = {\mnras},
         year = 2010,
        month = jul,
       volume = {405},
       number = {3},
        pages = {1690-1710},
          doi = {10.1111/j.1365-2966.2010.16620.x},
archivePrefix = {arXiv},
       eprint = {0912.2169},
 primaryClass = {astro-ph.CO},
       adsurl = {https://ui.adsabs.harvard.edu/abs/2010MNRAS.405.1690D}
}

@ARTICLE{2012Geha,
       author = {{Geha}, M. and {Blanton}, M.~R. and {Yan}, R. and {Tinker}, J.~L.},
        title = "{A Stellar Mass Threshold for Quenching of Field Galaxies}",
      journal = {\apj},
         year = 2012,
        month = sep,
       volume = {757},
       number = {1},
          eid = {85},
        pages = {85},
          doi = {10.1088/0004-637X/757/1/85},
archivePrefix = {arXiv},
       eprint = {1206.3573},
 primaryClass = {astro-ph.CO},
       adsurl = {https://ui.adsabs.harvard.edu/abs/2012ApJ...757...85G}
}

@ARTICLE{2021Dickey,
       author = {{Dickey}, Claire M. and {Starkenburg}, Tjitske K. and {Geha}, Marla and {Hahn}, ChangHoon and {Angl{\'e}s-Alc{\'a}zar}, Daniel and {Choi}, Ena and {Dav{\'e}}, Romeel and {Genel}, Shy and {Iyer}, Kartheik G. and {Maller}, Ariyeh H. and {Mandelker}, Nir and {Somerville}, Rachel S. and {Yung}, L.~Y. Aaron},
        title = "{IQ Collaboratory. II. The Quiescent Fraction of Isolated, Low-mass Galaxies across Simulations and Observations}",
      journal = {\apj},
         year = 2021,
        month = jul,
       volume = {915},
       number = {1},
          eid = {53},
        pages = {53},
          doi = {10.3847/1538-4357/abc014},
archivePrefix = {arXiv},
       eprint = {2010.01132},
 primaryClass = {astro-ph.GA},
       adsurl = {https://ui.adsabs.harvard.edu/abs/2021ApJ...915...53D}
}

@ARTICLE{2009Tolstoy,
       author = {{Tolstoy}, Eline and {Hill}, Vanessa and {Tosi}, Monica},
        title = "{Star-Formation Histories, Abundances, and Kinematics of Dwarf Galaxies in the Local Group}",
      journal = {\araa},
         year = 2009,
        month = sep,
       volume = {47},
       number = {1},
        pages = {371-425},
          doi = {10.1146/annurev-astro-082708-101650},
archivePrefix = {arXiv},
       eprint = {0904.4505},
 primaryClass = {astro-ph.CO},
       adsurl = {https://ui.adsabs.harvard.edu/abs/2009ARA&A..47..371T}
}

@ARTICLE{1980Larson,
       author = {{Larson}, R.~B. and {Tinsley}, B.~M. and {Caldwell}, C.~N.},
        title = "{The evolution of disk galaxies and the origin of S0 galaxies}",
      journal = {\apj},
         year = 1980,
        month = may,
       volume = {237},
        pages = {692-707},
          doi = {10.1086/157917},
       adsurl = {https://ui.adsabs.harvard.edu/abs/1980ApJ...237..692L}
}

@ARTICLE{1998Fujita,
       author = {{Fujita}, Yutaka},
        title = "{Quantitative Estimates of Environmental Effects on the Star Formation Rate of Disk Galaxies in Clusters of Galaxies}",
      journal = {\apj},
         year = 1998,
        month = dec,
       volume = {509},
       number = {2},
        pages = {587-594},
          doi = {10.1086/306518},
archivePrefix = {arXiv},
       eprint = {astro-ph/9807120},
 primaryClass = {astro-ph},
       adsurl = {https://ui.adsabs.harvard.edu/abs/1998ApJ...509..587F}
}

@ARTICLE{2020Rhee,
       author = {{Rhee}, Jinsu and {Smith}, Rory and {Choi}, Hoseung and {Contini}, Emanuele and {Jung}, S. Lyla and {Han}, San and {Yi}, Sukyoung K.},
        title = "{YZiCS: Unveiling the Quenching History of Cluster Galaxies Using Phase-space Analysis}",
      journal = {\apjs},
         year = 2020,
        month = apr,
       volume = {247},
       number = {2},
          eid = {45},
        pages = {45},
          doi = {10.3847/1538-4365/ab7377},
archivePrefix = {arXiv},
       eprint = {2002.04645},
 primaryClass = {astro-ph.GA},
       adsurl = {https://ui.adsabs.harvard.edu/abs/2020ApJS..247...45R}
}

@ARTICLE{2025Molina-Calzada,
       author = {{Molina-Calzada}, J.~A. and {Lara-L{\'o}pez}, M.~A. and {Gallego}, J. and {Hopkins}, A.~M. and {Holwerda}, B.~W. and {L{\'o}pez-S{\'a}nchez}, A.~R.},
        title = "{Galaxy And Mass Assembly (GAMA): From filaments to voids, how extreme environment affects gas metallicity and SFR in galaxies}",
      journal = {\aap},
         year = 2025,
        month = aug,
       volume = {700},
          eid = {A267},
        pages = {A267},
          doi = {10.1051/0004-6361/202555871},
archivePrefix = {arXiv},
       eprint = {2507.06781},
 primaryClass = {astro-ph.GA},
       adsurl = {https://ui.adsabs.harvard.edu/abs/2025A&A...700A.267M}
}

@ARTICLE{2013Marino,
       author = {{Marino}, R.~A. and {Rosales-Ortega}, F.~F. and {S{\'a}nchez}, S.~F. and {Gil de Paz}, A. and {V{\'\i}lchez}, J. and {Miralles-Caballero}, D. and {Kehrig}, C. and {P{\'e}rez-Montero}, E. and {Stanishev}, V. and {Iglesias-P{\'a}ramo}, J. and {D{\'\i}az}, A.~I. and {Castillo-Morales}, A. and {Kennicutt}, R. and {L{\'o}pez-S{\'a}nchez}, A.~R. and {Galbany}, L. and {Garc{\'\i}a-Benito}, R. and {Mast}, D. and {Mendez-Abreu}, J. and {Monreal-Ibero}, A. and {Husemann}, B. and {Walcher}, C.~J. and {Garc{\'\i}a-Lorenzo}, B. and {Masegosa}, J. and {Del Olmo Orozco}, A. and {Mour{\~a}o}, A.~M. and {Ziegler}, B. and {Moll{\'a}}, M. and {Papaderos}, P. and {S{\'a}nchez-Bl{\'a}zquez}, P. and {Gonz{\'a}lez Delgado}, R.~M. and {Falc{\'o}n-Barroso}, J. and {Roth}, M.~M. and {van de Ven}, G. and {CALIFA Team}},
        title = "{The O3N2 and N2 abundance indicators revisited: improved calibrations based on CALIFA and T$_{e}$-based literature data}",
      journal = {\aap},
         year = 2013,
        month = nov,
       volume = {559},
          eid = {A114},
        pages = {A114},
          doi = {10.1051/0004-6361/201321956},
archivePrefix = {arXiv},
       eprint = {1307.5316},
 primaryClass = {astro-ph.CO},
       adsurl = {https://ui.adsabs.harvard.edu/abs/2013A&A...559A.114M}
}

@ARTICLE{2009Kapferer,
       author = {{Kapferer}, W. and {Kronberger}, T. and {Breitschwerdt}, D. and {Schindler}, S. and {van Kampen}, E. and {Kimeswenger}, S. and {Domainko}, W. and {Mair}, M. and {Ruffert}, M.},
        title = "{Metal enrichment of the intra-cluster medium by thermally and cosmic-ray driven galactic winds. An analytical prescription for galactic outflows}",
      journal = {\aap},
         year = 2009,
        month = sep,
       volume = {504},
       number = {3},
        pages = {719-726},
          doi = {10.1051/0004-6361/200912099},
archivePrefix = {arXiv},
       eprint = {0907.3800},
 primaryClass = {astro-ph.CO},
       adsurl = {https://ui.adsabs.harvard.edu/abs/2009A&A...504..719K}
}

@ARTICLE{2006Kapferer,
       author = {{Kapferer}, W. and {Ferrari}, C. and {Domainko}, W. and {Mair}, M. and {Kronberger}, T. and {Schindler}, S. and {Kimeswenger}, S. and {van Kampen}, E. and {Breitschwerdt}, D. and {Ruffert}, M.},
        title = "{Simulations of galactic winds and starbursts in galaxy clusters}",
      journal = {\aap},
         year = 2006,
        month = mar,
       volume = {447},
       number = {3},
        pages = {827-842},
          doi = {10.1051/0004-6361:20053975},
archivePrefix = {arXiv},
       eprint = {astro-ph/0508107},
 primaryClass = {astro-ph},
       adsurl = {https://ui.adsabs.harvard.edu/abs/2006A&A...447..827K}
}

@ARTICLE{2005Schindler,
       author = {{Schindler}, S. and {Kapferer}, W. and {Domainko}, W. and {Mair}, M. and {van Kampen}, E. and {Kronberger}, T. and {Kimeswenger}, S. and {Ruffert}, M. and {Mangete}, O. and {Breitschwerdt}, D.},
        title = "{Metal enrichment processes in the intra-cluster medium}",
      journal = {\aap},
         year = 2005,
        month = may,
       volume = {435},
       number = {2},
        pages = {L25-L28},
          doi = {10.1051/0004-6361:200500107},
archivePrefix = {arXiv},
       eprint = {astro-ph/0504068},
 primaryClass = {astro-ph},
       adsurl = {https://ui.adsabs.harvard.edu/abs/2005A&A...435L..25S}
}

@ARTICLE{2001Vollmer,
       author = {{Vollmer}, B. and {Cayatte}, V. and {Balkowski}, C. and {Duschl}, W.~J.},
        title = "{Ram Pressure Stripping and Galaxy Orbits: The Case of the Virgo Cluster}",
      journal = {\apj},
         year = 2001,
        month = nov,
       volume = {561},
       number = {2},
        pages = {708-726},
          doi = {10.1086/323368},
archivePrefix = {arXiv},
       eprint = {astro-ph/0107237},
 primaryClass = {astro-ph},
       adsurl = {https://ui.adsabs.harvard.edu/abs/2001ApJ...561..708V}
}

@ARTICLE{1999Fujita,
       author = {{Fujita}, Yutaka and {Nagashima}, Masahiro},
        title = "{Effects of Ram Pressure from the Intracluster Medium on the Star Formation Rate of Disk Galaxies in Clusters of Galaxies}",
      journal = {\apj},
         year = 1999,
        month = may,
       volume = {516},
       number = {2},
        pages = {619-625},
          doi = {10.1086/307139},
archivePrefix = {arXiv},
       eprint = {astro-ph/9812378},
 primaryClass = {astro-ph},
       adsurl = {https://ui.adsabs.harvard.edu/abs/1999ApJ...516..619F}
}

@ARTICLE{1977Cowie,
       author = {{Cowie}, L.~L. and {Songaila}, A.},
        title = "{Thermal evaporation of gas within galaxies by a hot intergalactic medium}",
      journal = {\nat},
         year = 1977,
        month = apr,
       volume = {266},
        pages = {501-503},
          doi = {10.1038/266501a0},
       adsurl = {https://ui.adsabs.harvard.edu/abs/1977Natur.266..501C}
}

@ARTICLE{1972Gunn,
       author = {{Gunn}, James E. and {Gott}, III, J. Richard},
        title = "{On the Infall of Matter Into Clusters of Galaxies and Some Effects on Their Evolution}",
      journal = {\apj},
         year = 1972,
        month = aug,
       volume = {176},
        pages = {1},
          doi = {10.1086/151605},
       adsurl = {https://ui.adsabs.harvard.edu/abs/1972ApJ...176....1G}
}

@ARTICLE{2015Peng,
       author = {{Peng}, Y. and {Maiolino}, R. and {Cochrane}, R.},
        title = "{Strangulation as the primary mechanism for shutting down star formation in galaxies}",
      journal = {\nat},
         year = 2015,
        month = may,
       volume = {521},
       number = {7551},
        pages = {192-195},
          doi = {10.1038/nature14439},
archivePrefix = {arXiv},
       eprint = {1505.03143},
 primaryClass = {astro-ph.GA},
       adsurl = {https://ui.adsabs.harvard.edu/abs/2015Natur.521..192P}
}

@ARTICLE{2011MNRAS.416.1354D,
       author = {{Dav{\'e}}, Romeel and {Finlator}, Kristian and {Oppenheimer}, Benjamin D.},
        title = "{Galaxy evolution in cosmological simulations with outflows - II. Metallicities and gas fractions}",
      journal = {\mnras},
         year = 2011,
        month = sep,
       volume = {416},
       number = {2},
        pages = {1354-1376},
          doi = {10.1111/j.1365-2966.2011.19132.x},
archivePrefix = {arXiv},
       eprint = {1104.3156},
 primaryClass = {astro-ph.CO},
       adsurl = {https://ui.adsabs.harvard.edu/abs/2011MNRAS.416.1354D}
}

@ARTICLE{2018Kraljic,
       author = {{Kraljic}, K. and {Arnouts}, S. and {Pichon}, C. and {Laigle}, C. and {de la Torre}, S. and {Vibert}, D. and {Cadiou}, C. and {Dubois}, Y. and {Treyer}, M. and {Schimd}, C. and {Codis}, S. and {de Lapparent}, V. and {Devriendt}, J. and {Hwang}, H.~S. and {Le Borgne}, D. and {Malavasi}, N. and {Milliard}, B. and {Musso}, M. and {Pogosyan}, D. and {Alpaslan}, M. and {Bland-Hawthorn}, J. and {Wright}, A.~H.},
        title = "{Galaxy evolution in the metric of the cosmic web}",
      journal = {\mnras},
         year = 2018,
        month = feb,
       volume = {474},
       number = {1},
        pages = {547-571},
          doi = {10.1093/mnras/stx2638},
archivePrefix = {arXiv},
       eprint = {1710.02676},
 primaryClass = {astro-ph.GA},
       adsurl = {https://ui.adsabs.harvard.edu/abs/2018MNRAS.474..547K}
}

@ARTICLE{2023Hasan,
       author = {{Hasan}, Farhanul and {Burchett}, Joseph N. and {Abeyta}, Alyssa and {Hellinger}, Douglas and {Mandelker}, Nir and {Primack}, Joel R. and {Faber}, S.~M. and {Koo}, David C. and {Elek}, Oskar and {Nagai}, Daisuke},
        title = "{The Evolving Effect of Cosmic Web Environment on Galaxy Quenching}",
      journal = {\apj},
         year = 2023,
        month = jun,
       volume = {950},
       number = {2},
          eid = {114},
        pages = {114},
          doi = {10.3847/1538-4357/acd11c},
archivePrefix = {arXiv},
       eprint = {2303.08088},
 primaryClass = {astro-ph.GA},
       adsurl = {https://ui.adsabs.harvard.edu/abs/2023ApJ...950..114H}
}

@ARTICLE{2016MNRAS.455..127M,
       author = {{Mart{\'\i}nez}, H{\'e}ctor J. and {Muriel}, Hern{\'a}n and {Coenda}, Valeria},
        title = "{Galaxies infalling into groups: filaments versus isotropic infall}",
      journal = {\mnras},
         year = 2016,
        month = jan,
       volume = {455},
       number = {1},
        pages = {127-135},
          doi = {10.1093/mnras/stv2295},
archivePrefix = {arXiv},
       eprint = {1510.00390},
 primaryClass = {astro-ph.GA},
       adsurl = {https://ui.adsabs.harvard.edu/abs/2016MNRAS.455..127M}
}

@ARTICLE{2025Finn,
       author = {{Finn}, Rose A. and {Rudnick}, Gregory and {Jablonka}, Pascale and {Ramatsoku}, Mpati and {Nagaraj}, Gautam and {Vulcani}, Benedetta and {Koopmann}, Rebecca A. and {Fossati}, Matteo and {Agostino}, James and {Bah{\'e}}, Yannick and {Garcia-Burillo}, Santiago and {Castignani}, Gianluca and {Combes}, Francoise and {Conger}, Kim and {De Lucia}, Gabriella and {Desai}, Vandana and {Moustakas}, John and {Norman}, Dara and {Sperone-Longin}, Damien and {Townsend}, Melinda and {Xie}, Lizhi and {Zakharova}, Daria and {Zaritsky}, Dennis},
        title = "{Virgo Filaments. V. Disrupting the Baryon Cycle in the NGC 5364 Galaxy Group}",
      journal = {\apj},
         year = 2025,
        month = may,
       volume = {985},
       number = {1},
          eid = {81},
        pages = {81},
          doi = {10.3847/1538-4357/adc566},
archivePrefix = {arXiv},
       eprint = {2505.09782},
 primaryClass = {astro-ph.GA},
       adsurl = {https://ui.adsabs.harvard.edu/abs/2025ApJ...985...81F}
}

@ARTICLE{2019Sarron,
       author = {{Sarron}, F. and {Adami}, C. and {Durret}, F. and {Laigle}, C.},
        title = "{Pre-processing of galaxies in cosmic filaments around AMASCFI clusters in the CFHTLS}",
      journal = {\aap},
         year = 2019,
        month = dec,
       volume = {632},
          eid = {A49},
        pages = {A49},
          doi = {10.1051/0004-6361/201935394},
archivePrefix = {arXiv},
       eprint = {1903.02879},
 primaryClass = {astro-ph.GA},
       adsurl = {https://ui.adsabs.harvard.edu/abs/2019A&A...632A..49S}
}

@ARTICLE{2017Kuutma,
       author = {{Kuutma}, Teet and {Tamm}, Antti and {Tempel}, Elmo},
        title = "{From voids to filaments: environmental transformations of galaxies in the SDSS}",
      journal = {\aap},
         year = 2017,
        month = apr,
       volume = {600},
          eid = {L6},
        pages = {L6},
          doi = {10.1051/0004-6361/201730526},
archivePrefix = {arXiv},
       eprint = {1703.04338},
 primaryClass = {astro-ph.GA},
       adsurl = {https://ui.adsabs.harvard.edu/abs/2017A&A...600L...6K}
}

@ARTICLE{2017Chen,
       author = {{Chen}, Yen-Chi and {Ho}, Shirley and {Mandelbaum}, Rachel and {Bahcall}, Neta A. and {Brownstein}, Joel R. and {Freeman}, Peter E. and {Genovese}, Christopher R. and {Schneider}, Donald P. and {Wasserman}, Larry},
        title = "{Detecting effects of filaments on galaxy properties in the Sloan Digital Sky Survey III}",
      journal = {\mnras},
         year = 2017,
        month = apr,
       volume = {466},
       number = {2},
        pages = {1880-1893},
          doi = {10.1093/mnras/stw3127},
archivePrefix = {arXiv},
       eprint = {1509.06376},
 primaryClass = {astro-ph.GA},
       adsurl = {https://ui.adsabs.harvard.edu/abs/2017MNRAS.466.1880C}
}

@ARTICLE{2021Winkel,
       author = {{Winkel}, N. and {Pasquali}, A. and {Kraljic}, K. and {Smith}, R. and {Gallazzi}, A. and {Jackson}, T.~M.},
        title = "{The imprint of cosmic web quenching on central galaxies}",
      journal = {\mnras},
         year = 2021,
        month = aug,
       volume = {505},
       number = {4},
        pages = {4920-4934},
          doi = {10.1093/mnras/stab1562},
archivePrefix = {arXiv},
       eprint = {2105.13368},
 primaryClass = {astro-ph.GA},
       adsurl = {https://ui.adsabs.harvard.edu/abs/2021MNRAS.505.4920W}
}

@ARTICLE{2012Pasquali,
       author = {{Pasquali}, Anna and {Gallazzi}, Anna and {van den Bosch}, Frank C.},
        title = "{The gas-phase metallicity of central and satellite galaxies in the Sloan Digital Sky Survey}",
      journal = {\mnras},
         year = 2012,
        month = sep,
       volume = {425},
       number = {1},
        pages = {273-286},
          doi = {10.1111/j.1365-2966.2012.21454.x},
archivePrefix = {arXiv},
       eprint = {1206.3458},
 primaryClass = {astro-ph.CO},
       adsurl = {https://ui.adsabs.harvard.edu/abs/2012MNRAS.425..273P}
}

@ARTICLE{2011Cen,
       author = {{Cen}, Renyue},
        title = "{Environmentally Driven Global Evolution of Galaxies}",
      journal = {\apj},
         year = 2011,
        month = nov,
       volume = {741},
       number = {2},
          eid = {99},
        pages = {99},
          doi = {10.1088/0004-637X/741/2/99},
archivePrefix = {arXiv},
       eprint = {1104.5046},
 primaryClass = {astro-ph.CO},
       adsurl = {https://ui.adsabs.harvard.edu/abs/2011ApJ...741...99C}
}

@ARTICLE{2004DeLucia,
       author = {{De Lucia}, Gabriella and {Kauffmann}, Guinevere and {White}, Simon D.~M.},
        title = "{Chemical enrichment of the intracluster and intergalactic medium in a hierarchical galaxy formation model}",
      journal = {\mnras},
         year = 2004,
        month = apr,
       volume = {349},
       number = {3},
        pages = {1101-1116},
          doi = {10.1111/j.1365-2966.2004.07584.x},
archivePrefix = {arXiv},
       eprint = {astro-ph/0310268},
 primaryClass = {astro-ph},
       adsurl = {https://ui.adsabs.harvard.edu/abs/2004MNRAS.349.1101D}
}

@ARTICLE{2008Finlator,
       author = {{Finlator}, Kristian and {Dav{\'e}}, Romeel},
        title = "{The origin of the galaxy mass-metallicity relation and implications for galactic outflows}",
      journal = {\mnras},
         year = 2008,
        month = apr,
       volume = {385},
       number = {4},
        pages = {2181-2204},
          doi = {10.1111/j.1365-2966.2008.12991.x},
archivePrefix = {arXiv},
       eprint = {0704.3100},
 primaryClass = {astro-ph},
       adsurl = {https://ui.adsabs.harvard.edu/abs/2008MNRAS.385.2181F}
}

@ARTICLE{2015Maddox,
       author = {{Maddox}, Natasha and {Hess}, Kelley M. and {Obreschkow}, Danail and {Jarvis}, M.~J. and {Blyth}, S. -L.},
        title = "{Variation of galactic cold gas reservoirs with stellar mass}",
      journal = {\mnras},
         year = 2015,
        month = feb,
       volume = {447},
       number = {2},
        pages = {1610-1617},
          doi = {10.1093/mnras/stu2532},
archivePrefix = {arXiv},
       eprint = {1412.0852},
 primaryClass = {astro-ph.GA},
       adsurl = {https://ui.adsabs.harvard.edu/abs/2015MNRAS.447.1610M}
}

@ARTICLE{2012Huang,
       author = {{Huang}, Shan and {Haynes}, Martha P. and {Giovanelli}, Riccardo and {Brinchmann}, Jarle},
        title = "{The Arecibo Legacy Fast ALFA Survey: The Galaxy Population Detected by ALFALFA}",
      journal = {\apj},
         year = 2012,
        month = sep,
       volume = {756},
       number = {2},
          eid = {113},
        pages = {113},
          doi = {10.1088/0004-637X/756/2/113},
archivePrefix = {arXiv},
       eprint = {1207.0523},
 primaryClass = {astro-ph.CO},
       adsurl = {https://ui.adsabs.harvard.edu/abs/2012ApJ...756..113H}
}

@ARTICLE{2018Haynes,
       author = {{Haynes}, Martha P. and {Giovanelli}, Riccardo and {Kent}, Brian R. and {Adams}, Elizabeth A.~K. and {Balonek}, Thomas J. and {Craig}, David W. and {Fertig}, Derek and {Finn}, Rose and {Giovanardi}, Carlo and {Hallenbeck}, Gregory and {Hess}, Kelley M. and {Hoffman}, G. Lyle and {Huang}, Shan and {Jones}, Michael G. and {Koopmann}, Rebecca A. and {Kornreich}, David A. and {Leisman}, Lukas and {Miller}, Jeffrey and {Moorman}, Crystal and {O'Connor}, Jessica and {O'Donoghue}, Aileen and {Papastergis}, Emmanouil and {Troischt}, Parker and {Stark}, David and {Xiao}, Li},
        title = "{The Arecibo Legacy Fast ALFA Survey: The ALFALFA Extragalactic H I Source Catalog}",
      journal = {\apj},
         year = 2018,
        month = jul,
       volume = {861},
       number = {1},
          eid = {49},
        pages = {49},
          doi = {10.3847/1538-4357/aac956},
archivePrefix = {arXiv},
       eprint = {1805.11499},
 primaryClass = {astro-ph.GA},
       adsurl = {https://ui.adsabs.harvard.edu/abs/2018ApJ...861...49H}
}

@ARTICLE{2007Levy,
       author = {{Levy}, Lorenza and {Rose}, James A. and {van Gorkom}, Jacqueline H. and {Chaboyer}, Brian},
        title = "{The Effect of Cluster Environment on Galaxy Evolution in the Pegasus I Cluster}",
      journal = {\aj},
         year = 2007,
        month = mar,
       volume = {133},
       number = {3},
        pages = {1104-1124},
          doi = {10.1086/510723},
archivePrefix = {arXiv},
       eprint = {astro-ph/0611591},
 primaryClass = {astro-ph},
       adsurl = {https://ui.adsabs.harvard.edu/abs/2007AJ....133.1104L}
}

@ARTICLE{2001Solanes,
       author = {{Solanes}, Jos{\'e} M. and {Manrique}, Alberto and {Garc{\'\i}a-G{\'o}mez}, Carlos and {Gonz{\'a}lez-Casado}, Guillermo and {Giovanelli}, Riccardo and {Haynes}, Martha P.},
        title = "{The H I Content of Spirals. II. Gas Deficiency in Cluster Galaxies}",
      journal = {\apj},
         year = 2001,
        month = feb,
       volume = {548},
       number = {1},
        pages = {97-113},
          doi = {10.1086/318672},
archivePrefix = {arXiv},
       eprint = {astro-ph/0007402},
 primaryClass = {astro-ph},
       adsurl = {https://ui.adsabs.harvard.edu/abs/2001ApJ...548...97S}
}

@ARTICLE{2021Asplund,
       author = {{Asplund}, M. and {Amarsi}, A.~M. and {Grevesse}, N.},
        title = "{The chemical make-up of the Sun: A 2020 vision}",
      journal = {\aap},
         year = 2021,
        month = sep,
       volume = {653},
          eid = {A141},
        pages = {A141},
          doi = {10.1051/0004-6361/202140445},
archivePrefix = {arXiv},
       eprint = {2105.01661},
 primaryClass = {astro-ph.SR},
       adsurl = {https://ui.adsabs.harvard.edu/abs/2021A&A...653A.141A}
}

@ARTICLE{2008Kewley,
       author = {{Kewley}, Lisa J. and {Ellison}, Sara L.},
        title = "{Metallicity Calibrations and the Mass-Metallicity Relation for Star-forming Galaxies}",
      journal = {\apj},
         year = 2008,
        month = jul,
       volume = {681},
       number = {2},
        pages = {1183-1204},
          doi = {10.1086/587500},
archivePrefix = {arXiv},
       eprint = {0801.1849},
 primaryClass = {astro-ph},
       adsurl = {https://ui.adsabs.harvard.edu/abs/2008ApJ...681.1183K}
}

@ARTICLE{1999Kobulnicky,
       author = {{Kobulnicky}, Henry A. and {Kennicutt}, Jr., Robert C. and {Pizagno}, James L.},
        title = "{On Measuring Nebular Chemical Abundances in Distant Galaxies Using Global Emission-Line Spectra}",
      journal = {\apj},
         year = 1999,
        month = apr,
       volume = {514},
       number = {2},
        pages = {544-557},
          doi = {10.1086/306987},
archivePrefix = {arXiv},
       eprint = {astro-ph/9811006},
 primaryClass = {astro-ph},
       adsurl = {https://ui.adsabs.harvard.edu/abs/1999ApJ...514..544K}
}

@ARTICLE{1991McGaugh,
       author = {{McGaugh}, Stacy S.},
        title = "{H II Region Abundances: Model Oxygen Line Ratios}",
      journal = {\apj},
         year = 1991,
        month = oct,
       volume = {380},
        pages = {140},
          doi = {10.1086/170569},
       adsurl = {https://ui.adsabs.harvard.edu/abs/1991ApJ...380..140M}
}

@ARTICLE{2024Scholte,
       author = {{Scholte}, Dirk and {Saintonge}, Am{\'e}lie and {Moustakas}, John and {Catinella}, Barbara and {Zou}, Hu and {Dey}, Biprateep and {Aguilar}, J. and {Ahlen}, S. and {Anand}, A. and {Blum}, R. and {Brooks}, D. and {Circosta}, C. and {Claybaugh}, T. and {de la Macorra}, A. and {Doel}, P. and {Font-Ribera}, A. and {F{\"o}rster}, P.~U. and {Forero-Romero}, J.~E. and {Gazta{\~n}aga}, E. and {Gontcho A Gontcho}, S. and {Juneau}, S. and {Kehoe}, R. and {Kisner}, T. and {Koposov}, S.~E. and {Kremin}, A. and {Lambert}, A. and {Landriau}, M. and {Maraston}, C. and {Martini}, P. and {Meisner}, A. and {Mighty}, A.~S. and {Miquel}, R. and {Myers}, A.~D. and {Nie}, J. and {Poppett}, C. and {Prada}, F. and {Rezaie}, M. and {Rossi}, G. and {Sanchez}, E. and {Schubnell}, M. and {Silber}, J. and {Sprayberry}, D. and {Siudek}, M. and {Speranza}, F. and {Tarl{\'e}}, G. and {Tojeiro}, R. and {Weaver}, B.~A.},
        title = "{The atomic gas sequence and mass-metallicity relation from dwarfs to massive galaxies}",
      journal = {\mnras},
         year = 2024,
        month = dec,
       volume = {535},
       number = {3},
        pages = {2341-2356},
          doi = {10.1093/mnras/stae2477},
archivePrefix = {arXiv},
       eprint = {2408.03996},
 primaryClass = {astro-ph.GA},
       adsurl = {https://ui.adsabs.harvard.edu/abs/2024MNRAS.535.2341S}
}

@ARTICLE{2013MNRAS.430.3017B,
       author = {{Bah{\'e}}, Yannick M. and {McCarthy}, Ian G. and {Balogh}, Michael L. and {Font}, Andreea S.},
        title = "{Why does the environmental influence on group and cluster galaxies extend beyond the virial radius?}",
      journal = {\mnras},
         year = 2013,
        month = apr,
       volume = {430},
       number = {4},
        pages = {3017-3031},
          doi = {10.1093/mnras/stt109},
archivePrefix = {arXiv},
       eprint = {1210.8407},
 primaryClass = {astro-ph.CO},
       adsurl = {https://ui.adsabs.harvard.edu/abs/2013MNRAS.430.3017B}
}

@ARTICLE{2001Peebles,
       author = {{Peebles}, P.~J.~E.},
        title = "{The Void Phenomenon}",
      journal = {\apj},
         year = 2001,
        month = aug,
       volume = {557},
       number = {2},
        pages = {495-504},
          doi = {10.1086/322254},
archivePrefix = {arXiv},
       eprint = {astro-ph/0101127},
 primaryClass = {astro-ph},
       adsurl = {https://ui.adsabs.harvard.edu/abs/2001ApJ...557..495P}
}

@ARTICLE{2025Zarattini,
       author = {{Zarattini}, S. and {Aguerri}, J.~A.~L.},
        title = "{Galaxy transformation across the cosmic web: Evolution of stellar colours and star formation rates in filaments}",
      journal = {\aap},
         year = 2025,
        month = jun,
       volume = {698},
          eid = {A196},
        pages = {A196},
          doi = {10.1051/0004-6361/202453053},
archivePrefix = {arXiv},
       eprint = {2504.02026},
 primaryClass = {astro-ph.GA},
       adsurl = {https://ui.adsabs.harvard.edu/abs/2025A&A...698A.196Z}
}

@ARTICLE{2019Kraljic,
       author = {{Kraljic}, K. and {Pichon}, C. and {Dubois}, Y. and {Codis}, S. and {Cadiou}, C. and {Devriendt}, J. and {Musso}, M. and {Welker}, C. and {Arnouts}, S. and {Hwang}, H.~S. and {Laigle}, C. and {Peirani}, S. and {Slyz}, A. and {Treyer}, M. and {Vibert}, D.},
        title = "{Galaxies flowing in the oriented saddle frame of the cosmic web}",
      journal = {\mnras},
         year = 2019,
        month = mar,
       volume = {483},
       number = {3},
        pages = {3227-3254},
          doi = {10.1093/mnras/sty3216},
archivePrefix = {arXiv},
       eprint = {1810.05211},
 primaryClass = {astro-ph.GA},
       adsurl = {https://ui.adsabs.harvard.edu/abs/2019MNRAS.483.3227K}
}

@ARTICLE{2024Baker,
       author = {{Baker}, William M. and {Maiolino}, Roberto and {Bluck}, Asa F.~L. and {Belfiore}, Francesco and {Curti}, Mirko and {D'Eugenio}, Francesco and {Piotrowska}, Joanna M. and {Tacchella}, Sandro and {Trussler}, James A.~A.},
        title = "{Different regulation of stellar metallicities between star-forming and quiescent galaxies - insights into galaxy quenching}",
      journal = {\mnras},
         year = 2024,
        month = oct,
       volume = {534},
       number = {1},
        pages = {30-38},
          doi = {10.1093/mnras/stae2059},
archivePrefix = {arXiv},
       eprint = {2309.00670},
 primaryClass = {astro-ph.GA},
       adsurl = {https://ui.adsabs.harvard.edu/abs/2024MNRAS.534...30B}
}

@ARTICLE{2002Bekki,
       author = {{Bekki}, Kenji and {Couch}, Warrick J. and {Shioya}, Yasuhiro},
        title = "{Passive Spiral Formation from Halo Gas Starvation: Gradual Transformation into S0s}",
      journal = {\apj},
         year = 2002,
        month = oct,
       volume = {577},
       number = {2},
        pages = {651-657},
          doi = {10.1086/342221},
archivePrefix = {arXiv},
       eprint = {astro-ph/0206207},
 primaryClass = {astro-ph},
       adsurl = {https://ui.adsabs.harvard.edu/abs/2002ApJ...577..651B}
}

\begin{appendix}

\section{Comparison sample selection}\label{appendix0}

\begin{figure}
\centering
\includegraphics[width=\columnwidth]{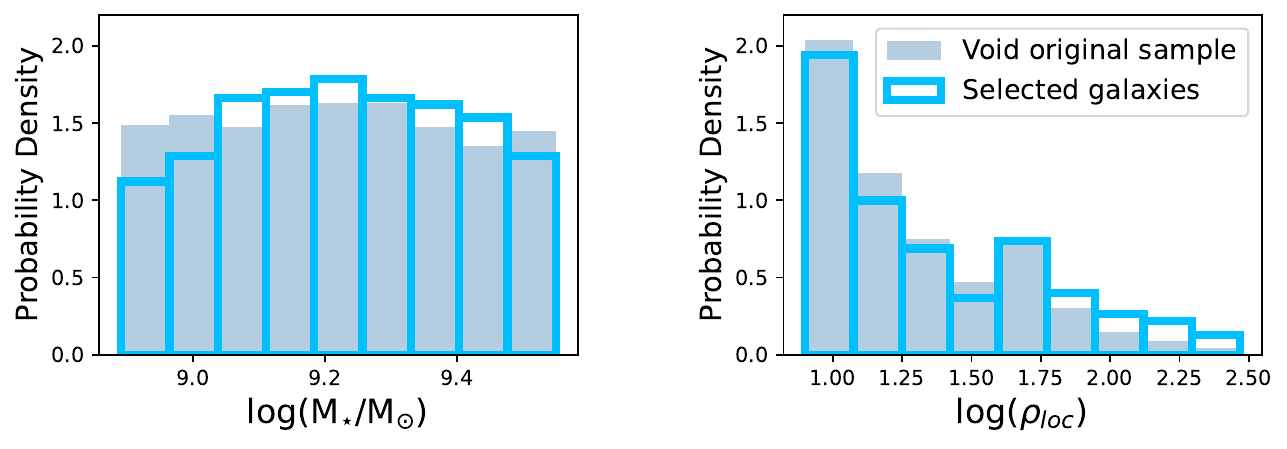}
\includegraphics[width=\columnwidth]{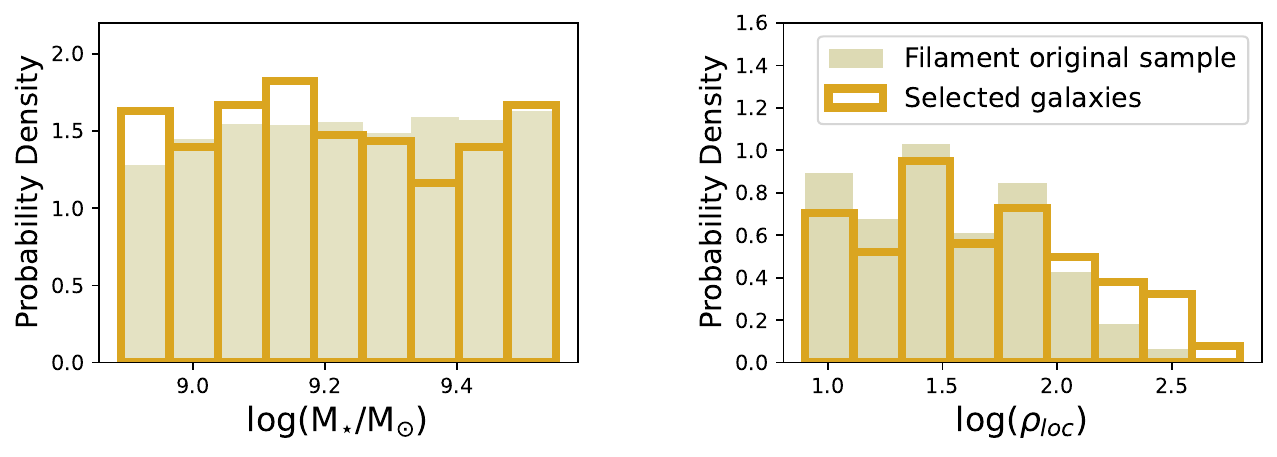}
\caption{Comparison of the distributions of non-isolated dwarf galaxies in voids and filaments with the selected targets. In each panel, filled histograms represent non-isolated dwarf galaxies, while empty histograms correspond to the selected targets. The left-hand panels display the stellar mass distributions, and the right-hand panels show the distributions in local density, defined as $\rho_{\rm loc}$\,=\,N$_{\rm neighbours}$ / V$_{\rm r=1.5 Mpc}$. Distributions for void galaxies are presented in the top panels, and those for filament galaxies are shown in the bottom panels.
} 
\label{Fig1}
\end{figure}

To create two representative subsamples of non-isolated dwarf galaxies in voids and filaments, we applied a stratified sampling approach. This method ensures that the final sample reflects the diversity and distribution of the full dataset in the two-dimensional parameter space of stellar mass and local density. Since the original samples presented in \citet{2023A&A...680A.111D} were already defined with considerations for key parameters such as redshift coverage, we do not base the classification at this stage on those parameters. We divided the stellar mass and local density of the galaxies into five bins each, creating a 5$\times$5 grid across the two variables, and performed stratified sampling by randomly selecting a fixed number of galaxies from each bin to ensure even coverage of the parameter space. If a bin had fewer galaxies than needed, we included all available ones. This procedure is only applied to the void and filament non-isolated dwarf samples. The cluster sample was sufficiently small that no additional sampling was necessary. In Fig.~\ref{Fig1}, we compare the stellar mass and local density of the non-isolated dwarf galaxies in voids and filaments with the selected ones. The stellar mass and local density distribution of non-isolated dwarf galaxies in voids and filaments are shown with filled histograms in the top and bottom panels, respectively. 

{To assess the impact of sampling variance introduced by the stratified selection of non-isolated galaxies, we performed a Monte Carlo resampling test. Starting from the full, non-isolated samples in voids and filaments, we repeatedly constructed stratified subsamples matched for stellar mass and local density, following the same procedure used in the main analysis. Each realisation was also constrained to match the size of the working samples (i.e. 400 and 450 galaxies in voids and filaments, respectively). For each realisation (1000 in total), we recomputed the MZR using the same fitting methodology as in the main analysis, including identical mass cuts, aperture corrections, and stellar mass normalisation. The only difference was that, in this case, we did not apply additional cleaning of the parent sample for potential outliers (e.g. merging systems or SDSS misclassifications). The mean intercepts and slopes of the fits for the MZR and SFR-$M_{\star}$ relations are summarised in the Table~\ref{MC_fits}. This test shows that both the slopes and the relative trends are stable across different realisations of the stratified subsample.}

\begin{table*}
\caption{\label{MC_fits} Mean intercepts ($\alpha$) and slopes ($\beta$) derived from 1000 Monte Carlo fits to the MZR and SFR--$M_{\star}$ relations; uncertainties denote the standard deviations of the fitted parameters.}
\centering
\begin{tabular}{c c c }
\hline
Sub ample & MZR &  SFR-$M_{\star}$ \\
\hline
\hline 
\texttt{Voids\_non-isolated}  & $\alpha$ = 8.47 $\pm$ 0.01, $\beta$ = 0.25 $\pm$ 0.02 &$\alpha$ = -0.69 $\pm$ 0.02, $\beta$ = 0.86 $\pm$ 0.11\\
\texttt{Filament\_non-isolated} &$\alpha$ = 8.47 $\pm$ 0.01, $\beta$ = 0.20 $\pm$ 0.02& $\alpha$ = -0.70 $\pm$ 0.02, $\beta$ = 0.85 $\pm$ 0.14\\
\hline
\end{tabular}\\
\vspace{0.2cm} 
\noindent
\end{table*}

\section{Other metallicity indicators and the effect of aperture correction}\label{appendix1}
In Table \ref{Other calibrators and indicators} we summarise the slopes and intercepts of the MZR that have been measured using various indicators and calibrations on the samples investigated in this study. In particular, we employed:
\begin{itemize}
    \item The O3N2 indicator, which is widely used for estimating gas-phase metallicity in dwarf galaxies. This index, defined as O3N2$\equiv$\,log$\left\{(\rm [OIII]\lambda5007/H\beta)/([NII]\lambda6584/H\alpha)\right\}$ \citep{2004Pettini}, is valid within the range  $-$1$<$O3N2$<$1.9. Some dwarf galaxies in the sample, particularly those at the lower-mass end, fall outside this validity range, which makes the corresponding metallicity estimates from this indicator less reliable. Overall, we observe a larger scatter around the fitted relations when using the O3N2 index. The metallicities derived from this indicator are based on the calibrations of \citet{2004Pettini} and \citet{2013Marino}, both of which yield consistent results to Section \ref{result1}.
    \item The N2 indicator, based on the calibration of \cite{2013Marino}, which represents an updated relation derived from CALIFA and electron temperature ($T_\mathrm{e}$)-based studies, yields results consistent with the \citet{2004Pettini} calibration on which the main discussion and analysis in this paper are based.
    
    \item {The R23 indicator, defined as $\rm R23 = ([O II] \lambda3727 + [O III] \lambda\lambda4959, 5007)/H\beta$, is calibrated using theoretical model grids from \cite{1991McGaugh}, with the parametrisation of \cite{1999Kobulnicky}. This indicator is independent of nitrogen abundance and is therefore not affected by potential variations or scatter in the N/O-O/H relation in the low-mass regime that may arise from environmental differences, among other factors. To break the well-known degeneracy between the upper and lower branches of the R23 indicator, we use the ratio [N II] $\lambda$6583 / [O II] $\lambda$3727, assigning galaxies to the upper branch when $log([N II]/[O II])>-1.2$ \citep[e.g.][]{2008Kewley}. However, the R23 indicator depends critically on the [O II] $\lambda$3727 line, which enters the SDSS spectral window only at redshifts z$\ge$0.027. Even at z $\simeq$ 0.02–0.03, this line lies near the blue end sensitivity limit of the SDSS spectra, where flux calibration and the S/N degrade significantly. As a result, [O II]$\lambda$3727 is not detected with sufficient S/N in all galaxies within our subsamples. In particular, applying the S/N\,$>$\,3 criterion described in Section~\ref{analysis1} reduces the cluster dwarf galaxy sample to only five objects, which is insufficient for robust statistical analysis. For completeness, the MZR derived using the R23 indicator for void and filament galaxies are reported in Table~\ref{Other calibrators and indicators}. These results are broadly consistent with those obtained using the N2 indicator. The main difference is observed in the slope of the MZR for isolated dwarf galaxies in voids, which appears steeper than that of their counterparts in filaments and non-isolated void galaxies. Consistent with our main findings, isolated filament galaxies and non-isolated void galaxies exhibit similar MZR slopes, both steeper than those of non-isolated dwarfs in filaments and clusters. }

\end{itemize}

{The metallicity gradients and star formation morphologies of dwarf galaxies may depend on environment. Therefore, aperture corrections may not be entirely environment-independent and could introduce systematic biases. To assess this effect, we repeat our analysis and construct the MZR and SFR–$M_\star$ relations without applying aperture corrections. In Table~\ref{Aperture_effect_tab}, we compare the slopes derived with and without aperture corrections. We find that the inferred trends across void, filament, and cluster environments remain consistent, indicating that our results are not significantly affected by aperture effects. Furthermore, we verified that the SDSS fiber covering fraction does not differ systematically across environments or between subsamples, indicating that fiber coverage does not bias our results.}

\setlength{\tabcolsep}{20.pt}
\begin{table*}
\caption{\label{Other calibrators and indicators} Best-fit slope ($\beta$) and normalisation ($\alpha$), evaluated at the mean stellar mass $<log(\rm M_{\star})>$, of fitted MZR, using different indicators. }
\centering
\begin{tabular}{c c c c c}
\hline
LSS Sample & $\alpha_{\rm MZR, <log(\rm M_{\star})>}$ &  $\beta_{\rm MZR}$& Indicator & Calibration\\
\hline
\hline
\texttt{Voids} & 8.48 $\pm$ 0.01 &  0.32 $\pm$ 0.03& O3N2 & \cite{2004Pettini}\\
\texttt{Filaments} & 8.48 $\pm$ 0.02 &  0.30 $\pm$ 0.04& O3N2& \cite{2004Pettini}\\
\texttt{Clusters} & 8.44 $\pm$ 0.03 &  0.24 $\pm$ 0.08 & O3N2 &\cite{2004Pettini}\\
\texttt{Voids\_isolated} & 8.48 $\pm$ 0.01 &  0.34 $\pm$ 0.04 & O3N2& \cite{2004Pettini}\\
\texttt{Voids\_non-isolated} & 8.47 $\pm$ 0.01 &  0.30 $\pm$ 0.05& O3N2&\cite{2004Pettini}\\
\texttt{Filament\_isolated} & 8.48 $\pm$ 0.01 &  0.36 $\pm$ 0.05 & O3N2&\cite{2004Pettini}\\
\texttt{Filament\_non-isolated} & 8.48 $\pm$ 0.02 &  0.23 $\pm$ 0.07& O3N2&\cite{2004Pettini}\\
\hline  
\texttt{Voids} & 8.37 $\pm$ 0.01 &  0.21 $\pm$ 0.02& O3N2 & \cite{2013Marino}\\
\texttt{Filaments} & 8.36 $\pm$ 0.01 &  0.19 $\pm$ 0.02& O3N2& \cite{2013Marino}\\
\texttt{Clusters} & 8.45 $\pm$ 0.02 &  0.16 $\pm$ 0.06 & O3N2 &\cite{2013Marino}\\
\texttt{Voids\_isolated} & 8.37 $\pm$ 0.01 &  0.22 $\pm$ 0.02 & O3N2& \cite{2013Marino}\\
\texttt{Voids\_non-isolated} & 8.37 $\pm$ 0.01 &  0.20 $\pm$ 0.04& O3N2&\cite{2013Marino}\\
\texttt{Filament\_isolated} & 8.38 $\pm$ 0.02 &  0.23 $\pm$ 0.03 & O3N2&\cite{2013Marino}\\
\texttt{Filament\_non-isolated} & 8.36 $\pm$ 0.03 &  0.15 $\pm$ 0.04& O3N2&\cite{2013Marino}\\ 
\hline  
\texttt{Voids} & 8.41 $\pm$ 0.01 &  0.21 $\pm$ 0.02& N2 & \cite{2013Marino}\\
\texttt{Filaments} & 8.41 $\pm$ 0.01 &  0.19 $\pm$ 0.02& N2& \cite{2013Marino}\\
\texttt{Clusters} & 8.46 $\pm$ 0.02 &  0.14 $\pm$ 0.06 & N2 &\cite{2013Marino}\\
\texttt{Voids\_isolated} & 8.42 $\pm$ 0.01 &  0.22 $\pm$ 0.02 & N2& \cite{2013Marino}\\
\texttt{Voids\_non-isolated} & 8.41 $\pm$ 0.02 &  0.21 $\pm$ 0.03& N2&\cite{2013Marino}\\
\texttt{Filament\_isolated} & 8.41 $\pm$ 0.01 &  0.23 $\pm$ 0.03 & N2&\cite{2013Marino}\\
\texttt{Filament\_non-isolated} & 8.40 $\pm$ 0.01 &  0.15 $\pm$ 0.03& N2&\cite{2013Marino}\\
\hline  
\texttt{Voids} & 8.90 $\pm$ 0.02 &  0.24 $\pm$ 0.02& R23 & \cite{1979Pagel}\\
\texttt{Filaments} & 8.90 $\pm$ 0.02 &  0.20 $\pm$ 0.03& R23 & \cite{1979Pagel}\\
\texttt{Clusters} & -- &  -- & R23 & \cite{1979Pagel}\\
\texttt{Voids\_isolated} & 8.91 $\pm$ 0.01 &  0.27 $\pm$ 0.02 & R23& \cite{1979Pagel}\\
\texttt{Voids\_non-isolated} & 8.92 $\pm$ 0.01 &  0.22 $\pm$ 0.03 & R23 &\cite{1979Pagel}\\
\texttt{Filament\_isolated} & 8.90 $\pm$ 0.01 &  0.22 $\pm$ 0.07 & R23&\cite{1979Pagel}\\
\texttt{Filament\_non-isolated} & 8.92 $\pm$ 0.02 &  0.20 $\pm$ 0.08 &  R23&\cite{1979Pagel}\\
\hline 
\hline
\end{tabular}\\
\vspace{0.2cm} 
\noindent
\end{table*}

\setlength{\tabcolsep}{20.pt}
\begin{table*}
\caption{\label{Aperture_effect_tab}Best-fit slope ($\beta$) and normalisation ($\alpha$), evaluated at the mean stellar mass $\langle \log(M_\star) \rangle$, for the fitted MZR. Values obtained with and without aperture corrections are compared. All the values reported are based on the N2 indicator. Those marked with a star are not aperture-corrected.}
\centering
\begin{tabular}{c c c c c}
\hline
LSS Sample & $\alpha_{\rm MZR, <log(\rm M_{\star})>}$ &  $\beta_{\rm MZR}$& $\alpha_{\rm MZR, <log(\rm M_{\star})>}^{\star}$  &  $\beta_{\rm MZR}^{\star}$ \\
\hline
\hline 
\texttt{Voids} &8.45$\pm$ 0.01 & 0.28$\pm$0.03 & 8.52 $\pm$ 0.01 &  0.27 $\pm$ 0.02\\
\texttt{Filaments} &8.45$\pm$0.01&0.25$\pm$0.04& 8.52 $\pm$ 0.01 &  0.23 $\pm$ 0.03\\
\texttt{Clusters} &8.55$\pm$0.02&0.16$\pm$0.09& 8.58 $\pm$ 0.01 &  0.19 $\pm$ 0.08 \\
\texttt{Voids\_isolated} &8.48$\pm$0.01&0.30$\pm$0.04& 8.52 $\pm$ 0.01 &  0.27 $\pm$ 0.03 \\
\texttt{Voids\_non-isolated} &8.49$\pm$0.02&0.24$\pm$0.08& 8.53 $\pm$ 0.02 &  0.26 $\pm$ 0.04\\
\texttt{Filament\_isolated} &8.48$\pm$0.01&0.28$\pm$0.06& 8.52 $\pm$ 0.01 &  0.28 $\pm$ 0.04 \\
\texttt{Filament\_non-isolated} &8.49$\pm$0.01&0.19$\pm$0.07& 8.51 $\pm$ 0.02 &  0.17 $\pm$ 0.05\\

\hline 
\hline
LSS Sample & $\alpha_{\rm SFR-M_{\star}, <log(M_{\star})>}$ &  $\beta_{\rm SFR-M_{\star}}$& $\alpha_{\rm SFR-M_{\star}, <log(M_{\star})>}^{\star}$  &  $\beta_{\rm SFR-M_{\star}}^{\star}$ \\
\hline
\hline 

\texttt{Voids} & -0.58$\pm$0.01 &  0.97$\pm$0.08& -1.04$\pm$ 0.02 & 0.94$\pm$0.12\\
\texttt{Filaments} & -0.55$\pm$0.01& 0.86$\pm$0.08 & -0.93$\pm$0.02& 0.68$\pm$0.11\\
\texttt{Clusters} & -0.56$\pm$0.06& 0.85$\pm$0.30& -0.87$\pm$0.06& 0.67$\pm$0.33\\
\texttt{Voids\_isolated}  & -0.57$\pm$0.02& 0.98$\pm$0.09 & -1.05$\pm$0.03 & 0.93 $\pm$0.14\\
\texttt{Voids\_non-isolated}  & -0.60$\pm$0.02& 0.93$\pm$0.14& -1.00$\pm$0.04 & 1.00$\pm$0.21\\
\texttt{Filament\_isolated}  & -0.52$\pm$0.02& 0.79$\pm$0.10& -0.94$\pm$0.02&0.57$\pm$0.14\\
\texttt{Filament\_non-isolated} & -0.60$\pm$0.02& 0.94$\pm$0.13& -0.91$\pm$0.03 & 0.89$\pm$0.17\\

\hline
\end{tabular}\\
\vspace{0.2cm} 
\noindent
\end{table*}

\section{Linear fits}\label{Appendix11}
To estimate the best-fit parameters and their uncertainties for the MZR and SFR–$M_{\star}$ relation, we used a bootstrap resampling method combined with linear regression. {Specifically, we generated 1000 bootstrap samples, each containing the same number of galaxies as the original dataset, drawn with replacement.} Each bootstrap sample preserved the original sample size but allowed repeated data points. A robust linear fit was then performed on each resampled dataset, yielding distributions of slope and intercept values. At each stellar mass bin, we derived the 2.5th and 97.5th percentiles of the predicted metallicity values across all bootstrap realisations, thereby constructing a 95\% confidence interval around the fitted line. This approach has the advantage of making no assumptions about the underlying error distribution and provides uncertainty estimates that naturally reflect the scatter and possible outliers of the data, making it particularly suitable for heterogeneous samples such as those studied here. Fits with other available routines, such as lmfit\footnote{\url{https://lmfit.github.io/lmfit-py/}} and LINmix\footnote{\url{https://linmix.readthedocs.io/en/latest/src/linmix.html}} yielded similar results, albeit with larger uncertainties.

\section{MZR and SFR-$M_{\star}$ relation based on statistically larger samples}\label{AppendixAB}
\begin{figure*}
\centering
\includegraphics[width=0.9\textwidth]{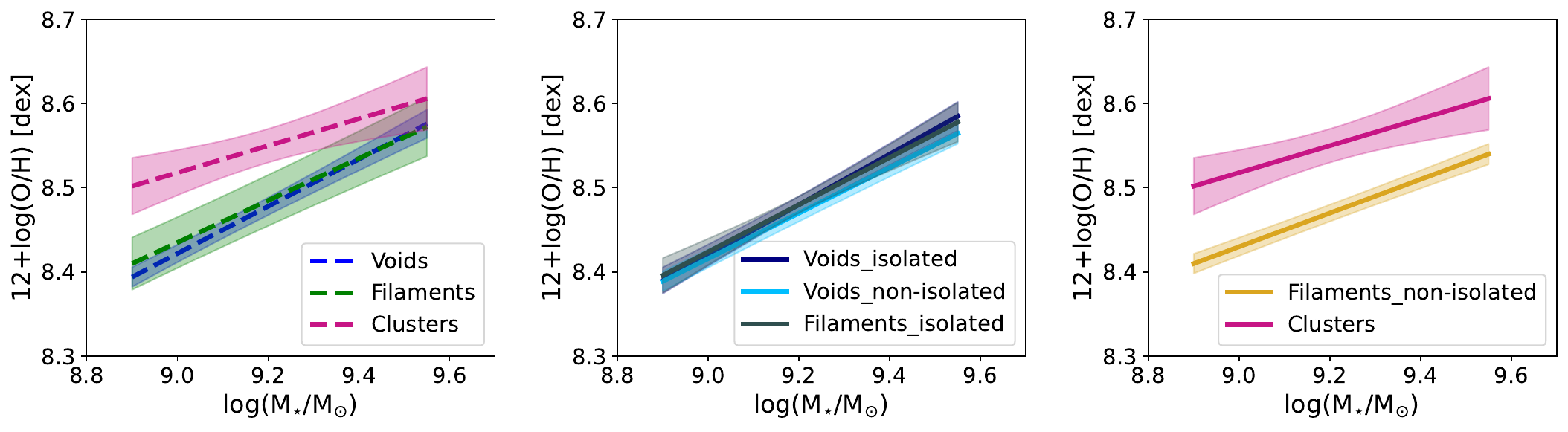}
\includegraphics[width=0.9\textwidth]{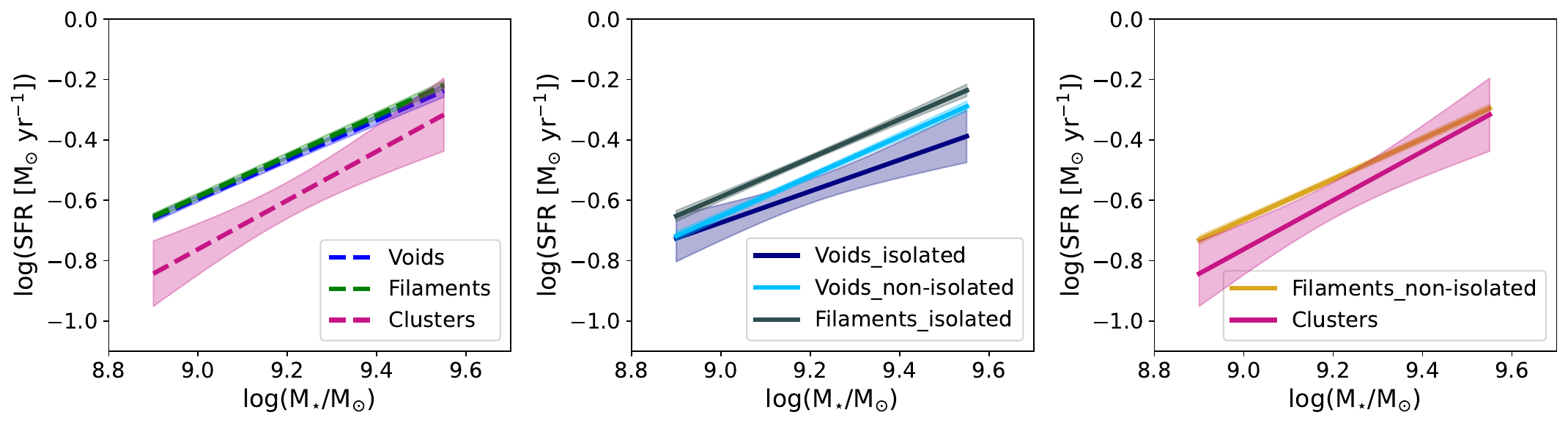}
\includegraphics[width=0.8\textwidth]{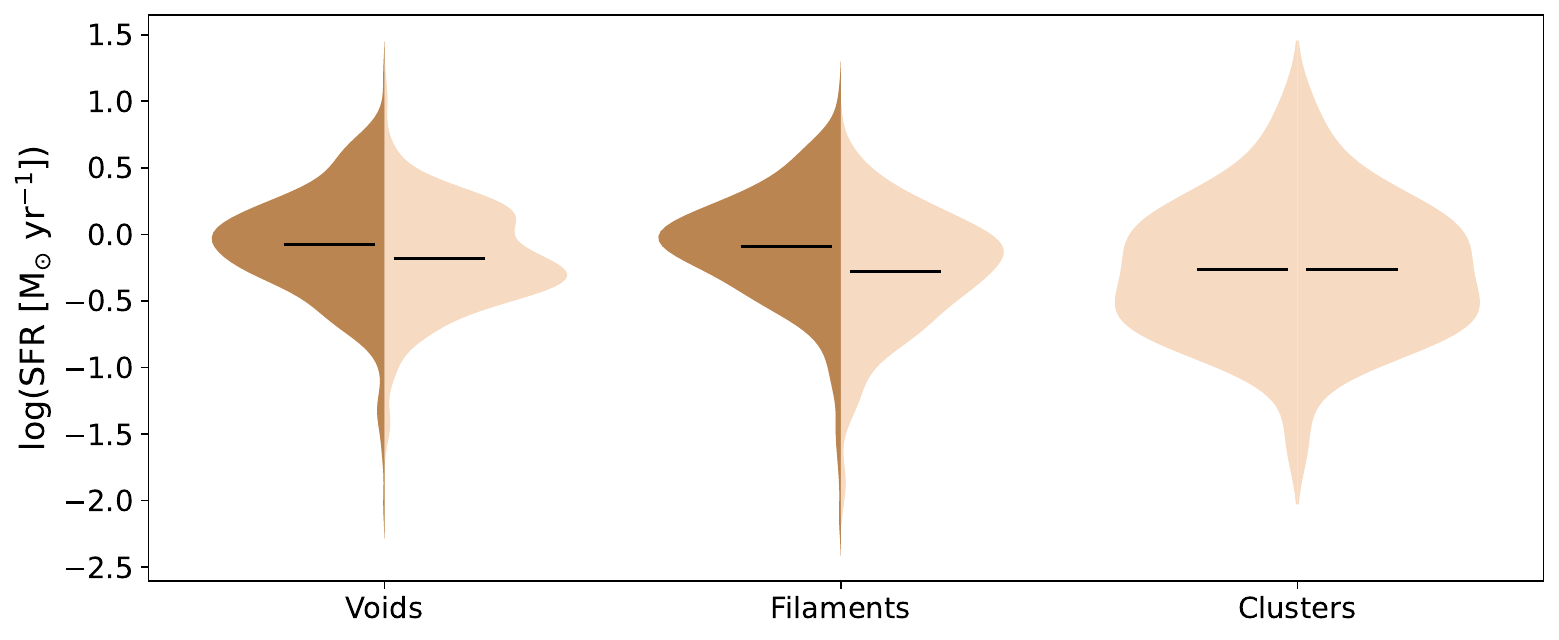}
\caption{{\textit{{Top and middle panels: }}Same as Fig.~\ref{Fig4_slopes} but for larger comparison samples, containing 1267 and 4090 star-forming non-isolated dwarf galaxies in voids and filaments, respectively. \textit{Bottom panel:} Comparing the SFR values distribution among samples. The dark shades represent isolated populations in voids and filaments, while the lighter shades indicate non-isolated dwarf galaxies. The horizontal black lines indicate the medians of the distributions.}}
\label{SFR_complete_samples_dwarfs}
\end{figure*}

{We repeated the analysis on the {MZR and} SFR–$M_{\star}$ relation, considering the full comparison samples (i.e. 1267 and 4090 star-forming non-isolated dwarf galaxies in voids and filaments, respectively) to check for possible impacts of limited statistics on the reported relations in Section~\ref{result2}.  By definition, the isolated subsamples cannot be expanded, as they already include all isolated dwarf galaxies identified in the dataset. Likewise, the cluster sample cannot be statistically improved, since the selection criteria were applied to the entire cluster dataset rather than to selected subsamples. For this test, we did not visually inspect galaxies (and thus allowed for potential contamination from interacting or merging systems), and we imposed no requirements on the S/N of individual emission lines. Instead, we used the SDSS MPA–JHU value-added catalogueue \citep{2004Brinchmann}. {The metallicity and SFR values used in this test are aperture corrected and defined using the same indicator and relation as discussed in Section~\ref{analysis2}}. Linear fits are obtained similarly to Section~\ref{result1}, and the results of the fits for the MZR and SFR–$M_{\star}$ relation are reported in Table~\ref{Fits_entire sample}. {We show the resulting MZR and SFR–$M_{\star}$ relations based on these large samples in Fig.~\ref{SFR_complete_samples_dwarfs}.}

With improved statistics for the non-isolated dwarf galaxies in voids and filaments, the results in the left-hand panel of Fig.~\ref{SFR_complete_samples_dwarfs} show that dwarf galaxies in voids and filaments follow {the MZR and} SFR–$M_{\star}$ relation that lies systematically above the cluster relation, as expected. In the middle panel of Fig.~\ref{SFR_complete_samples_dwarfs} we show results for star-forming dwarf galaxies in voids and isolated ones in filaments. As observed for the MZR, star-forming dwarf galaxies in voids exhibit nearly identical relations regardless of their local environment, once the scatter among isolated void dwarfs is taken into account. With the larger sample, we still see that isolated filament dwarfs have higher SFRs {and similar MZR slope} compared to non-isolated dwarfs in both voids and filaments, mainly due to their different underlying stellar mass distribution. Moreover, we found that non-isolated dwarf galaxies in voids have similar SFRs {and MZR slope} to their counterparts in filaments. These results are consistent with the discussion in Section~\ref{result2}. The expanded statistics do not alter the MZR trends reported in Section~\ref{result1} (see Table~\ref{Fits_entire sample}).

\setlength{\tabcolsep}{19.pt}
\begin{table*}
\caption{\label{Fits_entire sample} Best-fit slope ($\beta$) and normalisation ($\alpha$), evaluated at the mean stellar mass $<$$\rm log(M_{\star})$$>$, describing the MZR and the SFR-$M_{\star}$ relations for statistically larger comparison samples. }
\centering
\begin{tabular}{c c c c c}
\hline
LSS Sample & $\alpha_{\rm MZR, <log(M_{\star})>}$ &  $\beta_{\rm MZR}$& $\alpha_{\rm SFR-M_{\star}, <log(M_{\star})>}$ &$\beta_{\rm SFR-M_{\star}}$\\
\hline
\hline
\texttt{Voids} & 8.45 $\pm$ 0.01 &  0.28 $\pm$ 0.03& -0.53 $\pm$ 0.01 & 0.65 $\pm$ 0.04\\
\texttt{Filaments} & 8.46 $\pm$ 0.03 &  0.25 $\pm$ 0.04& -0.52 $\pm$ 0.01& 0.67 $\pm$ 0.02\\
\texttt{Clusters} & 8.55 $\pm$ 0.02 &  0.16 $\pm$ 0.09& -0.60 $\pm$ 0.06 &0.81 $\pm$ 0.30\\
\hline 
\texttt{Voids\_isolated} & 8.48 $\pm$ 0.01 &  0.30 $\pm$ 0.04& -0.57 $\pm$ 0.04& 0.52
$\pm$0.22\\
\texttt{Voids\_non-isolated} & 8.47 $\pm$ 0.01 &  0.27 $\pm$ 0.02& -0.52 $\pm$ 0.01&  0.66 $\pm$ 0.04\\
\texttt{Filament\_isolated} & 8.48 $\pm$ 0.01 &  0.28 $\pm$ 0.06& -0.46 $\pm$ 0.01 &0.64 $\pm$ 0.05\\
\texttt{Filament\_non-isolated} & 8.47 $\pm$ 0.01 &  0.20 $\pm$ 0.02& -0.53 $\pm$ 0.01 & 0.67 $\pm$ 0.03\\

\hline
\end{tabular}\\
\vspace{0.2cm} 
\noindent

\end{table*}

\section{The H\,\textsc{i} mass}\label{AppendixB}
Neutral hydrogen (H\,\textsc{i}) is the primary reservoir of cold gas from which molecular clouds form, ultimately fueling ongoing star formation. In dense environments, RPS can strip both the halo gas reservoir and the H\,\textsc{i} disk, while processes such as thermal evaporation and strangulation suppress the accretion of pristine gas \citep[e.g.][]{2001Solanes, 2014Boselli}. Observational evidence of these mechanisms includes the increasing fraction of H\,\textsc{i}-deficient dwarf galaxies towards cluster centres \citep{1984Haynes}, as well as the presence of galaxies with long, one-sided H\,\textsc{i} tails \citep{2009Chung} and truncated H\,\textsc{i} disks \citep{2007Levy}. Similar trends, although investigated in fewer studies, have also been reported in low-density environments. For example, \cite{1997Huchtmeier} reported that dwarf galaxies located nearer to void centres have higher H\,\textsc{i} than those close to void walls. 

We quantified H\,\textsc{i} masses using the Arecibo Legacy Fast ALFA Survey \citep[ALFALFA;][]{2018Haynes} catalogue of extragalactic H\,\textsc{i} sources. A cone search with a 0.01 degree radius was used to cross-match the three environmental samples with ALFALFA detections, yielding 71 isolated and 471 non-isolated dwarf galaxies in voids, and 35 isolated and 752 non-isolated dwarf galaxies in filaments. We could find reported H\,\textsc{i} flux only for three dwarf galaxies in the cluster sample. Due to the very limited number of data points in the latter, which precludes robust analysis, we do not discuss them. 
 
\begin{figure*}
\centering
\includegraphics[width=0.9\textwidth]{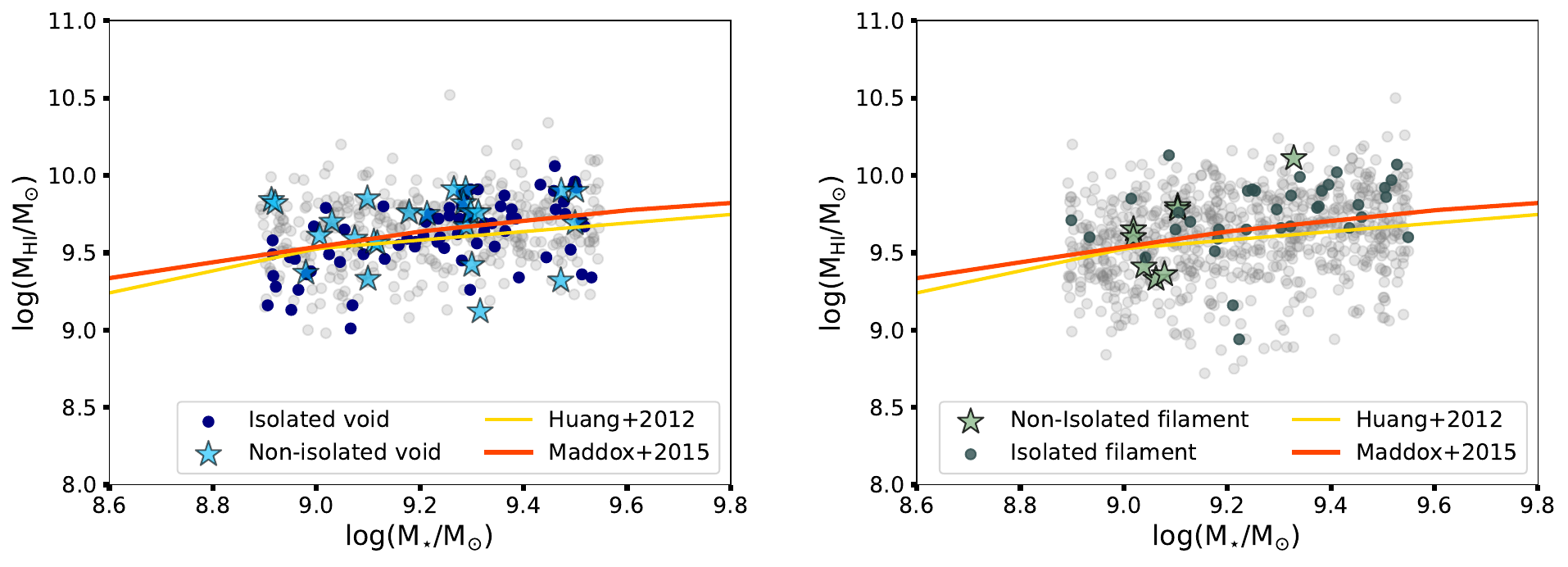}
\caption{The M$_{\rm HI}$–M$_{\star}$ scaling relation for dwarf galaxies in voids and filaments detected in H\,\textsc{i} is shown in the left- and right-hand panels, respectively. Grey data points are non-isolated dwarf galaxies in voids and filaments with H\,\textsc{i} detection. Isolated systems are plotted as dark blue (voids) and green (filaments) circles, while non-isolated systems studied here are shown as star symbols in light blue (voids) and light green (filaments). For reference, the relations reported by \citet{2012Huang} and \citet{2015Maddox} are overplotted as yellow and red lines, respectively. } 
\label{HI_dwarfs}
\end{figure*}

In Fig. \ref{HI_dwarfs} we present the H\,\textsc{i} mass (M$_{\rm HI}$) as a function of stellar mass for dwarf galaxies in voids (left-hand panel) and filaments (right-hand panel). In these plots, we are showing isolated dwarfs with H\,\textsc{i} detected in voids and filaments in dark blue and green, respectively. Grey data points are all the non-isolated dwarf galaxies from the original samples with H\,\textsc{i} detections, and data points in light-blue and green are non-isolated dwarfs in voids and filaments for which their MZR and SFR-$M{_\star}$ are investigated here. The H\,\textsc{i} masses are estimated using distances from \cite{2018Haynes}. Overplotted in each panel are two reference relations illustrating the typical HI–stellar mass trends. The yellow line shows the relation from \cite{2012Huang}, derived for 9417 SDSS galaxies using stellar masses from SDSS, and the red line corresponds to the relation from \cite{2015Maddox} based on 9153 SDSS galaxies at $z$=0, utilising ALFALFA H\,\textsc{i} data, as in this study.

We find no significant difference in the H\,\textsc{i} content between isolated and non-isolated dwarf galaxies in voids, at least among those with H\,\textsc{i} detections. These galaxies exhibit similar ranges in metallicity, SFR, and M$_{\rm HI}$. The isolated dwarfs with H\,\textsc{i} detections in voids also show comparable M$_{\rm HI}$ ranges to those in filaments. However, isolated dwarfs in filaments tend to lie above the M$_{\rm HI}$ relations reported by previous studies from \cite{2012Huang} and \cite{2015Maddox}, which could be attributed to better accessibility of gas inflow within filamentary structures and intrinsic scatter in the true distribution which is likely underrepresented by the sample of isolated filament galaxies. 

The number of dwarf galaxies with detected H\,\textsc{i} in the sample is significantly smaller than the total number of galaxies considered. This is partly attributable to the detection limit of the ALFALFA survey. The limiting H\,\textsc{i} mass can be estimated using the standard relation from \cite{2018Haynes}:
\begin{equation}
    M_{\rm HI} = 2.356\times10^{5}D_{L}^{2}S_{int}
\end{equation}

\setlength{\parindent}{0pt}where D$_{L}$ is the luminosity distance in Mpc and S$_{int}$ is the integrated H\,\textsc{i} flux in Jy km/s. Assuming a survey sensitivity of 0.72 Jy km/s for W${50}$ = 200 km/s at a  5$\sigma$ detection threshold \citep{2018Haynes}, we estimated the ALFALFA detection limit to be $M_{HI}$ = 3$\times$10$^{8}$ M$_{\odot}$ and 8$\times$10$^{9}$ M$_{\odot}$ at z = 0.01 and 0.05, respectively. While the detection limit depends on line width, recalculations for a narrower, more representative W${50}$ of dwarfs (100 km/s) yield only modest changes towards lower values. Therefore, a likely explanation for the small number of H\,\textsc{i} detections in void and filament dwarf galaxies is that their H\,\textsc{i} content lies below ALFALFA’s sensitivity at their respective distances. Another possible factor contributing to the limited H\,\textsc{i} detections in the sample is the partial sky coverage of the ALFALFA survey relative to the SDSS footprint \citep{2018Haynes}.

Comparing the numbers of isolated and non-isolated dwarf galaxies with (H\,\textsc{i}) detections reveals a possible systematic difference between the two populations, suggesting that non-isolated dwarf galaxies in voids and filaments may have better access to neutral gas. However, this interpretation must be treated with caution. Within the redshift range considered in this study, the 3.5 arcmin angular resolution of the ALFALFA beam corresponds to a projected diameter of approximately 0.05 to 0.2 Mpc. For isolated dwarf galaxies, defined here as having no neighbouring galaxies within 1.5 Mpc, this ensures that the detected (H\,\textsc{i}) is very likely associated with the target galaxy itself. In contrast, for non-isolated dwarfs, we cannot be sure that the detected (H\,\textsc{i}) originates from the dwarf galaxy rather than from a nearby companion within the beam. We acknowledge that these limitations, combined with the survey’s inherent bias towards detecting the most gas-rich and least obscured systems, introduce non-negligible biases into any comparison of environments, particularly for non-isolated dwarf galaxies.}
\end{appendix}

\end{document}